\documentclass[manuscript,twocolappendix,numberedappendix]{emulateapj}

\usepackage{epsfig}
\usepackage{epstopdf}
\usepackage{graphicx}
\usepackage{color,soul}
\usepackage{rotating}
\usepackage{soul}
\usepackage{fancyvrb}
\usepackage{lineno}
\usepackage{amssymb}
\usepackage{amsmath}
\usepackage{xspace}
\usepackage{hyperref}
\usepackage{url}

\begin{document}

\title{ Models and Simulations for the 
    Photometric LSST Astronomical \\ Time Series Classification Challenge (\acro) }
   

\def\andname{}

\author{
R.~Kessler\altaffilmark{1,2},
G.~Narayan\altaffilmark{3},
A.~Avelino\altaffilmark{4},
E.~Bachelet\altaffilmark{5},
R.~Biswas\altaffilmark{6},
P.~J.~Brown\altaffilmark{7},
D.~F.~Chernoff\altaffilmark{8},
A.~J.~Connolly\altaffilmark{9},
M.~Dai\altaffilmark{10},
S.~Daniel\altaffilmark{9},
R.~Di~Stefano\altaffilmark{4},
M.~R.~Drout\altaffilmark{11},
L.~Galbany\altaffilmark{12},
S.~Gonz\'alez-Gait\'an\altaffilmark{13},
M.~L.~Graham\altaffilmark{9},
R.~Hlo\v{z}ek\altaffilmark{11,14},
E.~E.~O.~Ishida\altaffilmark{15},
J.~Guillochon\altaffilmark{4},
S.~W.~Jha\altaffilmark{10},
D.~O.~Jones\altaffilmark{16},
K.~S.~Mandel\altaffilmark{17,18},
D.~Muthukrishna\altaffilmark{17},
A.~O'Grady\altaffilmark{11,14},
C.~M.~Peters\altaffilmark{14},
J.~R.~Pierel\altaffilmark{19},
K.~A.~Ponder\altaffilmark{20},
A.~Pr\v{s}a\altaffilmark{21},
S.~Rodney\altaffilmark{19},
V.~A.~Villar\altaffilmark{4}
\\ \vspace{0.1cm} (The LSST Dark Energy Science Collaboration and the \\ 
    Transient and Variable Stars Science Collaboration) \\
}

\affil{$^{1}$ Kavli Institute for Cosmological Physics, University of Chicago, Chicago, IL 60637, USA}
\affil{$^{2}$ Department of Astronomy and Astrophysics, University of Chicago, Chicago, IL 60637, USA}
\affil{$^{3}$ Space Telescope Science Institute, 3700 San Martin Drive, Baltimore, MD 21218, USA}
\affil{$^{4}$ Harvard-Smithsonian Center for Astrophysics, 60 Garden St., Cambridge, MA 02138, USA}
\affil{$^{5}$ Las Cumbres Observatory, 6740 Cortona Drive, Suite 102, Goleta, CA 93117, USA}
\affil{$^{6}$ The Oskar Klein Centre for Cosmoparticle Physics, Department of Physics, Stockholm University, AlbaNova, Stockholm SE-10691, Sweden}
\affil{$^{7}$ George P. and Cynthia Woods Mitchell Institute for Fundamental Physics \& Astronomy, Texas A\&M University, College Station, TX 77843, USA}
\affil{$^{8}$ Department of Astronomy, Cornell University, Ithaca, NY 14853, USA}
\affil{$^{9}$ Department of Astronomy, University of Washington, Box 351580, U.W., Seattle, WA 98195, USA}
\affil{$^{10}$ Department of Physics and Astronomy, Rutgers, The State University of New Jersey, 136 Frelinghuysen Road, Piscataway, NJ 08854, USA}
\affil{$^{11}$ Department of Astronomy \& Astrophysics, University of Toronto, 50 St. George St., Toronto, ON, M5S3H4, Canada}
\affil{$^{12}$ PITT PACC, Department of Physics and Astronomy, University of Pittsburgh, Pittsburgh, PA 15260, USA}
\affil{$^{13}$ CENTRA, Instituto Superior T\'ecnico, Universidade de Lisboa, Av. Rovisco Pais 1, 1049-001 Lisboa, Portugal}
\affil{$^{14}$ Dunlap Institute for Astronomy and Astrophysics, University of Toronto, Toronto, ON M5S 3H4, Canada}
\affil{$^{15}$ Universit\'e Clermont Auvergne, CNRS/IN2P3, LPC, F-63000 Clermont-Ferrand, France}
\affil{$^{16}$ Department of Astronomy and Astrophysics, University of California, Santa Cruz, CA 92064,USA}
\affil{$^{17}$ Institute of Astronomy and Kavli Institute for Cosmology, Madingley Road, Cambridge, CB3 0HA, UK}
\affil{$^{18}$ Statistical Laboratory, DPMMS, University of Cambridge, Wilberforce Road, Cambridge, CB3 0WB, UK}
\affil{$^{19}$ Department of Physics and Astronomy, University of South Carolina, 712 Main St., Columbia, SC 29208, USA}
\affil{$^{20}$ Berkeley Center for Cosmological Physics, Campbell Hall 341, University of California Berkeley, Berkeley, CA 94720,USA}
\affil{$^{21}$ Villanova University, Department of Astrophysics and Planetary Science, 800 E Lancaster Avenue, Villanova PA 19085, USA}


\newcommand{\URLMODELCALL}{\url{https://plasticcblog.files.wordpress.com/2017/05/noi.pdf}}
\newcommand{\URLSNANA}{\url{http://snana.uchicago.edu}}
\newcommand{\URLMOSFIT}{\url{https://github.com/guillochon/MOSFiT}}
\newcommand{\URLKAGGLE}{\url{https://www.kaggle.com/c/PLAsTiCC-2018}}
\newcommand{\URLMODELS}{\url{https://doi.org/10.5281/zenodo.2612896}}
\newcommand{\URLOPSIM}{\url{http://opsim.lsst.org/runs/minion_1016/data/minion_1016_sqlite.db.gz}}
\newcommand{\URLSextractor}{\url{https://www.astromatic.net/software/sextractor}}
\newcommand{\URLKEYNUM}{\url{https://www.lsst.org/scientists/keynumbers}}
\newcommand{\URLHEALPIX}{\url{http://healpix.sourceforge.net}}
 
\newcommand{\headerOverview}{ Overview of }
\newcommand{\headerDetails}{Technical Details for}
\newcommand{\textEventGenSED}[1]{For event generation, each of the {#1} SED time series was given equal weight.}
\newcommand{\textEventGenLC}[1]{For event generation, each of the {#1} model light curves was given equal weight.}

\newcommand{\HOSTLIB}{{\tt HOSTLIB}}
\newcommand{\SALTII}{{\sc SALT-II}}
\newcommand{\SDSS}{SDSS-II}
\newcommand{\PS}{Pan-STARRS1}
\newcommand{\Diff}{{\tt DiffImg}}
\newcommand{\autoScan}{{\tt autoScan}}
\newcommand{\DESSN}{DES-SN}
\newcommand{\wCDM}{$w$CDM}
\newcommand{\lowz}{low-$z$}
\newcommand{\mosfit}{{\tt MOSFiT}\xspace}
\newcommand{\bands}{$ugrizy$}

\newcommand{\PLASTICC}{Photometric LSST Astronomical Time Series Classification Challenge}
\newcommand{\acro}{{\tt PLAsTiCC}}
\newcommand{\SNANA}{{\tt SNANA}}
\newcommand{\LSST}{Large Synoptic Survey Telescope}
\newcommand{\DES}{Dark Energy Survey}
\newcommand{\OPSIM}{OpSim}

\newcommand{\Spec}{Spectroscopic}
\newcommand{\spec}{spectroscopic}
\newcommand{\specy}{spectroscopically}
\newcommand{\obs}{observation}
\newcommand{\obss}{observations}
\newcommand{\eff}{efficiency}
\newcommand{\ineff}{inefficiency}
\newcommand{\effs}{efficiencies}
\newcommand{\unc}{uncertainty}
\newcommand{\uncs}{uncertainties}
\newcommand{\Prob}{Probability}
\newcommand{\prob}{probability}
\newcommand{\probs}{probabilities}
\newcommand{\biasCor}{BiasCor}
\newcommand{\con}{contamination}
\newcommand{\Con}{Contamination}
\newcommand{\exgal}{extragalactic}
\newcommand{\Exgal}{Extragalactic}
\newcommand{\Gal}{Galactic}

\newcommand{\MJD}{{\bf\tt MJD}}
\newcommand{\IDEXPT}{{\bf\tt IDEXPT}}
\newcommand{\FLT}{{\bf\tt FLT}}
\newcommand{\GAIN}{{\bf\tt GAIN}}
\newcommand{\RDNOISE}{{\bf\tt RDNOISE}}
\newcommand{\SKYSIG}{{\bf\tt SKYSIG}}
\newcommand{\PSF}{{\bf\tt PSF}}
\newcommand{\ZPTADU}{{\bf\tt ZPTADU}}
\newcommand{\ZPTpe}{{\bf\tt ZPTpe}}
\newcommand{\NEA}{{\bf\tt NEA}}
\newcommand{\PEAKMJD}{{\bf\tt PEAKMJD}}

\newcommand{\ZPHOT}{{\bf\tt ZPHOT}}
\newcommand{\ZPHOTERR}{{\bf\tt ZPHOTERR}}

\newcommand{\Sersic}{S\'ersic}

\newcommand{\LCDM}{\Lambda{\rm CDM}}
\newcommand{\OL}{\Omega_{\Lambda}}
\newcommand{\OM}{\Omega_{\rm M}}
\newcommand{\zcmb}{z_{\rm cmb,true}}
\newcommand{\zcmbObs}{z_{\rm cmb,obs}}
\newcommand{\zhel}{z_{\rm hel,true}}
\newcommand{\zhelObs}{z_{\rm hel,obs}}
\newcommand{\mutrue}{\mu_{\rm true}}
\newcommand{\SNRtrue}{{\rm S/N}_{\rm true}}

\newcommand{\RV}{R_V}
\newcommand{\Msun}{M_{\sun}}
\newcommand\omicron{o}
\newcommand{\DF}{\Delta F} 

\newcommand{\muwCDM}{\mu_{\rm wCDM}}
\newcommand{\muLens}{\mu_{\rm lens}}
\newcommand{\vpec}{v_{\rm pec}}
\newcommand{\vcor}{v_{\rm pec,cor}}
\newcommand{\verr}{v_{\rm pec,err}}
\newcommand{\sigmavpec}{\sigma_{\rm vpec}}
\newcommand{\sigz}{\sigma_{z}}
\newcommand{\dzNoise}{\delta z_{\rm noise}}
\newcommand{\RateUnit}{{\rm yr}^{-1}{\rm Mpc}^{-3}}
\newcommand{\RateUnitGpc}{{\rm yr}^{-1}{\rm Gpc}^{-3}}

\newcommand{\mtrue}{m_{\rm true}}
\newcommand{\Ftrue}{F_{\rm true}}
\newcommand{\FSMP}{F_{\rm SMP}}
\newcommand{\FSIM}{F_{\rm sim}}
\newcommand{\sigF}{\sigma_{F}}
\newcommand{\sigFP}{\sigma_{F}^{\prime}}
\newcommand{\sigFtrue}{\sigma_{\rm Ftrue}}
\newcommand{\sigHOST}{\sigma_{\rm host}}
\newcommand{\sigZPT}{\sigma_{\rm ZPT}}
\newcommand{\sigZERO}{\hat{\sigma}_{0}}
\newcommand{\ERRSCALESIM}{\hat{S}_{\rm sim}}
\newcommand{\ERRSCALESMP}{\hat{S}_{\rm SMP}}
\newcommand{\zSN}{z_{\rm SN}}
\newcommand{\zHOST}{z_{\rm HOST}}
\newcommand{\zSpec}{z_{\rm spec}}
\newcommand{\EFFspec}{E_{\rm spec}}  
\newcommand{\ipeak}{i_{\rm peak}}
\newcommand{\Bpeak}{B_{\rm peak}}
\newcommand{\Trest}{{\rm T}_{\rm rest}}
\newcommand{\Dmb}{\Delta m_b}   

\newcommand{\sigplus}{\sigma_{+}}
\newcommand{\sigminus}{\sigma_{-}}
\newcommand{\cpeak}{\bar{c}}

\newcommand{\SFR}{\rm SFR}
\newcommand{\Ovec}{\vec{\cal O}_{\sigma}}
\newcommand{\mSB}{m_{\rm SB}}
\newcommand{\sigPIPE}{\sigma_{\rm PhotPipe}}
\newcommand{\sigSMP}{\sigma_{\rm SMP}}


\newcommand{\NSIMGEN}{1.1{\times}10^8}
\newcommand{\NSIMTRAIN}{3333}
\newcommand{\NSIMTEST}{3.5{\times}10^6}
\newcommand{\NSIMOBS}{453}  
\newcommand{\NCLASSTOT}{18}
\newcommand{\NCLASSOBS}{14}
\newcommand{\NCLASSOTHER}{4}
\newcommand{\NEXTRAGALMODEL}{15}
\newcommand{\NMOONLSST}{35}
\newcommand{\NMOONZTF}{170}
\newcommand{\NSKYLOCDDF}{133}  
\newcommand{\NSKYLOCWFD}{50,000}  
\newcommand{\NOBSAVGDDF}{330}  
\newcommand{\NOBSAVGWFD}{130}  

\newcommand{\DATECALL}{2017 May 1}   
\newcommand{\DATESTART}{2018 September 28}   
\newcommand{\DATESTOP}{2018 December 17}    
\newcommand{\NDAY}{78}                  
\newcommand{\NTEAM}{1,094}
\newcommand{\NPARTICIPANT}{1,384}
\newcommand{\NENTRY}{22,895}
\newcommand{\NMISTAKE}{6}   


\newcommand{\NMODELTRAIN}{14} 
\newcommand{\NMODELOTHER}{5} 

\newcommand{\Nevent}{N_{\rm event}}

\newcommand{\NtemplateSNIIT}{20}  
\newcommand{\NtemplateSNIINMF}{384}     
\newcommand{\NtemplateSNIIn}{839}     
\newcommand{\NtemplateSNIbcT}{13}  
\newcommand{\NtemplateSNIbcMOSFIT}{699}     
\newcommand{\NtemplateSNIa}{-1}     
\newcommand{\Ntemplatebg}{35}     
\newcommand{\NtemplateSNIax}{1,001}     
\newcommand{\NtemplateKN}{329}     
\newcommand{\NtemplateSLSN}{960}     
\newcommand{\NtemplatePISN}{1,000}    
\newcommand{\NtemplateILOT}{385}    
\newcommand{\NtemplateCART}{225}    
\newcommand{\NtemplateTDE}{745}     
\newcommand{\NtemplateAGN}{5490}     
\newcommand{\NtemplateRRL}{49,130}     
\newcommand{\NtemplateMdwarf}{1,846}     
\newcommand{\NtemplatePHOEBE}{500}     
\newcommand{\NtemplateMIRA}{3,000}     
\newcommand{\NtemplateuLensBinary}{11,860}    
\newcommand{\NtemplateuLensSingle}{19,360} 
\newcommand{\NtemplateuLensSinglePylima}{7,500} 
\newcommand{\NtemplateuLensSingleGenlens}{11,860}     
\newcommand{\NtemplateuLensSTRING}{846}    

\newcommand{\modelNameTOTAL}{TOTAL}     
\newcommand{\modelNameSNII}{SNII}     
\newcommand{\modelNameSNIbc}{SNIbc}     
\newcommand{\modelNameSNIa}{SNIa}     
\newcommand{\modelNamebg}{SNIa-91bg}     
\newcommand{\modelNameSNIax}{SNIax}     
\newcommand{\modelNameKN}{KN}     
\newcommand{\modelNameSLSN}{SLSN-I}     
\newcommand{\modelNamePISN}{PISN}    
\newcommand{\modelNameILOT}{ILOT}    
\newcommand{\modelNameCART}{CaRT}    
\newcommand{\modelNameTDE}{TDE}     
\newcommand{\modelNameAGN}{AGN}     
\newcommand{\modelNameRRL}{RRL}     
\newcommand{\modelNameMdwarf}{M-dwarf}     
\newcommand{\modelNamePHOEBE}{EB}     
\newcommand{\modelNameMIRA}{Mira} 
\newcommand{\modelNameuLensBinary}{$\mu$Lens-Binary}    
\newcommand{\modelNameuLensSingle}{$\mu$Lens-Single}     
\newcommand{\modelNameuLensSTRING}{$\mu$Lens-String}    

\newcommand{\modelNumTOTAL}{0}     
\newcommand{\modelNumSNII}{42}     
\newcommand{\modelNumSNIbc}{62}     
\newcommand{\modelNumSNIa}{90}     
\newcommand{\modelNumbg}{67}     
\newcommand{\modelNumSNIax}{52}     
\newcommand{\modelNumKN}{64}     
\newcommand{\modelNumSLSN}{95}     
\newcommand{\modelNumPISN}{994}    
\newcommand{\modelNumILOT}{992}    
\newcommand{\modelNumCART}{993}    
\newcommand{\modelNumTDE}{15}     
\newcommand{\modelNumAGN}{88}     
\newcommand{\modelNumRRL}{92}     
\newcommand{\modelNumMdwarf}{65}     
\newcommand{\modelNumPHOEBE}{16}     
\newcommand{\modelNumMIRA}{53}     
\newcommand{\modelNumuLensBinary}{991}    
\newcommand{\modelNumuLensSingle}{6}     
\newcommand{\modelNumuLensSTRING}{995}    

\newcommand{\NgenTOTAL}{117,128,700}     
\newcommand{\NgenSNII}{59,198,660}     
\newcommand{\NgenSNIbc}{22,599,840}     
\newcommand{\NgenSNIa}{16,353,270}     
\newcommand{\Ngenbg}{1,329,510}     
\newcommand{\NgenSNIax}{8,660,920}     
\newcommand{\NgenKN}{43,150}     
\newcommand{\NgenSLSN}{90,640}     
\newcommand{\NgenPISN}{5,650}    
\newcommand{\NgenILOT}{4,521,970}    
\newcommand{\NgenCART}{2,834,500}    
\newcommand{\NgenTDE}{58,550}     
\newcommand{\NgenAGN}{175,500}     
\newcommand{\NgenRRL}{200,200}     
\newcommand{\NgenMdwarf}{800,800}     
\newcommand{\NgenPHOEBE}{220,200}     
\newcommand{\NgenMIRA}{1,490}     
\newcommand{\NgenuLensBinary}{1,010}    
\newcommand{\NgenuLensSingle}{2,820}     
\newcommand{\NgenuLensSTRING}{30,020}    

\newcommand{\NtrainTOTAL}{7,846}     
\newcommand{\NtrainSNII}{1,193}     
\newcommand{\NtrainSNIbc}{484}     
\newcommand{\NtrainSNIa}{2,313}     
\newcommand{\Ntrainbg}{208}     
\newcommand{\NtrainSNIax}{183}     
\newcommand{\NtrainKN}{100}     
\newcommand{\NtrainSLSN}{175}     
\newcommand{\NtrainPISN}{0}    
\newcommand{\NtrainILOT}{0}    
\newcommand{\NtrainCART}{0}    
\newcommand{\NtrainTDE}{495}     
\newcommand{\NtrainAGN}{370}     
\newcommand{\NtrainRRL}{239}     
\newcommand{\NtrainMdwarf}{981}     
\newcommand{\NtrainPHOEBE}{924}     
\newcommand{\NtrainMIRA}{30}     
\newcommand{\NtrainuLensBinary}{0}    
\newcommand{\NtrainuLensSingle}{151}     
\newcommand{\NtrainuLensSTRING}{0}    

\newcommand{\NtestTOTAL}{3,492,888}     
\newcommand{\NtestSNII}{1,000,150}     
\newcommand{\NtestSNIbc}{175,094}     
\newcommand{\NtestSNIa}{1,659,831}     
\newcommand{\Ntestbg}{40,193}     
\newcommand{\NtestSNIax}{63,664}     
\newcommand{\NtestKN}{131}     
\newcommand{\NtestSLSN}{35,782}     
\newcommand{\NtestPISN}{1,172}    
\newcommand{\NtestILOT}{1,702}    
\newcommand{\NtestCART}{9,680}    
\newcommand{\NtestTDE}{13,555}     
\newcommand{\NtestAGN}{101,424}     
\newcommand{\NtestRRL}{197,155}     
\newcommand{\NtestMdwarf}{93,494}     
\newcommand{\NtestPHOEBE}{96,572}     
\newcommand{\NtestMIRA}{1,453}     
\newcommand{\NtestuLensBinary}{533}    
\newcommand{\NtestuLensSingle}{1,303}     
\newcommand{\NtestuLensSTRING}{0}    

\newcommand{\zmaxTOTAL}{0}     
\newcommand{\zmaxSNII}{2.0}     
\newcommand{\zmaxSNIbc}{1.3}     
\newcommand{\zmaxSNIa}{1.6}     
\newcommand{\zmaxbg}{0.9}     
\newcommand{\zmaxSNIax}{1.3}     
\newcommand{\zmaxKN}{0.3}     
\newcommand{\zmaxSLSN}{3.4}     
\newcommand{\zmaxPISN}{1.9}    
\newcommand{\zmaxILOT}{0.4}    
\newcommand{\zmaxCART}{0.9}    
\newcommand{\zmaxTDE}{2.6}     
\newcommand{\zmaxAGN}{3.4}     
\newcommand{\zmaxRRL}{0}     
\newcommand{\zmaxMdwarf}{0}     
\newcommand{\zmaxPHOEBE}{0}     
\newcommand{\zmaxMIRA}{0}     
\newcommand{\zmaxuLensBinary}{0}    
\newcommand{\zmaxuLensSingle}{0}     
\newcommand{\zmaxuLensSTRING}{0}    


\email{kessler@kicp.uchicago.edu}

\begin{abstract}
We describe the simulated data sample for  the ``{\PLASTICC}"  ({\acro}),
a publicly available challenge to classify transient and variable events that will be
observed by the \LSST\ (LSST), a new facility expected to start in the early 2020s.
The challenge was hosted by Kaggle, ran from \DATESTART\ to \DATESTOP,
and included \NTEAM\ teams competing for prizes.
Here we provide details of the \NCLASSTOT\ transient and variable source models,  
which were not revealed until after the challenge,
and release the model libraries at \URLMODELS.
We describe  the LSST Operations Simulator used to predict 
realistic observing conditions, and we describe the publicly available
\SNANA\ simulation code used to transform the 
models into observed  fluxes and \uncs\ in the LSST passbands ({\bands}).
Although \acro\ has finished, the publicly available models and simulation
tools are being used within the astronomy community to further improve classification, 
and to study contamination in photometrically identified samples of type Ia supernova
used to measure properties of dark energy.
Our simulation framework will continue serving as a platform to 
improve the \acro\ models, and to develop new models.

\keywords{techniques: cosmology, supernovae}
\end{abstract}

\section{Introduction}
\label{sec:intro}

The study of sources with variable brightness in the night sky has captured human imagination
for millennia, and this fascination continues today in the era of large telescopes.
There are two classes of sources whose brightness changes on time scales less than a year.
The first class is called ``transients," which brighten and fade over a well-defined
time period, and are never seen again. 
The second class is called ``variables,'' which brighten and fade repeatedly.
We can categorize transients and variables based
on their brightness and a time scale, such as duration of the event (e.g., supernova)
or time between peak brightness (e.g., RR Lyrae). 
With modern telescopes and computers, our ability to categorize has improved 
dramatically through the use of additional features such as 
colors (brightness ratio between two wavelength bands), 
shape of brightness-versus-time (light curve), and host-galaxy environment.
In addition to improving how these sources are characterized,
our theoretical understanding has also improved, 
such as explaining mechanisms for stellar explosions,
for the variability associated with supermassive black holes (SMBHs), 
and for stellar physics.

The study of one particular class of transients, 
known as type Ia supernovae (SNe~Ia),
led to the discovery of cosmic acceleration \citep{Riess98,Saul99}, 
which could be the result of a mysterious repulsive fluid called dark energy.
This discovery motivated astronomical surveys to collect larger 
SN~Ia samples to improve measurements of cosmic acceleration, 
and these surveys have included many other types of transients as well.

Optimizing a search for transients and variables is difficult  
because of two conflicting goals:
(1) to repeatedly search the sky over a large area and 
(2) to allocate significant exposure time over each small 
sky patch at each repeat observation. 
For a given instrument, increasing the sky area or number of passbands
reduces the exposure time and vice versa. 
A recently commissioned project called Zwicky Transient Factory 
(ZTF; \citealt{ZTF2019}) searches nearly 1/10 of the entire sky every hour to a 
depth of 20.5 mag ($R$ band).
This search takes place at the Palomar Observatory using a new camera
with a 47 square-degree field of view (${\sim}170\times$ moon area)
for each exposure.  
Another project under construction, called the ``{\LSST}" 
(LSST; \citealt{LSST_SciBook,2008arXiv0805.2366I},
is scheduled to start in the early 2020s and will observe half the night sky every
week to a depth of  24th magnitude. 
While ZTF {\obss} repeat much more often than LSST,
LSST will be sensitive to sources that are 25 times fainter than ZTF can find,
and LSST will observe in six different filters ({\bands}), compared with two for ZTF.
LSST expects to find millions of transient and variable sources every night,
and processing this incredible volume data is a major challenge.

There are two distinct issues related to this data processing challenge.
The first is to identify a subset of interesting transients sources quickly, 
before they fade, so that other instruments can make more precise 
{\spec} \obss\ while the source is still bright enough 
\citep[e.g.,][]{Howell2005,Zheng2008,Ishida2019}. 
The second issue, and the focus of this challenge, 
is to classify all events using the six filters and their entire light curve. 
While high-resolution spectroscopy is much more reliable for classifying events, 
the necessary \spec\ \obs\ time greatly exceeds current and planned resources.
LSST is therefore obligated to classify transient and variable events with the 
compressed filter data, and with the aid of a small ``\spec\ training set."

To motivate development of
classification methods from a broad range of disciplines,
we began optimizing a full light-curve analysis (second issue above) 
with a   ``{\PLASTICC}''  ({\acro}).
On \DATECALL, the \acro\ team issued a 
call\footnote{\URLMODELCALL}
for members of the astronomy community
to develop and deliver models of transients and variables.
This request resulted in a contribution of \NCLASSTOT\ models used in \acro, 
\NCLASSOBS\ of which are based on enough \obss\ to be represented in the training set. 
The remaining four classes have not been convincingly observed,
or have never been observed but are predicted to exist; 
these four classes were combined into a single 
(15th) class for the challenge.

While the planned LSST survey duration is 10 years,
we restricted the \acro\ data set to 3 years 
to limit data volume and computational resources.
Using the {\NCLASSTOT} models, their expected rates, and 3 years of LSST \obss,
more than 100 million transient and variable sources were generated to cover
the southern sky and explore distances reaching out billions of light years.
Most of these generated sources are too distant and faint to be detected with LSST, 
but 3.5~million of them satisfied the detection criteria 
(\S\ref{subsec:trigger_model}).
The resulting set of 3.5~million \bands\ light curves includes 
\NSIMOBS\ million \obss, and were provided in the \acro\ data set.
We also modeled \spec\ classification on prescaled subsets
to provide a training set of ${\sim}8000$ labeled events.
Each model in the training set was defined by an integer tag instead of a descriptive 
string. Random tag numbers (e.g., 90 for {\modelNameSNIa}) were used to avoid 
detectable patterns such as sequential numbers for the SN types.

The \acro\ challenge was formally announced {\DATESTART} through a 
competition-hosting platform called Kaggle\footnote{\label{fn:kaggle}\URLKAGGLE}.
The challenge ended \DATESTOP\ with \NTEAM\  teams, and \NENTRY\ classification entries.
Classifications were evaluated using a weighted log-loss metric \citep{Malz2018},
and background astronomy information for the general public was provided in \citet{PLASTICC}.
Classification results will be described in R.Hlo\v{z}ek et al (2019, in preparation).
The unblinded challenge data are available in \citet{PLASTICC_DATA_UNBLIND},
and the model libraries are in \citet{PLASTICC_MODEL_LIBS}.

To transform these models into realistic light-curve \obss,  
we used the simulation code from the publicly available SuperNova ANAlysis package,
{\SNANA}\footnote{\URLSNANA} \citep{SNANA}.
This simulation program has been under development for more than a decade,
and has been used primarily to simulate \modelNameSNIa\ distance-bias corrections 
in cosmology analyses focused on measuring properties of dark energy \citep{K09,JLA,Pantheon,DES-SNKP}.
The LSST Operations Simulator, hereafter referred to as ``\OPSIM''
\citep{LSST_OPSIM,LSST_OPSIM2,LSST_OPSIM3}, was used to model
variations in depth and seeing based on
detailed modeling of weather and instrument performance.
The {\SNANA} simulation is designed to work for arbitrary surveys, 
which means that the models developed for \acro\ can be applied to other surveys.

There are a few particularly challenging aspects of {\acro}.
First is the wide distribution of class sizes, spanning from ${\sim}10^2$
for the Kilonova class to ${\sim}10^6$ for a few supernova types.
Another difficulty is the training set
determined from estimates of future \spec\ resources;
the training set is small (0.2\% of the test set),
biased toward brighter events,
and not a representative subsample of the full test set.
Finally, many of the light curves are truncated 
(e.g., 2nd panel of Fig.~1 in \citealt{PLASTICC})
because any given sky location is not visible (at night) from the 
LSST site for several months of the year.

Another goal for \acro\ is to develop simulation tools for studies far beyond
this initial challenge. As indicated above, there is a need to develop {\it early epoch}
classification based on a handful of \obss\ so that \spec\ \obss\ can be 
scheduled on interesting subsets. Another important use of simulations is to
optimize the LSST observing strategy, which defines the time between 
visits in each filter band for each region of the sky.
To measure volumetric rates, simulations are crucial for characterizing the \eff\ 
and contamination for each class of events. 
Finally, for the cosmology analysis using photometrically identified SNe~Ia,
models of core-collapse (CC) SNe and other transients are needed to model
contamination.

To prevent the astronomy community from acquiring information beyond what 
is provided on the Kaggle platform, only a small number of astronomers were 
allowed to review the models prior to the challenge,
and each model developer agreed to keep their contribution anonymous 
until the end of the challenge.
We therefore caution that some of the model assumptions and 
choices are approximations, but we are confident
that the model quality is more than adequate for our challenge goals.
While we prepare for LSST operations, we anticipate that some of these models 
will be improved, and that new models will be developed.

The outline of this paper is as follows.
We begin with an overview of LSST in \S\ref{sec:LSST}.
In \S\ref{sec:models_overview} and \S\ref{sec:models_source} we reveal details about the 
transient and variable source models used in \acro.
In \S\ref{sec:model_zphot} we describe our model of photometric redshifts of 
host galaxies, which were included in the \acro\ data set.
In \S\ref{sec:sim} we describe how the \SNANA\ simulation uses these models
to produce realistic light curves in the LSST passbands.
Discussion and conclusions are in \S\ref{sec:discuss}.

\section{Overview of LSST}
\label{sec:LSST}

\newcommand{\URLCFHT}{\url{http://www.cfht.hawaii.edu/Science/CFHTLS}}
\newcommand{\URLPS}{\url{https://panstarrs.stsci.edu}}
\newcommand{\URLDES}{\url{https://www.darkenergysurvey.org }}
\newcommand{\URLLSST}{\url{https://www.lsst.org}}
\newcommand{\URLASASSN}{\url{http://www.astronomy.ohio-state.edu/~assassin/index.shtml}}
\newcommand{\URLATLAS}{\url{https://www.fallingstar.com/home.php}}
\newcommand{\URLPTF}{\url{https://www.ptf.caltech.edu}}
\newcommand{\URLZTF}{\url{https://www.ztf.caltech.edu}}

The era of wide-area CCD astronomy began in the late 1990s with the
2.5~m Sloan Digital Sky Survey \citep{York2000}, which imaged 8,000 deg$^2$
in five passbands ($ugriz$). Many wide-area surveys followed with increasing
area and/or depth, and some examples include the
Canada-France-Hawaii Telescope Legacy Survey (CFHTLS),\footnote{\URLCFHT}
Palomar Transient Factory (PTF),\footnote{\URLPTF} 
All-Sky Automated Supernova Survey (ASASSN),\footnote{\URLASASSN}
Panoramic Survey Telescope and Rapid Response System (Pan-STARRS1),\footnote{\URLPS} and
Dark Energy Survey (DES).\footnote{\URLDES}

LSST\footnote{\URLLSST} \citep{LSST_SciBook,2008arXiv0805.2366I}
will be a revolutionary step in large surveys with an 
8.4~m primary mirror, a nearly 10~deg$^2$ field of view (size of \NMOONLSST\ moons), 
and a 3.2 Giga-pixel camera. 
Over 10 years, LSST will make a slow-motion movie of 
half the sky, visiting each location roughly twice per week in at
least one of the six passbands, \bands. 
Each night LSST will produce 15 Terabytes of imaging data, 
and up to ${\sim}10^7$ transient detections
for the community to sift through
and find interesting candidates to analyze and to target for \spec\ \obss.
Additional key numbers for LSST can be found online.\footnote{\URLKEYNUM}

The current version of the LSST observing strategy includes five distinct 
components, two of which are simulated for \acro.
The primary component is called Wide-Fast-Deep (WFD), which covers almost half the sky. 
The second component is a specialized mini survey called
Deep-Drilling-Fields (DDF), a set of 5 telescope pointings covering almost 50~deg$^2$.
Compared with WFD, the DDF observations are 
 ${\times}20$ more frequent with the same exposure time. 
For \acro, all \obss\ within the same night are coadded
as a simplification, 
and therefore compared with WFD, the DDF nightly visits are ${\sim}2.5$
more frequent and ${\sim}1.5$~mag deeper.
The remaining three mini surveys were not considered useful
for transient science and were therefore not included in {\acro}: 
Southern Celestial Pole (SCP), Galactic Plane (GP), and Northern Ecliptic Spur (NES).

Next, we describe four broad categories of science goals for LSST.
While all science goals are used to determine the observing strategy,
only the first two goals are part of \acro. 

\medskip{\bf Nature of Dark Matter and Dark Energy:}
LSST will probe dark matter and dark energy properties with unprecedented
precision by mapping billions of galaxies as a function of cosmic time
and spatial clustering. Large numbers of Type Ia supernovae, which are included in \acro,
will be used as cosmic distance indicators to measure dark energy properties 
with improved precision.

\medskip{\bf Transients \& Variables:}
As described above, LSST will revolutionize time-domain astronomy with millions 
of new detections every night. 
This science goal is the driving motivation for \acro.

\medskip{\bf Solar System Objects:}
LSST will find millions of moving objects, and gain new insights 
into planet formation and evolution of our solar system.
These moving objects include asteroids and comets (which are not part of \acro),
and those passing relatively close to Earth
are commonly referred to as near-Earth objects (NEOs).
LSST has the potential to find most of the potentially hazardous asteroids (PHAs) 
larger than 140 meters.\footnote{In 2005, Congress directed NASA to find at 
least 90\% of potentially hazardous NEOs sized 140 meters or larger by the end of 2020.}

\medskip{\bf Milky Way Structure \& Formation:}
LSST will measure colors and brightness for billions of stars within
our own Milky Way galaxy, covering a volume that is ${\sim}1000$
larger than in previous surveys. This data set will be used to
probe Milky Way structure, 
study its history of satellite galaxy mergers over cosmic time,
and search for faint dwarf galaxies that store dense 
volumes of dark matter.

\section{Overview of Models}
\label{sec:models_overview}

\begin{figure*}  
  \begin{center}
    \includegraphics[scale=0.8]{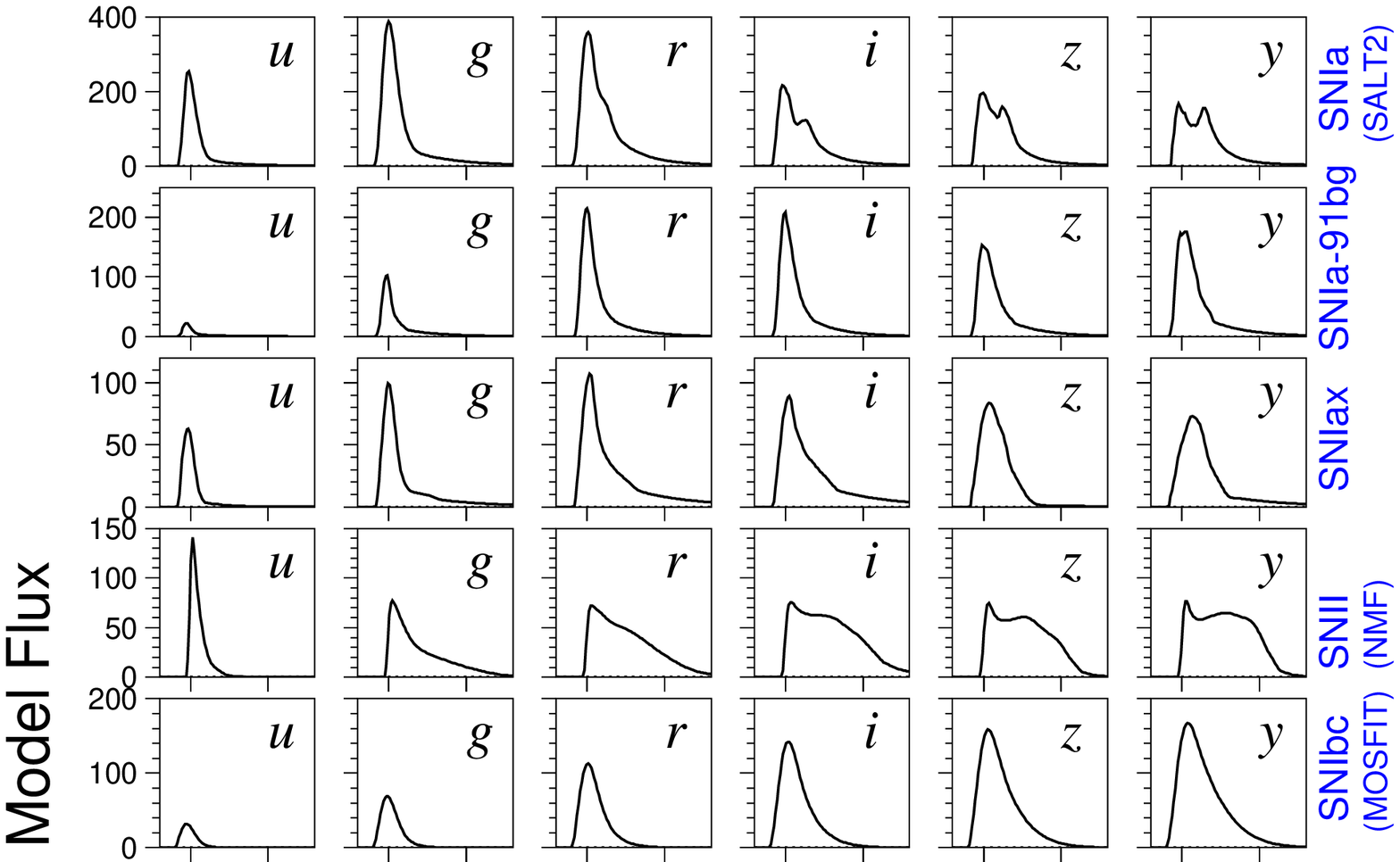}
    \includegraphics[scale=0.8]{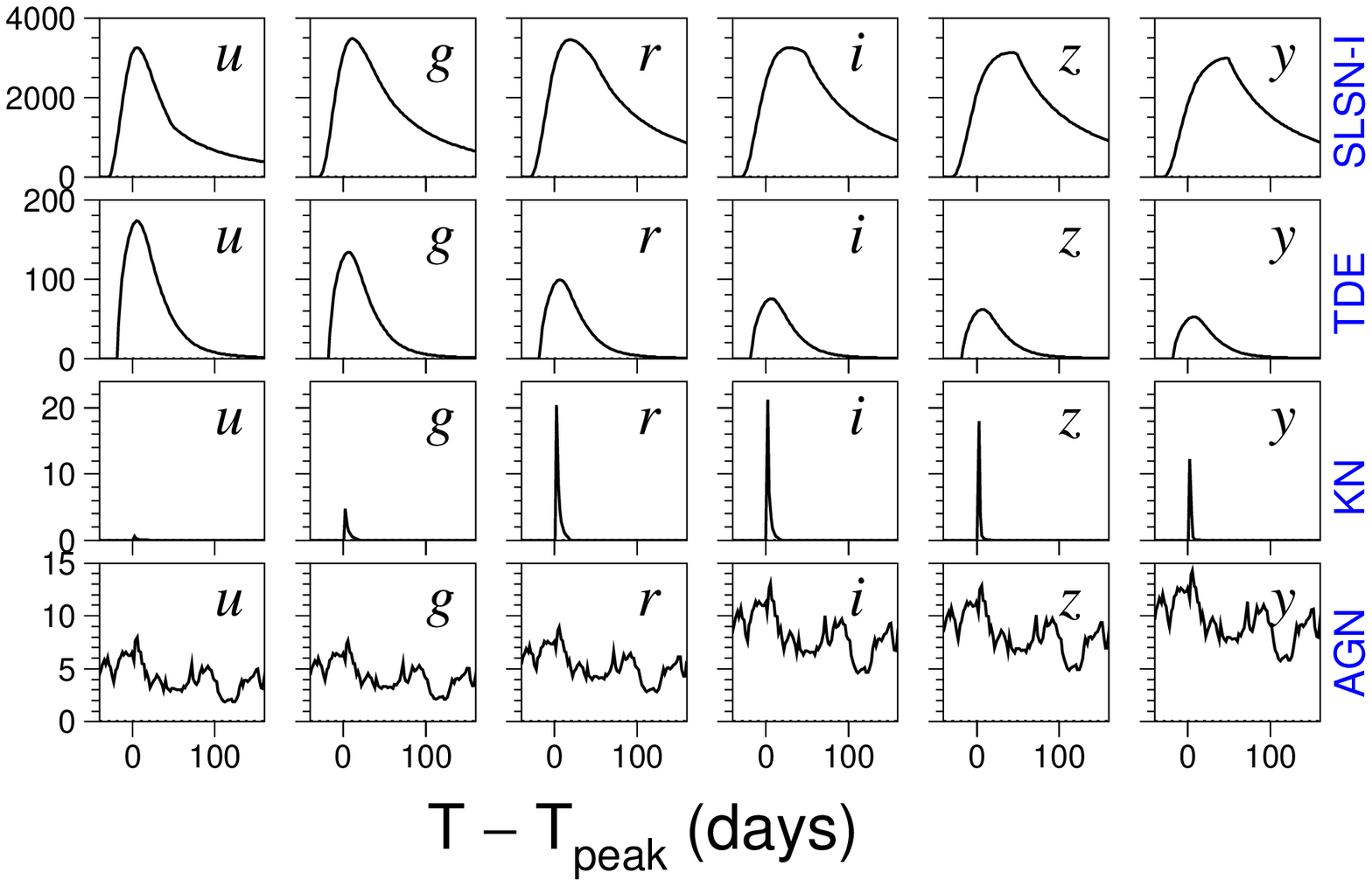}
  \end{center}
 \vspace{-0.2in}
  \caption{ 
    Each row shows model light curves in the \bands\ passbands for the \exgal\ models 
    (Table~\ref{tb:models}) shown on the right. 
    The time reference $T_{\rm peak}$ corresponds to the peak bolometric flux.
    For each model, the vertical flux axis is the same for each passband;
    the flux axis is different for each model.
    The transient model fluxes are computed for redshift $z=0.02$ ($\mu=34.70$) and 
    AB zero point of 22; thus for ${\rm Flux}=100$, 
    the AB magnitude is $m = 22 - 2.5\log_{10}(100) = 17$.
    For the AGN model (bottom), $z=0.12$ and a random 200~day time-window is shown.
	} 
\label{fig:LCmodel_exgal}  \end{figure*}

\begin{figure*}  
\begin{center}
    \includegraphics[scale=0.8]{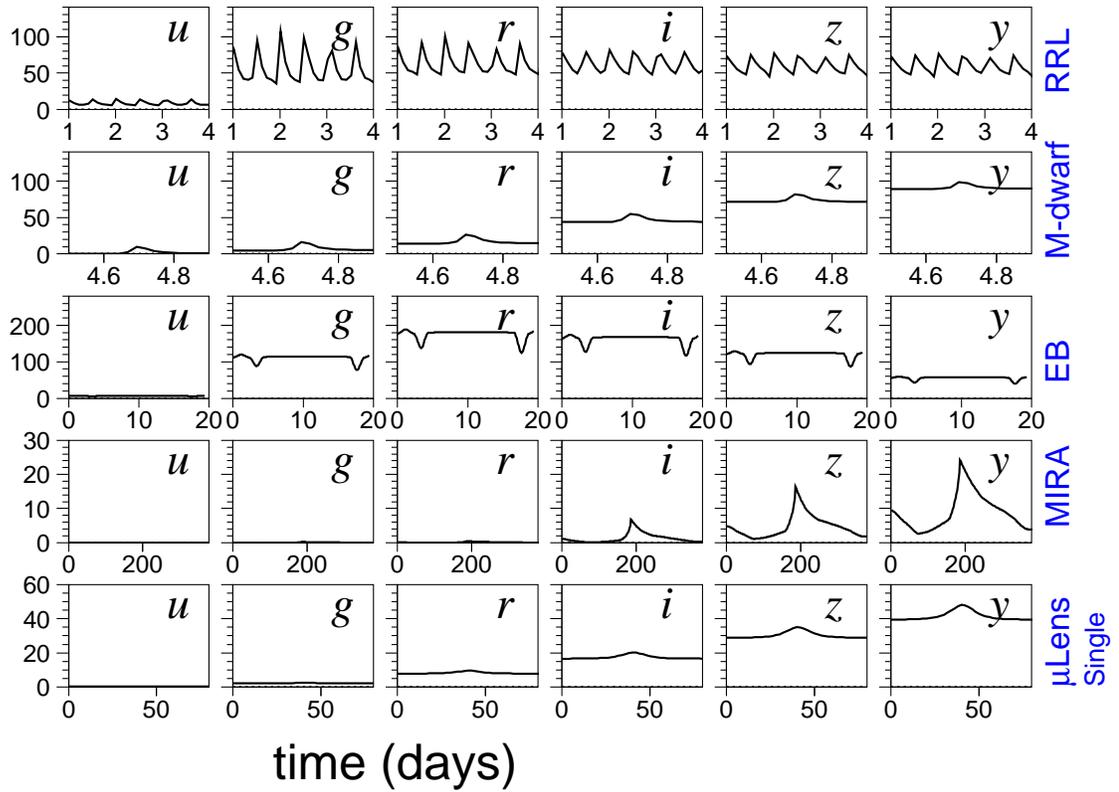}
\end{center}
 \vspace{-0.2in}
  \caption{ 
    Each row shows model light curves in the \bands\ passbands for the \Gal\ models 
    (Table~\ref{tb:models}) shown on the right. 
    The time axis is different for each model.
    For each model, the vertical flux axis is the same for each passband;
    the flux axis is different for each model.
    The flux zero point is given in Fig.~\ref{fig:LCmodel_exgal} caption.
	} 
\label{fig:LCmodel_gal}  \end{figure*}

\begin{figure*}  
\begin{center}
   \includegraphics[scale=0.8]{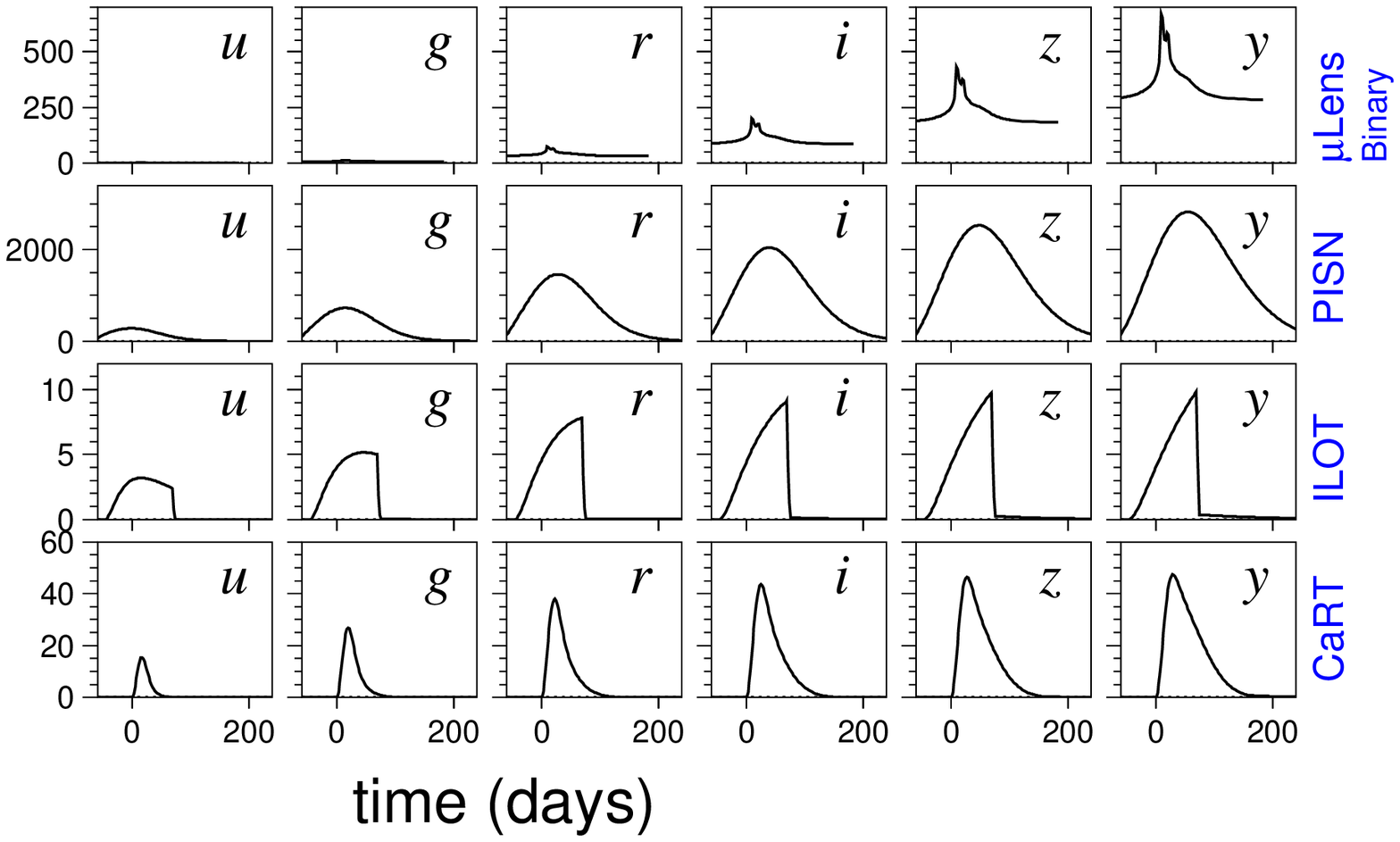}
\end{center}
  \vspace{-0.4in}
  \caption{ 
    For models not included in the training set (class ${>}990$ in Table~\ref{tb:models}),
    each row shows light curves in the \bands\ passbands for the model 
    shown on the right.
    For each model, the vertical flux axis is the same for each passband;
    the flux axis is different for each model.
    For transient models (lower 3), the redshift, distance, and flux zero point 
    are given in Fig.~\ref{fig:LCmodel_exgal} caption. 
  }   
\label{fig:LCmodel_class99}  \end{figure*}

A summary of the models used in \acro\ is shown in Table~\ref{tb:models}.
The first 9 models are \exgal, based on events occurring in distant galaxies,
and they have non-zero redshifts in Table~\ref{tb:models}.
Fig.~\ref{fig:LCmodel_exgal} shows an example model light curve for
each passband and each \exgal\ model in the training set.
The next 5 models are \Gal, corresponding to events occurring within our own 
Galaxy, and they have zero redshift in Table~\ref{tb:models}.
Fig.~\ref{fig:LCmodel_gal} shows an example model light curve for each passband 
and each \Gal\ model in the training set.
The remaining 5 {\it unknown} models (model num${>}990$) are based on 
theoretical expectations, 
or there are too few \obss\ to construct a reliable training set.
Fig.~\ref{fig:LCmodel_class99} shows an example model light curve for
each passband and each unknown model in the test set.

There are a total of \NCLASSOBS\ models in the training set,
and $\NCLASSTOT$ models in the test set. 
A 19th model (\modelNameuLensSTRING) was simulated, but was not included 
in the test set because it brightens for no more than a few minutes 
and it never satisfies the 2-detection trigger requirement 
(\S\ref{subsec:trigger_model}).

\begin{table*} 
\caption{  Summary of transient and variable models for \acro. }     
\vspace{-0.15in}
\begin{center}
\begin{tabular}{ | l | l l | r| r r | l |}
\tableline
   Model Class       & Model  &                &  $\Nevent$                             
         & $\Nevent$ & $\Nevent$  & Redshift \\
   Num\tablenotemark{a}: Name  &   Description  & Contributor(s)\tablenotemark{b} &  Gen\tablenotemark{c}     
            & Train\tablenotemark{d}  & Test\tablenotemark{e}  & Range\tablenotemark{f} \\
    \hline
   \modelNumSNIa: \modelNameSNIa\  & WD detonation, Type Ia SN & RK
            &  \NgenSNIa\  & \NtrainSNIa\ & \NtestSNIa\  & $<\zmaxSNIa$ \\
   \modelNumbg: \modelNamebg\    &  Peculiar type Ia: 91bg  & SG,LG
            & \Ngenbg\   & \Ntrainbg\ & \Ntestbg\   & $<\zmaxbg$  \\
   \modelNumSNIax:  \modelNameSNIax\  & Peculiar SNIax  & SJ,MD 
            &  \NgenSNIax\  & \NtrainSNIax\ &  \NtestSNIax\   & $<\zmaxSNIax$ \\
   \modelNumSNII: \modelNameSNII\    &  Core collapse, Type II SN & SG,LG:RK,JRP:VAV  
            &  \NgenSNII\  & \NtrainSNII\  & \NtestSNII\  & $<\zmaxSNII$ \\
   \modelNumSNIbc: \modelNameSNIbc\  &  Core collapse, Type Ibc SN & VAV:RK,JRP
            &  \NgenSNIbc\  & \NtrainSNIbc\  & \NtestSNIbc\   & $<\zmaxSNIbc$ \\
   \modelNumSLSN: \modelNameSLSN\ & Super-lum. SN (magnetar) & VAV 
            &  \NgenSLSN\ & \NtrainSLSN\ & \NtestSLSN\  & $<\zmaxSLSN$\\
   \modelNumTDE: \modelNameTDE\    & Tidal disruption event  & VAV
            &  \NgenTDE\ & \NtrainTDE\ & \NtestTDE\  & $<\zmaxTDE$  \\  
   \modelNumKN: \modelNameKN\       & Kilonova (NS-NS merger) & DK,GN 
            &  \NgenKN\ & \NtrainKN\  & \NtestKN\  &  $<\zmaxKN$  \\
   \modelNumAGN: \modelNameAGN\   &  Active galactic nuclei & SD
            &  \NgenAGN\  & \NtrainAGN\ & \NtestAGN\  & $<\zmaxAGN$ \\   
   \modelNumRRL: \modelNameRRL\ & RR Lyrae  & SD
            &    \NgenRRL\ & \NtrainRRL\ & \NtestRRL\  & 0 \\
   \modelNumMdwarf: \modelNameMdwarf\ & M-dwarf stellar flare & SD
            &    \NgenMdwarf\ & \NtrainMdwarf\ & \NtestMdwarf\  & 0 \\
   \modelNumPHOEBE: \modelNamePHOEBE\  & Eclipsing binary stars & AP     
            &   \NgenPHOEBE\ & \NtrainPHOEBE\ & \NtestPHOEBE\  & 0 \\  
   \modelNumMIRA: \modelNameMIRA\  & Pulsating variable stars & RH
            &   \NgenMIRA\ & \NtrainMIRA\ & \NtestMIRA\  & 0 \\    
   \modelNumuLensSingle: \modelNameuLensSingle\ & $\mu$-lens from single lens & RD,AA:EB,GN
			& \NgenuLensSingle\   & \NtrainuLensSingle\ & \NtestuLensSingle\   & 0 \\         
\hline
   \modelNumuLensBinary: \modelNameuLensBinary\ & $\mu$-lens from binary lens & RD,AA
			& \NgenuLensBinary\   & \NtrainuLensBinary\ & \NtestuLensBinary\   & 0 \\
   \modelNumILOT: \modelNameILOT\   & Intermed. Lum. Optical Trans. & VAV 
            &  \NgenILOT\ & \NtrainILOT\ & \NtestILOT\  & $<\zmaxILOT$ \\
   \modelNumCART: \modelNameCART\ & Calcium-rich  Transient & VAV           
            & \NgenCART\ & \NtrainCART\ & \NtestCART\   & $<\zmaxCART$\\
   \modelNumPISN: \modelNamePISN\ &  Pair-instability SN & VAV   
            & \NgenPISN\     & \NtrainPISN\    & \NtestPISN\   & $<\zmaxPISN$  \\  
   \modelNumuLensSTRING: \modelNameuLensSTRING\ & $\mu$-lens from cosmic strings & DC
			& \NgenuLensSTRING\  & \NtrainuLensSTRING\ & \NtestuLensSTRING\   & 0 \\ 
     \hline
   \modelNameTOTAL\  & Sum of all models &  
            & \NgenTOTAL\  &  \NtrainTOTAL\ & \NtestTOTAL\  & ---   \\
   \hline
\end{tabular}
\end{center} 
  \tablenotetext{1}{num${<}99$ were randomly chosen to avoid detectable patterns.
    num${>}990$ were in unknown class 99 during the competition;
    an extra digit is added here to distinguish each model.}
  \tablenotetext{2}{Co-author initials. Colon separates independent methods. 
      See author contributions in \S\ref{sec:Ack}.}
  \tablenotetext{3}{Number of generated events, corresponding to 
        the true population without observational selection bias.}
  \tablenotetext{4}{Labeled subset from \spec\ classification. 
     $0\to$  predicted from theory, not convincingly observed, or very few \obss.}
  \tablenotetext{5}{Unlabeled sample. \acro\ goal is to label this sample.}
  \tablenotetext{6}{Redshift$>0$ for \exgal\ models; Redshift$=0$ for Galactic models.}
  \label{tb:models}
\end{table*}

In addition to modeling light curves, we also modeled the rates.
For \exgal\ models, our goal was to model physically motivated 
volumetric rates vs. redshift, $R_V(z)$, to generate realistic sample sizes. 
We achieved this goal for all but the \modelNameAGN\ model.
For \Gal\ models we did not receive rate models, and realistic rates
would likely have resulted in a data sample too large for a public
challenge. We therefore selected arbitrary rates so that \Gal\ models
would comprise ${\sim}10$\% of the \acro\ sample.

\subsection{\Exgal\ Models}
\label{subsec:overvlew_exgal}

Most of the \exgal\ models are exploding stars called supernovae (`SN' in the name),
and the peak brightness varies by almost 2 orders of magnitude.
The kilonova (\modelNameKN) model is an explosive event from 
two colliding neutron stars, and thought to be a primary source of
elements heavier than iron. 
The remaining two \exgal\ models are based on interactions with a SMBH
at the center of a galaxy:
tidal disruption events ({\modelNameTDE}) from stars being shredded
due to their proximity to a SMBH,
and active galactic nuclei ({\modelNameAGN}) driven by gas falling into
a SMBH.

Fig.~\ref{fig:LCmodel_exgal} illustrates some model features, but beware
that there can be significant feature variations within each model class.
The SNIa models ({\modelNameSNIa}, {\modelNamebg}, {\modelNameSNIax})
are brightest in the $g$ and $r$ bands, 
while {\modelNameSNII} is brightest in the $u$-band, but only for a short time. 
{\modelNameSNIbc} is faint in the bluer bands ($u,g$), and 
{\modelNameSLSN} is bright in all bands, about an order of magnitude
brighter than the other SNe.
{\modelNameTDE} are brightest in the blue bands, and 
{\modelNameKN} are very short-lived.
{\modelNameAGN} is the only recurring \exgal\ model, and can show activity
over arbitrary time scales.

Each \exgal\ model is defined as a spectral energy distribution (SED) at
discrete rest-frame time intervals, and as a function of several parameters
characterizing the model. The volumetric rate (per year per cubic Mpc) is 
described as an analytical function of redshift ($R_V(z)$), and  is based on 
measurements, theory, or a combination of both.
A summary of rate models is given in Table~\ref{tb:rates}.
For rates proportional to star formation with a $z$-dependence
from \citet[hereafter MD14]{MD14},
$R_V(0)$ is specified in \S\ref{sec:models_source}.
The other rate models include $R_V(0)$. 
\Exgal\ events are assumed to be isotropically distributed over the sky,
and therefore the DDF and WFD sky area, combined with $R_V(z)$,
are used to determine the number of generated events.

\begin{table}  
\caption{  Summary of \Exgal\ Rate Models for \acro. }     
\vspace{-0.15in}
\begin{center}
\begin{tabular}{ | l | r | r | l | }
\tableline
   Model  &                 &                   &                       \\
   Name   & $R(0)$\tablenotemark{a}  & $R(1)/R(0)\tablenotemark{b}$  & $z$-dependence   \\
    \hline
   \modelNameSNIa\  & 25 & 2.8 & D08\tablenotemark{c} and H18\tablenotemark{d} \\
   \modelNamebg\    &  3 & 2.8 & D08 \\
  \modelNameSNIax\  & 6 & 5.6 & MD14\tablenotemark{e} \\
  \modelNameSNII\   &  45 & 4.9 & S15\tablenotemark{f} \\
  \modelNameSNIbc\  &  19 & 4.9 & S15 \\
  \modelNameSLSN\   &  0.02 & 5.6 & MD14 \\
  \modelNameTDE\    &  1 & 0.15 & K16\tablenotemark{g} \\
 \modelNameKN\      &  6 & 1.0 & flat \\
%
\hline
 \modelNameILOT\   & 3.9 & 4.9 & S15 \\
 \modelNameCART\ & 2.3 & 5.6 & MD14 \\
  \modelNamePISN\ & 0.002 & 5.4 & \citet{Pan2012_PISN} \\
   \hline
\end{tabular}
\end{center} 
  \tablenotetext{1}{Volumetric rate at redshift $z=0$, $(\times 10^{-6}\RateUnit)$.}
  \tablenotetext{2}{Ratio of rate at $z=1$ divided by rate at $z=0$.}
   \tablenotetext{3}{D08: $z<1$ SNIa rate from \citet{Dilday2008}: $(1+z)^{1.5}$}
   \tablenotetext{4}{H18:  $z>1$ SNIa rate from \citet{Hounsell2018}: $(1+z)^{-0.5}$.}
  \tablenotetext{5}{MD14: star-formation rate from \citet{MD14}.}
  \tablenotetext{6}{S15:  core collapse rate from \citet{CC_S15}.}
    \tablenotetext{7}{K16: TDE rate from \citet{Koch2016}: $10^{(-5z/6)}$.}
\label{tb:rates}
\end{table}

We do not provide rate \uncs\ because they are not explicitly used in 
the simulation. For each model, however, we provide an estimate for the number 
of observed events used to construct the model,
and thus statistical rate \unc\ can be estimated.
For science applications, note that there is an implicit
\unc\ on the number of simulated events:
$\sigma_N/N = \sigma_{R_V}/R_V$.

\begin{figure*}  
\begin{center}
      \includegraphics[scale=0.7]{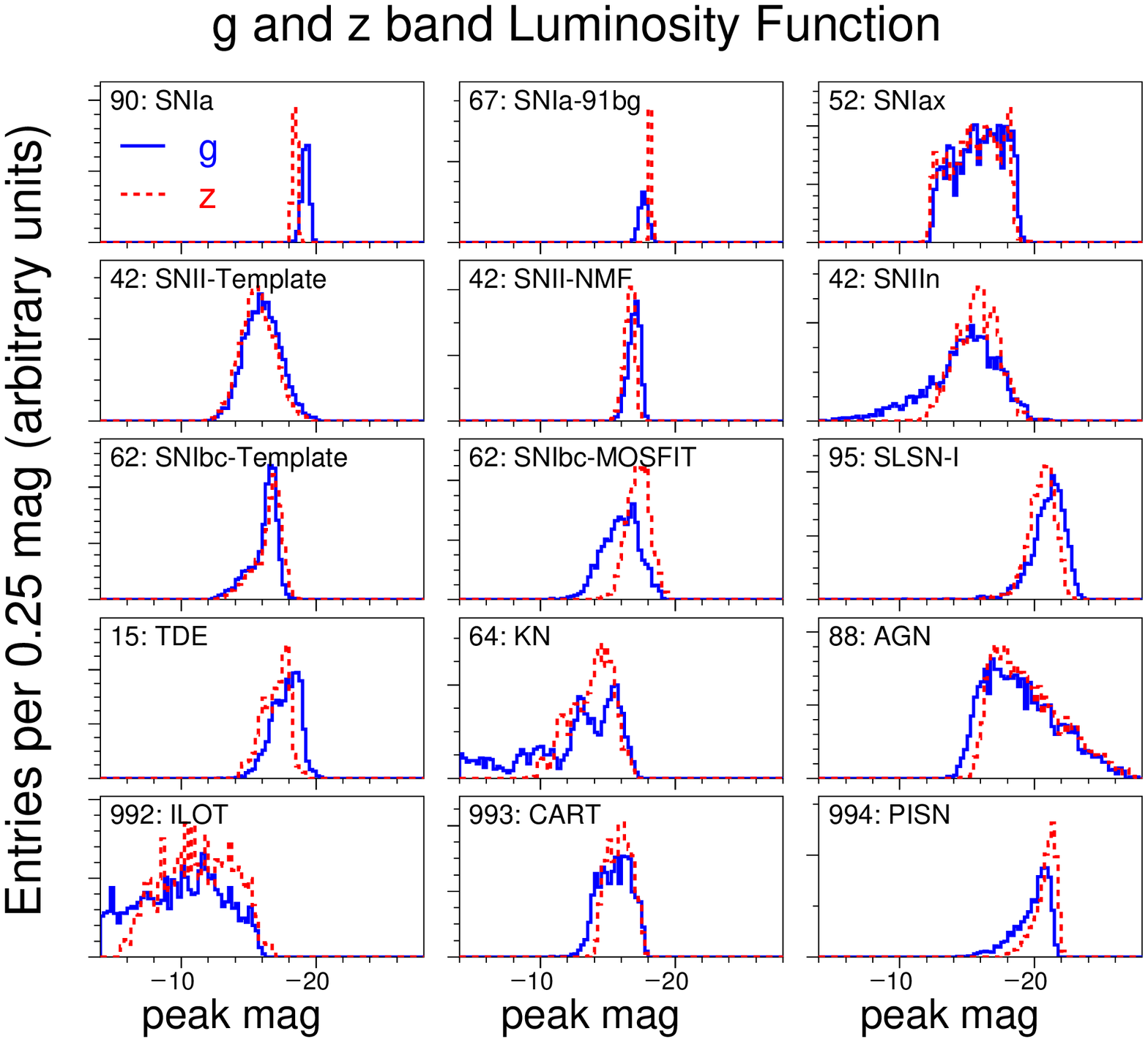}
\end{center}
 \vspace{-0.2in}
  \caption{ Peak $g$ and $z$ band magnitude distributions
   in rest-frame (luminosity function) for \exgal\ models.  
  	Each panel shows a different set of 4000 model light curves.
  	Models appearing twice show independent implementations.
	} 
\label{fig:LF}  \end{figure*}

\begin{figure*}  
\begin{center}
      \includegraphics[scale=0.7]{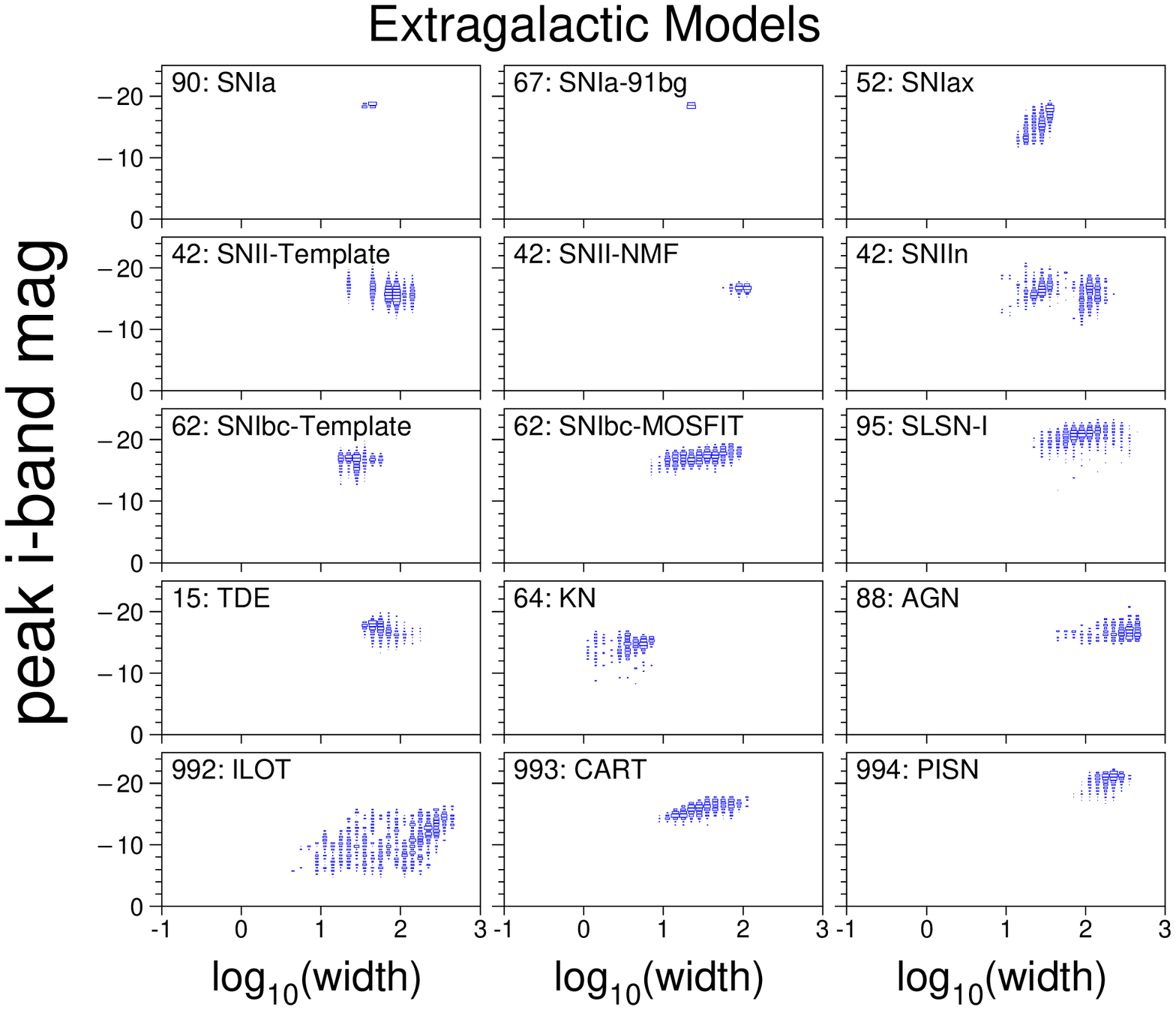}
\end{center}
 \vspace{-0.2in}
  \caption{ 
   For \exgal models, two-dimensional histograms of
   peak $i$-band magnitude (rest-frame) vs. $\log_{10}({\rm width})$,
   where width is the FWHM 
     (days), or time duration where the flux is greater than half the peak flux.
  Each panel shows a different set of 4000 model light curves.
  Models appearing twice show independent implementations.
  Within each panel, the box sizes are proportional to number of events in the 
  two-dimensional bin.
	} 
\label{fig:LFwidth}  \end{figure*}

Next we illustrate some global properties of \exgal\ models.
Fig.~\ref{fig:LF} shows the rest-frame luminosity function in the $g$ and $z$ bands;
note that \modelNameSNIa\ are bright and have a narrow magnitude distribution, 
making them excellent standard candles for measuring cosmic distances.
The brightest models are superluminous supernova ({\modelNameSLSN})
and pair-instability supernova ({\modelNamePISN}), both exceeding $-22$~mag.
Fig.~\ref{fig:LFwidth} shows peak magnitude ($i$-band) vs. FWHM width of the light curve.
The duration varies from a few days ({\modelNameKN}) to ${\sim}$year
({\modelNameSLSN},{\modelNamePISN}). There is significant interest in 
searching unpopulated regions of the mag-versus-width plane.

\begin{figure*}  
\begin{center}    
      \includegraphics[scale=0.8]{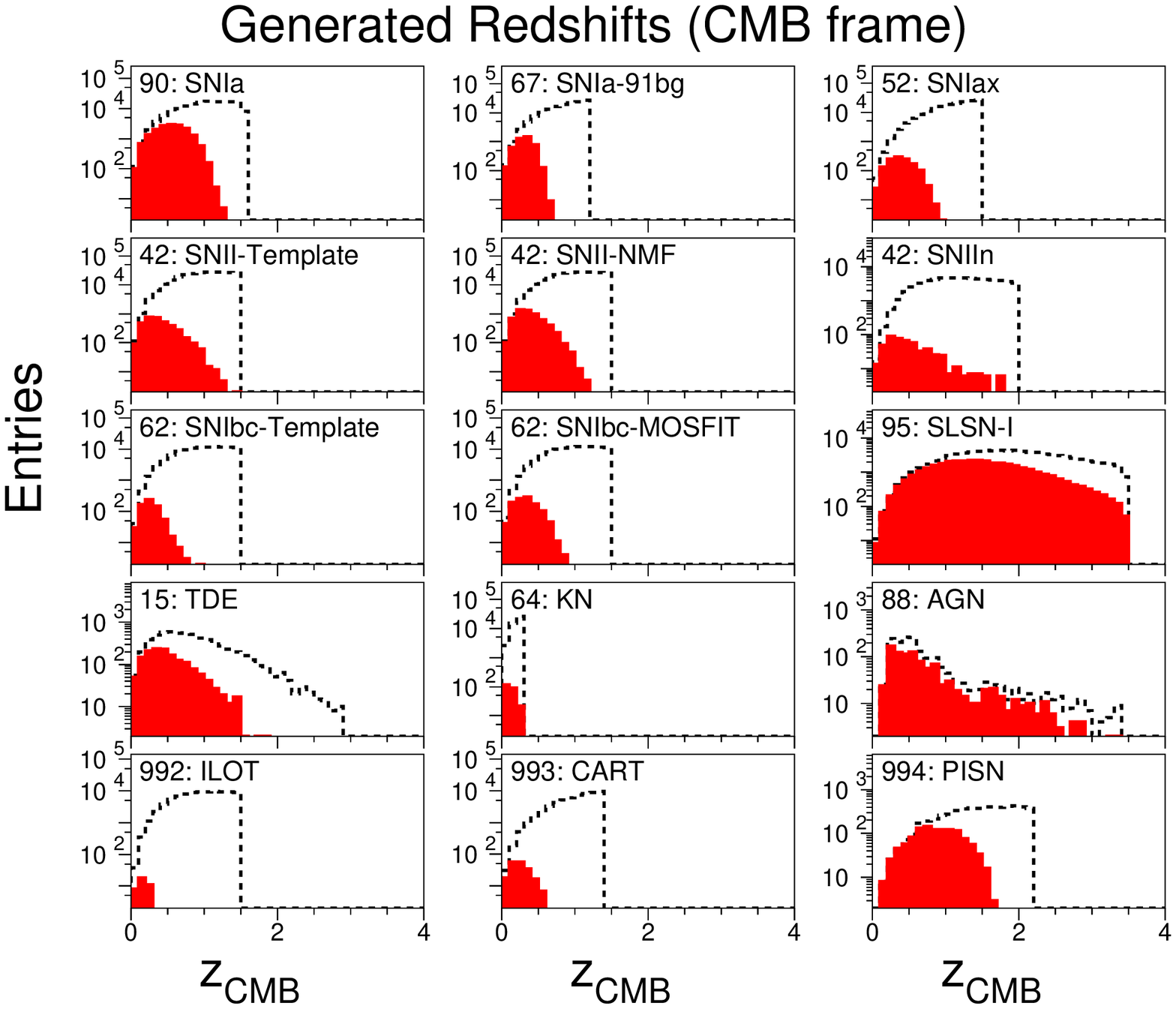}
\end{center}
    \vspace{-0.3in}
  \caption{ 
  	For \exgal\ models, CMB-frame redshifts for all generated events (dashed histogram) 
   	and for events passing 2-detection trigger (red shaded).
    Each panel shows a different model;
    models appearing twice show independent implementations.
	}   
\label{fig:zCMB}  \end{figure*}

Fig.~\ref{fig:zCMB} shows the redshift distribution for generated events 
using the rate model, and for the subset satisfying the 
2-detection trigger (\S\ref{subsec:trigger_model})
and included in the challenge (red shade). 
Each distribution depends on the rate model and the luminosity 
function in each passband. An apparent paradox is the significant
difference between the \modelNameSLSN\ and \modelNamePISN\ redshift 
distributions (after trigger), even though they both have similar
peak brightness in the rest-frame 
(see \modelNameSLSN\ in Fig.~\ref{fig:LCmodel_exgal}, and 
\modelNamePISN\ in Fig.~\ref{fig:LCmodel_class99}).
While the \modelNameSLSN\ model is bright in all LSST passbands,
the \modelNamePISN\ model is bright only in the redder bands,
and thus at high redshift the brightest wavelength region 
is outside the wavelength sensitivity of LSST.

\subsection{\Gal\ Models}
\label{subsec:overvlew_gal}

Three of the \Gal\ models in Fig.~\ref{fig:LCmodel_gal} are recurring 
({\modelNameRRL}, {\modelNamePHOEBE}, {\modelNameMIRA}),
with time time scales of $\sim$day ({\modelNameRRL}) to a year ({\modelNameMIRA}).
The two nonrecurring models are \modelNameMdwarf\ flares,
with time scales less than a day,
and \modelNameuLensSingle\ with time scales from days to years.
In addition to recurring and nonrecurring subclasses,
there are two distinct mechanisms of variability.
The first mechanism is intrinsic, where the stellar brightness varies 
without interacting with other objects: 
these intrinsically variable models are
{\modelNameRRL}, {\modelNameMIRA}, and {\modelNameMdwarf}.
The second mechanism involves an effect between two objects:
eclipsing binary (\modelNamePHOEBE) from a pair of stars blocking each other's light,
and microlensing (\modelNameuLensSingle) from a background star that is
magnified by a foreground star.

\begin{figure}  
\begin{center}    
      \includegraphics[scale=0.4]{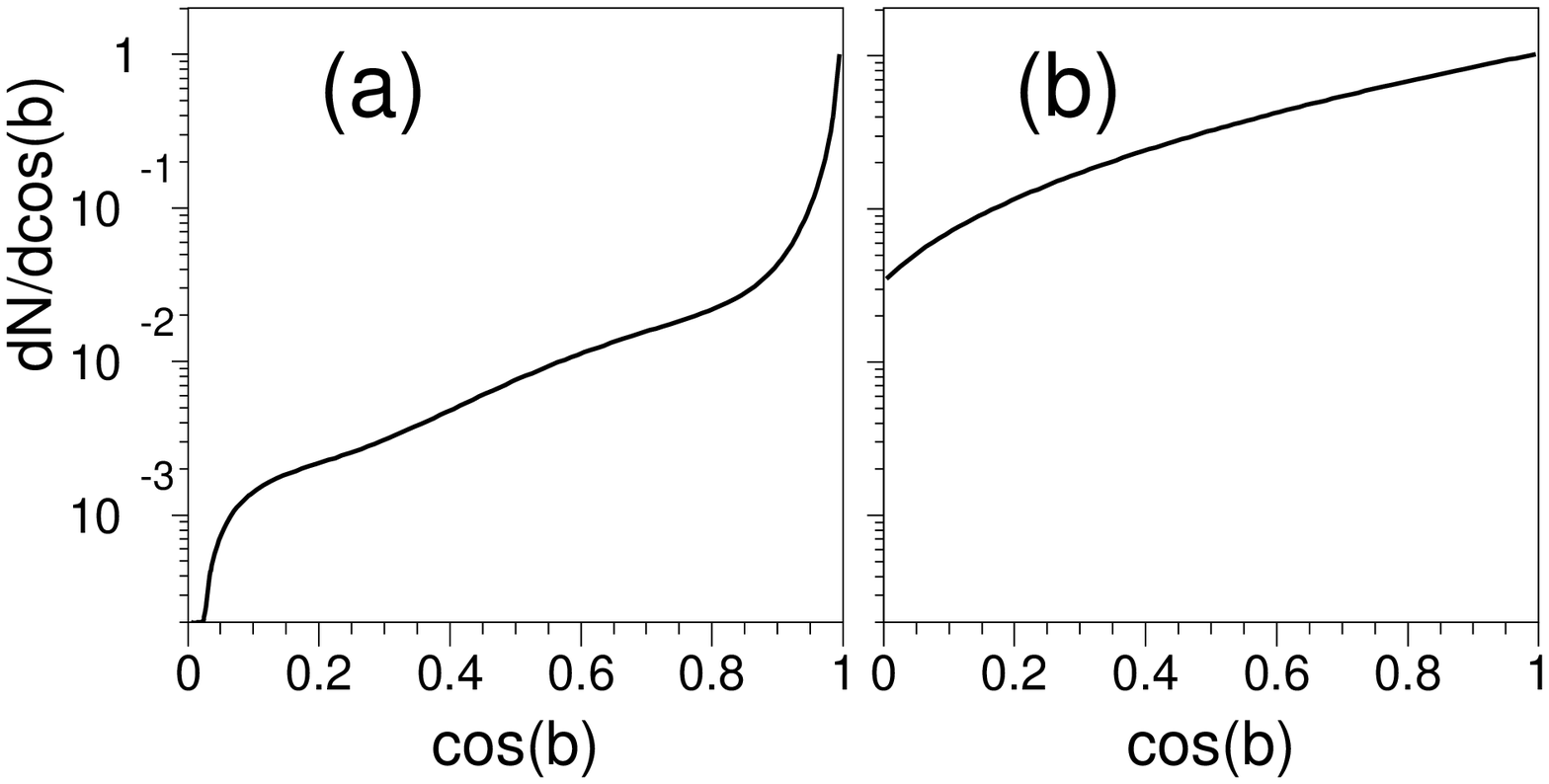}
\end{center}
       \vspace{-0.2in}
  \caption{  $dN/d\cos(b)$ profiles used for \Gal\ models:
  (a) based on Gaia data, and (b) smoother profile for \modelNameMdwarf\ model.}
  A flat distribution corresponds to isotropy. 
\label{fig:dndb}  \end{figure}

For \Gal\ models there is no need to store the SEDs, and they are instead 
defined as a 4-year time sequence of true magnitudes in the \bands\ filter bands. 
The rate model has two components. The first component is the dependence on 
\Gal\ latitude, $b$. For all \Gal\ models except \modelNameMdwarf,
we use the profile in Fig.~\ref{fig:dndb}a which is based on a fit to stellar 
data from the Gaia data release 2 \citep{Gaia2018:DR2}.
Fig.~\ref{fig:dndb}b shows a smoother profile used for the \modelNameMdwarf\ model.
We do not account for Galactic structures such as the 
Large Magellanic Cloud.
The second rate component is the absolute number of generated events,
but since we did not obtain \Gal\ rate models (except for {\modelNameMIRA}),
arbitrary rate values were used. 
The \Gal\ rates described in \S\ref{sec:models_source}
are cited for WFD; the number generated in 
DDF is 0.083\%\footnote{\label{fn:Mdwarf} There is
  a DDF rate bug for the \modelNameMdwarf\ model: 
  here we used the DDF/WFD ratio from  Fig.~\ref{fig:dndb}a instead of
  Fig.~\ref{fig:dndb}b. The WFD profile was simulated correctly.} 
of the WFD number, where this DDF/WFD ratio was determined from the 
profile in Fig.~\ref{fig:dndb}a.

\subsection{Unknown Models}
\label{subsec:overvlew_99}

For models not included in the training set 
(Fig.~\ref{fig:LCmodel_class99} and class ${>}990$ in Table~\ref{tb:models}),
one is a \Gal\ model where a background star is lensed by a binary star system ({\modelNameuLensBinary}). 
The remaining three models are \exgal\  supernova explosions. 
{\modelNameILOT} and {\modelNameCART} have been observed with low statistics.
{\modelNamePISN} events have never been observed,
and they are predicted to be extremely bright, red, and rare;
a high-redshift survey enables the best prospects for discovery.

\section{Models-I: Transients and Variables}
\label{sec:models_source}

The subsections below describe each model as follows.
First, we give a general overview describing the physical mechanism of the process
(e.g., thermonuclear explosion for SNIa),
and \spec\ features which are typically used to classify
these objects for training sets.
Next, we give implementation details geared for experts,
with specific references to methods, data samples, and software packages.
Finally, the rate model is given: 
volumetric rate vs. redshift for \exgal\ models,
and Galactic latitude dependence for \Gal\ models.
As described in the subsections below,
some of the \exgal\ models are based on publicly available data from
the Sloan Digital Sky Survey \citep[hereafter SDSS]{Sako2018},
the Carnegie Supernova Project \citep[hereafter CSP]{Kris2017},  and
the Supernova Legacy Survey \citep[hereafter SNLS]{Gonz2015}.

Ideally, each model would be characterized by observations that have
been corrected for survey selection effects in order to model the 
true underlying populations. However, only the \modelNameSNIa\ model 
accounts for survey selection, and thus the other model populations 
are less accurate. In addition, several models are based on very
low statistics (e.g., 1 observed kilonova event),
and thus the true diversity is not fully realized in \acro.


\subsection{Type Ia Supernova (\modelNameSNIa)    }
\label{ssec:model_SNIa}

\subsubsection{ {\headerOverview}  {\modelNameSNIa} }
\label{sss:overview_SNIa}

A \modelNameSNIa\ event is thought to be the thermonuclear explosion 
of a carbon-oxygen white dwarf (WD) star, the dense exposed core of a former low-mass star.
WDs are typically stable, supported by electron degeneracy pressure, 
but can explode under certain conditions when they are in binary systems.
Leading models for the progenitor systems of \modelNameSNIa\ 
\citep{Moaz2014} include 
(1) a WD plus a main-sequence or giant companion star, from which the WD accretes 
material, or 
(2) the merger of two WDs in a close binary system. 
In addition to the nature of the companion star in the first progenitor model, 
other aspects of this process remain uncertain, including the composition of the 
accreted material, the mass at which the WD explodes 
(expected to be near the Chandrasekhar limit of 1.4~$\Msun$), 
and the explosion mechanism \citep{WW1986,LA1995,GCD2004,Shen2018}.

The thermonuclear fusion of carbon and oxygen results in the formation of iron-group elements (like iron, cobalt, and nickel) and intermediate-mass elements 
(like magnesium, silicon, sulfur, and calcium). This fusion releases a tremendous 
amount of energy, ${\sim}10^{51}$~erg in a few seconds, blowing apart the entire WD.

The explosion energy goes into the kinetic energy of the explosion debris 
(called the ejecta), which flies away at tremendous speeds ($\sim$10,000 km/s) 
and rapidly cools. Such rapidly cooling debris would not emit much light, 
except for the fact that some of the newly created elements are radioactive. 
The radioactive decay of the isotopes $^{56}$Ni (half-life of 6.1~days) 
and $^{56}$Co (half-life of 77~days)
deposits energy into the ejecta over a longer time scale. 
Shortly after the explosion when the material is very dense, 
this heat energy cannot quickly diffuse out and thus remains trapped until the 
ejecta expands and rarefies. This heat-trapping leads to visible light emission 
that rises to a peak luminosity approximately three weeks after the explosion 
and fades thereafter over the next few months. The \modelNameSNIa\ peak luminosity 
is about 10~billion times brighter than our Sun, 
and therefore using optical telescopes these events 
can be viewed from billions of light years away.

The type I classification refers to spectra which have no hydrogen lines.
The type Ia classification is associated with the presence of silicon, 
and in particular,  the strong Si~II $\lambda$6355 absorption feature.
For high-redshift \modelNameSNIa\ where the Si~II feature is too red
for typical spectrographs, there are several bluer features
(Ca~II, Fe~II, Fe~III) that are commonly used for identification.

\modelNameSNIa\ are probably most well known as ``standardizable" candles used to 
study the expansion history of the universe. Observationally we find that each event 
has a similar luminosity, and small variations in luminosity are correlated with other
observable properties such as the timescale of the light curve 
\citep{Rust1974,Phillips1993}
and the color of the supernova \citep{Riess1996,Tripp1998}.
Using \modelNameSNIa\ to probe cosmic distances, accelerating cosmic expansion was 
discovered 20 years ago by \citet{Riess98} and \citet{Saul99}.

\subsubsection{ {\headerDetails}  {\modelNameSNIa} }
\label{sss:details_SNIa}

We used the SALT-II light-curve model from \citet{Guy2010}, 
and the training parameters determined from nearly 500 well-measured light curves 
in the ``Joint Lightcurve Analysis"  (JLA; \citealt{JLA}).
These training parameters describe a time-dependent SED, the SED-dependence on 
light-curve width, and a color law.
The SED model is extended into the ultraviolet (UV) and 
near infrared (NIR) as described in \citet{Pierel2018},
and we use the extended wavelength model from 
WFIRST\footnote{\tt https://wfirst.gsfc.nasa.gov} simulations \citep{Hounsell2018}.
We extrapolated the SED model beyond rest-frame phase $+50$ days using
exponential fits to the late-time flux data of SN~2003hv 
\citep{Leloudas2009} and SN~2012fr \citep{Contreras2018}.

Each rest-frame SED model depends on a randomly chosen color ($c$) and stretch ($x_1$)
from the populations in \citet{SK2016}. The amplitude parameter ($x_0$) is computed
from $c$, $x_1$, and the distance modulus.
Intrinsic scatter is implemented with the ``G10'' 
SED variation model described in
\citet{Kessler2013}.


\medskip {\bf Rate Model:}

The volumetric rate versus redshift, $R(z)$, is based on measurements:
\begin{eqnarray}
     R(z)  & = & 2.5  \times 10^{-5} (1+z)^{1.5}~{\RateUnit} ~~(z<1) \\
     R(z) & = & 9.7  \times 10^{-5} (1+z)^{-0.5}~{\RateUnit} ~~(z>1)~.
\end{eqnarray}
For redshifts $z<1$, the rate is from \citet{Dilday2008}. 
For $z>1$ we follow \citet{Hounsell2018}.
The anonymous journal reviewer noticed a mistake: 
$R(z=1)$ has a 3\% discontinuity.



\subsection{Peculiar \modelNameSNIa\ subtype (\modelNamebg) }
\label{subsec:model_91bg}

\newcommand{\sbg}{s_{\rm 91bg}}
\newcommand{\cbg}{\mathcal{C}_{\rm 91bg}}
\newcommand{\SIFTO}{{\tt SiFTO}}

\subsubsection{ {\headerOverview}  {\modelNamebg}  }

The faintest end of the thermonuclear \modelNameSNIa\ population is
composed of SN1991bg-like objects \citep[e.g.][]{91bg}. 
This subgroup is characterized by the following properties:
(1) under-luminous,  with rest-frame $B$ band magnitude $m_B \gtrsim -18$, 
(2) somewhat red  with $B-V \gtrsim 0.3$, 
(3) fast lived with light-curve width less than 70\% of the average
\modelNameSNIa\ width, 
(4) lack of  a secondary maximum in the infrared bands,
(5) light-curve width does not correlate with peak magnitude
\citep{Phillips1993},
and
(6)  {Ti~II} lines in their spectra.

This subclass comprises 15-20\% of the \modelNameSNIa\ class \citep{Li2011,graur2017loss},
and they occur mostly in old environments \citep[e.g.][]{Gonz2011}.  
Although highly debated, recent theoretical studies suggest that their explosion 
mechanism is the prolongation of normal \modelNameSNIa\ with less $^{56}$Ni 
powering the light-curve, and lower temperature that leads to an earlier recombination 
of ionized elements \citep{Hoeflich2017,Shen2018,Polin2018}.
In contrast to normal \modelNameSNIa, \modelNamebg\ 
do not follow the stretch-brightness relation \citep{Phillips1993}
and are therefore not typically used to measure cosmological distances.

\subsubsection{ {\headerDetails} {\modelNamebg} }
\label{sss:91bg_details}

To model 91bg-like type Ia supernovae, we start with the  SED template series based 
on \citet{Nugent2002}.\footnote{\tt https://c3.lbl.gov/nugent/nugent\_templates.html} 
The near-UV regions are extended using synthetic spectra from \citet{2009MNRAS.399.1238H},
which are warped to match light-curves of four subluminous SNe~Ia 
(SN2005ke, SN2006mr, SN2007on, SN2010cr) measured with 
Swift\footnote{\tt https://www.nasa.gov/mission\_pages/swift/main}  \citep{Brown2009}.
This extended SED template series is used with the \SIFTO\  light curve fitting model \citep{SIFTO2008}, 
which provides best-fit parameters for stretch ($\sbg$) and  color ($\cbg$).
We fit a sample of  {\specy}-confirmed 91bg-like objects at low redshift from
\citet{Gonzalez2014}. 

These fitted parameters are used to determine the ranges for
stretch (7 bins, $0.65\leq \sbg \leq 1.25$) and
color (5 bins, $0\leq \cbg \leq 1$), 
resulting in a set of 35  SED template series.
Each SED template series spans 1000-12000~\AA\  (10~\AA\ bins), and 
$-13$ to $+100$ days (1~day bins).
The stretch and color are drawn from Gaussian distributions with
means of $0.975$ and $0.557$, respectively, and $\sigma$ values
of 0.096 and 0.175, respectively.
$\sbg$ and $\cbg$ are generated with a reduced correlation of $-0.656$.

While preparing this manuscript we noticed a modeling mistake.
Only a single stretch value was used instead of a continuous range,
and therefore the variation among the 35 SEDs corresponds to only 5 SEDs.
This mistake does not result in leakage, but would result in
data-simulation discrepancies if real data were available.


\medskip {\bf Rate Model:} 
Since {\modelNamebg} are found in more passive (and massive) galaxies 
compared with \modelNameSNIa\ (\S5.3 of \citealt{Gonz2011}), 
we expect the {\modelNamebg} rate to have a smaller dependence on the 
host-galaxy star formation rate.
For simplicity, however, we model the {\modelNamebg} volumetric
rate to be 12\% of the \modelNameSNIa\ rate: 
\begin{equation}
     R(z)  = 3 \times 10^{-6} (1+z)^{1.5}~\RateUnit\
\end{equation}



\subsection{Peculiar SN (\modelNameSNIax)    }
\label{subsec:model_SNIax}

\subsubsection{ {\headerOverview}  {\modelNameSNIax}  }

\newcommand{\SNIaxURLmodel}{See {\tt SED-Iax-0000.dat} in \citet{PLASTICC_MODEL_LIBS} }
\newcommand{\SNIaxURLcode}{\url{https://github.com/RutgersSN/SNIax-PLAsTiCC}}
\newcommand{\DMXV} {\Delta m_{15}}
\newcommand{\trise}{t_{\rm rise}}

Transient surveys have uncovered a wide range of diversity in supernovae, and LSST will continue this revolution, discovering many thousands of ``peculiar'' exploding stars. 
Objects that had been \spec\ outliers  to known classes will become distinct classes. 
With this in mind we chose to broaden the range of supernovae in \acro\ with the aim of photometrically identifying peculiar objects, and also to examine how much confusion they cause for identifying the ``standard'' supernova types (e.g., SN Ia, Ib/c, II).

The largest class of peculiar white dwarf (thermonuclear) supernovae are Type Iax supernovae, denoted ``{\modelNameSNIax}" \citep{Foley2013,Jha2017}, which are based on the prototype SN~2002cx \citep{Li2003}.
\modelNameSNIax\ show some similarities to normal \modelNameSNIa, but in general \modelNameSNIax\ have lower luminosity, lower ejecta velocity (measured from spectra), and more variation in these parameters and in their overall photometric behavior compared to normal \modelNameSNIa. 
The brightest \modelNameSNIax\ could be a contaminant in \modelNameSNIa\ samples used to measure cosmological parameters.

\subsubsection{ {\headerDetails} {\modelNameSNIax} }

To generate light curves that mimic the diverse class of \modelNameSNIax, we began with an 
SED time-series model generated from spectroscopic and photometric observations 
of a single well-measured event: SN~2005hk.
We used the Open Supernova Catalog \citep[OSC]{Gui2017} to collect from various sources near-UV to near-IR photometry \citep{Stanishev2007,Holtzman2008,Sahu2008,Brown2014,Friedman2015,Sako2018,Kris2017} and optical spectroscopy \citep{Chornock2006,Phillips2007,Matheson2008,Silverman2012,Blondin2012}. 

Three spectra of SN~2011ay \citep{Foley2013} were added to the collection to fill the 
phase gap of SN~2005hk spectra between 0 and 10 days after the time of peak brightness.
All spectra were warped so that synthetic photometry 
matches the observed photometry, and the warped SEDs are interpolated 
in phase and wavelength space to create the full SED time series.
Our SN~2005hk SED model is publicly available\footnote{\SNIaxURLmodel}.

We inferred a luminosity function for \modelNameSNIax\ based on the observed
sample of $\sim 50$ events presented in Table~1 of \citet{Jha2017}.
There are strong selection effects for these objects as they span a wide range of absolute magnitude, but we find that a linear luminosity function between $-18 \le M_V \le -13$ with Gaussian roll-offs ($\sigma =$ 0.5 and 0.4 mag at the bright and faint ends, respectively) is adequate to match the observed distribution for a limiting apparent discovery magnitude of $m_V = 20.3$.

Given an absolute magnitude ($M_V$), we estimate a rise time ($\trise$) and decline rate ($\DMXV$) in the $B$ and $R$ bands using correlations based on \citet{Stritz2015} and \citet{Magee2016}, as shown in Fig.~2 of \citet{Jha2017}. 
We define distributions for each of four light-curve parameters ($M_V$, $\trise$, $\DMXV(B)$, $\DMXV(R)$) that capture their correlations and observed scatter.
To create a \modelNameSNIax\ SED time series, we draw a random sample from these
light-curve parameter distributions and ``warp'' our SN 2005hk SED so that the photometric light curve properties correspond to the four selected parameters. 
The code for this process is publicly available.\footnote{\SNIaxURLcode}

\medskip {\bf Rate Model:} 
The volumetric \modelNameSNIax\ rate was set to $6{\times}10^{-6}\ \RateUnit$ at $z = 0$ \citep{Foley2013,Miller2017}, corresponding to 24\% of the normal \modelNameSNIa\ rate. 
The redshift evolution of the \modelNameSNIax\ rate was chosen to follow the
star-formation rate \citep{MD14}
because \modelNameSNIax\ environments and host galaxies suggest a young progenitor population
\citep{Foley2009,Valenti2009,Perets2010,Lyman2013,Lyman2018}.
\textEventGenSED{\NtemplateSNIax}



\subsection{Type II Supernova (\modelNameSNII) }
\label{subsec:model_SNII}
\subsubsection{ {\headerOverview}  {\modelNameSNII} }

Type II supernovae (\modelNameSNII) are explosions of massive stars 
typically with main-sequence masses in the range $8\lesssim M \lesssim18~\Msun$ 
\citep{Smartt2009}. 
The explosion results when the core of the star has fused to form the element iron, 
from which no further nuclear energy can be extracted. 
The cessation of fusion energy release in the stellar core removes the
thermal pressure required to support the star against its own gravity. 
Without this pressure, the core rapidly (in milliseconds) collapses 
in a ``core collapse" (CC) event, to form either a neutron star or a black hole. 
Most of the gravitational energy released in the CC goes into enormous 
emission of neutrinos that mostly escape into space; 
this neutrino burst was observed more than 30 years ago when
about a dozen CC neutrinos were detected from SN 1987A \citep{SN1987A}. 
The surrounding material of the star rebounds off the inner core, 
and a small fraction (${\sim}$1\%) of the gravitational energy released in the 
CC is transferred to this surrounding material, 
causing it to unbind from the core and be expelled into space. 
Some of this kinetic energy is thermalized as heat causing the supernova to shine. 
The optical brightness of CC supernovae can be significantly fainter than \modelNameSNIa, 
even though the total energy release is about one hundred times more.

If the dying star has retained a significant amount of hydrogen in its outer layer 
at the time of explosion, that hydrogen can be seen in the spectrum and we classify 
this as a \modelNameSNII.
The amount of hydrogen and the density structure of the outer layers affects the
supernova light curve in a continuous range from long-lasting brightness
plateaus (type IIP) to more linearly declining (type IIL) light curves.
%
Type IIn supernovae are a subtype ($<10$\%) that have narrow lines of hydrogen 
emission in spectra, implying dense pre-existing circumstellar material 
(CSM) prior to the explosion.  
These IIn events are thought to be powered by the interaction of hydrogen-rich CSM 
surrounding the star and the supernova ejecta, converting more of the kinetic
energy of the explosion debris into light.

Since \modelNameSNII\ are much more abundant than \modelNameSNIa,
(\S\ref{ssec:model_SNIa}), there are efforts to standardize the \modelNameSNII\
brightness and use them to measure cosmic distances to redshifts $z{\sim}0.5$
\citep{Hamuy2002,de_Jaeger2015ApJ}.

\subsubsection{ {\headerDetails}  {\modelNameSNII} }
\label{sss:SNII_details}

This class includes type II SNe and corresponds to 70\% of the CC rate, 
while the \modelNameSNIbc\ class (\S\ref{ssec:model_SNIbc})
accounts for the remaining 30\%.
\modelNameSNII\ are generated and combined from three distinct models: 
two models of type II SNe with equal rate, 
and a 3rd IIn model with a much smaller rate.
Approximately 100 well-measured light curves were used to develop these models, 
and each of these models is described below.

\medskip{\bf \modelNameSNII-Templates:}
We use a time series of SEDs that has been warped such that
synthetic photometry matches observed light curves from SDSS and CSP.
Each warped SED time series is called a template, and the 
original templates are from a decade-old classification challenge \citep{K10_SNPCC}.
For \acro, the warping beyond 8000~\AA\ has been updated 
as described in \citet{Pierel2018}.
There are \NtemplateSNIIT\ templates after discarding those resulting in 
artifacts in the $z$ and $Y$ band light curves.
To match the mean and rms peak brightness in \citet{Li2011},
a magnitude offset (1.5~mag) and Gaussian scatter (1.05~mag)
are applied.

\newcommand{\NMFSYM}{S}
\newcommand{\NMFARG}{\lambda,t}
\medskip{\bf \modelNameSNII-NMF:}
We include a newer model of \modelNameSNII\ with an empirical 
SED that is a linear combination of three `eigenvector' components.
To build the model we apply a dimensionality reduction technique known as 
Non-negative matrix factorization (NMF) to a large sample of 
\modelNameSNII\ multi-band light-curves. 
This sample includes events used to search for progenitors \citep{Anderson2014},
a compilation of several surveys \citep{Galbany2016}, 
the SDSS \citep{Sako2018} and 
the SNLS \citep{Gonz2015}. 
The NMF input is a large matrix of observed photometry  (SN $\times$ fluxes)
and the three resulting light-curve eigenvectors that represent the data are 
always positive (as opposed to Principal Component Analysis,
where eigenvectors may be negative).

Next, we take a large sample of \modelNameSNII\ spectra and calculate a single 
weighted-average \modelNameSNII\ spectral time series. 
These spectra are warped so that their synthetic photometry matches
each of the three multi-band light-curve eigenvectors obtained previously.
The output of this procedure is a three-component SED basis from which any 
given SED time series, $\NMFSYM(\NMFARG)$, can be obtained as
\begin{equation}
     \NMFSYM(\NMFARG) = 
     a_1\NMFSYM_1(\NMFARG) +
     a_2\NMFSYM_2(\NMFARG)+
     a_3\NMFSYM_3(\NMFARG)
\end{equation}
where $\NMFSYM_{1,2,3}(\NMFARG)$ are the three warped SED eigenvectors and 
$a_{1,2,3}$ are the projections, 
i.e. the factors that multiply the eigenvectors for each SN. 
The empirical ranges of projections for these eigenvectors are
$0.0<a_1<0.5$ in 0.1 steps, 
$0.0<a_2<0.07$ in 0.01 steps and 
$0.0<a_3<0.07$ in 0.01 steps.
The number of templates in this 3D space is $6{\times}8{\times}8 = \NtemplateSNIINMF$.
For each simulated \modelNameSNII\ event, $a_{1,2,3}$ are
drawn from correlated Gaussian distributions  measured from the data:
$\sigma_{1,2,3} = 0.0854,0.020,0.025$, and reduced correlations
$\rho_{1,2,3} = 0.241,0.052,-0.74$.
Since the $a_{1,2,3}$ values are randomly selected from a continuous 
distribution, linear 3D interpolation is used to ensure a
continuous distribution of SEDs.

While the SNII-Templates include magnitude scatter to match \obss, 
the SNII-NMF scatter was not checked prior to the challenge.
This mistake resulted in a luminosity function that is too narrow
(Fig.~\ref{fig:LF}).

\medskip {\bf SNIIn-{\mosfit}:}
We use the \mosfit\ software package  (Appendix~\ref{app:mosfit})
to simulate the {\tt csm} model 
using the parameter range described in \cite{Villar2017} 
for Type IIn SNe. In this model, the transient 
is powered by the forward and reverse shocks which convert their kinetic energy 
into radiation \citep{wanderman2015rate}. 
A number of parameters affect the SEDs, including the CSM density, 
the CSM mass, the ejecta mass and the ejecta velocity. 
We assume that the photosphere is stationary and within the CSM. 
We generate a set of \NtemplateSNIIn\ SED time series by sampling 
physical parameters as described in \citet{Villar2017}.
We use rejection sampling to match the 
luminosity function found in \citet{Richardson2014},
and require rest-frame $M_r < -10$~mag.
The faint tail in the $g$-band luminosity function (Fig.~\ref{fig:LF})
is an artifact of the model.

\medskip {\bf Rate Model:} 
The total CC volumetric rate versus redshift is given by 
Fig.~6 (green line) in \citet{CC_S15}.
The Type II fraction of the total CC rate is 70\% 
\citep{Smartt2009}, 
and is consistent with the 75\% estimate in \cite{Li2011}. 
The rate is split equally among the 
\NtemplateSNIIT\ \modelNameSNII-Template SED time series, and the 
\modelNameSNII-NMF model.
The IIn fraction is 6\%, and equal weight was given to each of
the \NtemplateSNIIn\ SED time series.


\subsection{Stripped Envelope Core Collapse Supernova (\modelNameSNIbc) }
\label{ssec:model_SNIbc}

\subsubsection{ {\headerOverview}  {\modelNameSNIbc} }

Supernovae types Ib and Ic, also known as `stripped envelope SNe,'
are a distinct class of core collapse SNe characterized by spectra which 
lack hydrogen features, and the Ic subclass spectra lack helium.
These spectral characteristics imply a progenitor star that has been 
stripped of its hydrogen and helium envelope before the explosion.
While massive stars are the likely progenitor, 
there is evidence of binary system progenitors 
\citep{Eldridge2016,Folatelli2016,VanDyk2017}.
This transient is likely powered by the radioactive decay of $^{56}$Ni 
formed in the  supernova ejecta. 

\modelNameSNIbc\ photometric light curves are similar to those from 
\modelNameSNIa\ (\S\ref{ssec:model_SNIa}),
but they are fainter and redder \citep{Galbany2017}. 
In an effort to use photometrically identified SNe~Ia to measure
cosmic distances and cosmological parameters \citep{Jones2017}, 
\modelNameSNIbc\ events are an expected source of contamination
because the brightest \modelNameSNIbc\ events overlap the 
\modelNameSNIa\ luminosity function (Fig.~\ref{fig:LF}),
and the \modelNameSNIbc\ and \modelNameSNIa\ colors are similar.

\subsubsection{ {\headerDetails}  {\modelNameSNIbc} }
\label{sss:SNIbc_details}

Type Ibc SNe are generated and combined from two distinct models: 
templates and \mosfit\ parameterization.
A few dozen well-measured light curves were used to develop these models, 
and each of these models is described below.

{\bf {\modelNameSNIbc}-Templates:}
This is the same procedure as for \modelNameSNII-Templates in \S\ref{sss:SNII_details},
except the observed \modelNameSNII\ light curves are replaced with 
\modelNameSNIbc\ events.
There are \NtemplateSNIbcT\ SED time-series templates (7~Ib plus 6~Ic) after discarding 
those resulting in artifacts in the $z$ and $y$ band light curves.

{\bf {\modelNameSNIbc}-{\mosfit}:}
We use the \mosfit\ {\tt default} model (Appendix~\ref{app:mosfit}), 
using the \modelNameSNIbc\ parameter ranges and distributions described in \cite{Villar2017}. 
We use rejection sampling to match the luminosity function found in \citet{Richardson2014}.
\textEventGenSED{\NtemplateSNIbcMOSFIT}

\medskip {\bf Rate Model:} 
We are not aware of studies that explicitly measure the \modelNameSNIbc\
volumetric rate as a function of redshift,
but measurements of the CC rate at high redshift often assume 
constant Ibc/CC fractions
when calculating their detection efficiencies.
However, for both single and binary star progenitors, 
the relative Ibc/CC fraction is expected to
decline with metallicity. This effect is observed in 
low-redshift populations when examining the 
fraction of hydrogen-poor SNe~Ibc as a function of host galaxy mass or metallicity. 
\citet{graur2017loss} find a ratio of hydrogen-poor to hydrogen-rich CC SNe 
that decreases by a factor of 
$\sim3.5$ between $8.8 < 12+\log(\mathrm{O/H})_\mathrm{T04} < 9.3$.
Since we do not model host galaxies, we do not model a
metallicity-dependent rate.

The total CC volumetric rate versus redshift is given by Fig.~6 (green line) 
in \citet{CC_S15}. The Type Ibc rate is 30\% of the total CC rate \citep{Smartt2009}, 
and is split equally among the two 
\modelNameSNIbc\ submodels (Templates and {\mosfit}).
To generate events with each submodel, equal weight was given to each of the
\NtemplateSNIbcT\ Template SED time series, and also to each of the
\NtemplateSNIbcMOSFIT\ \mosfit\ SED time series.


\subsection{Type I Superluminous Supernova ({\modelNameSLSN})  }

\label{ssec:model_SLSN}

\subsubsection{ {\headerOverview} {\modelNameSLSN} }

\modelNameSLSN\ events are among the brightest optical transients,
with peak absolute brightness $\lesssim-21$~mag.
Their spectra are blue and lack hydrogen,
and their light curves last several months
\citep{Chomiuk2011_SLSN,Quimby2011_SLSN}. 
They tend to be found in metal-poor dwarf host galaxies \citep{Lunnan2014,Angus2016},
and a significant fraction are well described by a central engine known as a ``magnetar:''
a neutron star with a strong magnetic field ($B\gtrsim10^{13}$~G).
These rare transients (${\sim}0.1$\% of SNIa rate)
are a relatively new discovery \citep{Quimby2011_SLSN}, 
largely due to the rise in wide-field surveys. 
Since these events can be up to 50 times brighter than \modelNameSNIa\ 
(\S\ref{ssec:model_SNIa}), there are efforts to standardize their
brightness and use them to measure cosmic distances to redshifts $z{\sim}3$
\citep{Scov2016}.

\subsubsection{ {\headerDetails} {\modelNameSLSN} }
\label{sss:model_SLSN}

Based on a few dozen well-measured light curves,
we model the central engine as a newly born magnetar,
which transfers rotational energy into the surrounding environment 
as it spins down from dipole radiation. 
The magnetar's strength depends on the initial spin period, 
the mass of the newly born neutron star, and the magnetic field of the system. 
Recent work (e.g., \citealt{Nicholl2017,Villar2018_SLSN}) has shown that the magnetar 
model can largely reproduce the diversity of UV through NIR light curves. 
However, our model neglects pre-peak bumps seen in a number of events 
(e.g., \citealt{Nicholl2015_lsq14bdq,Smith2016_des14x3taz,Angus2018}).
The power source and basic properties of these bumps is currently unknown.

We use the \mosfit\ {\tt slsn} model (Appendix~\ref{app:mosfit}), 
which assumes a magnetar engine and blackbody SED with a linear cutoff for 
$\lambda<3000$~\AA\ (see Fig.~1 in \citealt{Nicholl2017}).
To generate light curves consistent with current observations, we fit a set of 58 well-observed Type I SLSNe to our magnetar model \citep{Nicholl2017,Villar2018_SLSN}. 
In short, we use the fitted physical parameters 
(e.g., ejecta mass, velocity, magnetic field, initial magnetar spin period, etc.) 
to generate a multivariate Gaussian which represents the distribution of 
physical parameters for the underlying progenitor population. 
We draw sets of physical parameters from this multivariate Gaussian to produce a set of \modelNameSLSN\ light curves. 
The visible kink in the light curve (Fig.~\ref{fig:LCmodel_exgal}) 
is due to a temperature floor in the model.
Some of the models result in a peak luminosity fainter than $-21$~mag
(Fig.~\ref{fig:LF}), and we mistakenly included these faint events.

During the Kaggle competition, a recent analysis of 21 \modelNameSLSN\ light curves
from DES \citep{Angus2018} suggests that the magnetar model is not sufficient to 
describe all of these events. 
To describe the full \modelNameSLSN\ population,
other models may be needed such as interactions with circumstellar material
(e.g., \citealt{CI2011_SLSN,Chatz2013,Chatz2016}).

\medskip {\bf Rate Model:} 
\modelNameSLSN\ events are observed to occur at a rate of approximately 
$10^{-8}$ to $10^{-7}~\RateUnit$ \citep{Quimby:2013a, McCrum:2015a, Prajs:2017a}.
Spectroscopically confirmed SLSNe have been discovered as far as redshift 
$z=1.998$ \citep{Smith2018_SLSN}, 
and the evolution of their rate with redshift is consistent with the cosmic 
star-formation history \citep{Prajs:2017a}. 
We therefore model the redshift-dependent rate using the star formation history from \citet{MD14}, with $R(0) = 2{\times}10^{-8}~\RateUnit$.
\textEventGenSED{\NtemplateSLSN}


\subsection{Tidal Disruption Events (\modelNameTDE) }

\subsubsection{ {\headerOverview} {\modelNameTDE} }

A {\modelNameTDE} occurs when a star passes near a SMBH,
and the strong tidal fields tidally disrupt the star.
Roughly half of the stellar mass is pulled into the SMBH,
and the relativistic speed of the in-falling material powers 
a transient light curve \citep{Rees1988}.
The observed \modelNameTDE\ properties depend on the SMBH mass, 
the stellar properties, and the local interstellar medium \citep{Mockler2018}. 
The expected SMBH mass range is $10^6$-$10^7\Msun$; larger masses
have a Schwarzschild radius too large to disrupt a star, and instead would
swallow the entire star without leaving a visible signal.

The observed characteristics of a \modelNameTDE\ are based
on the following theoretical expectations:
(1) they have a hot, blue continuum,
(2) they occur near the center of galaxies, and 
(3) some have the predicted $t^{-5/3}$ power law for the bolometric 
  light curve \citep{Evans1989}.
While the light-curve luminosity is expected to peak at UV and X-ray
wavelengths, a dusty environment near the black hole can result in 
absorption of UV photons and re-radiation in the NIR \citep{Jiang2016}.

\subsubsection{ {\headerDetails} {\modelNameTDE} }

We use \mosfit\ (Appendix~\ref{app:mosfit}) to simulate the {\tt tde} model,
which assumes that the luminosity traces the fallback rate of the stellar 
material onto the black hole. To generate light curves  consistent with
current observations, we fit a set of 11 well-observed TDEs to our model. 
We use these fitted physical parameters 
(e.g., the stellar mass, black hole mass, impact parameters, etc.) 
to generate a multivariate Gaussian, accounting for observational 
volume associated with each event. 
We draw sets of physical parameters from this multivariate Gaussian to produce 
a set of \modelNameTDE\ light curves.
With this small sample of observed events, the distribution \uncs\ are large.

\medskip {\bf Rate Model:} 
The volumetric rate at redshift $z=0$ is from \citet{Velzen2018},
and the rate vs. redshift is from \citet{Koch2016}:
\begin{equation}
    R(z) = (1.0{\times}10^{-6}) \times 10^{-(5z/6)}~\RateUnit
\end{equation}
\textEventGenSED{\NtemplateTDE}



\subsection{Kilonova (\modelNameKN)    }
\label{subsec:model_KN}

\subsubsection{ {\headerOverview}  {\modelNameKN}  }

A Kilonova ({\modelNameKN}) event is from the merger of a
compact binary system containing at least one neutron star:
a black hole and a neutron star (BH-NS), or binary neutron star (BNS) system.
The two objects collide at roughly half the speed of light,
releasing enormous energy in the ejecta and in gravitational waves (GWs).
A neutron star is slightly heavier than the sun,
and is packed into a small volume with a radius of ${\sim}10$~km;
a tea spoon of this dense neutron material has a mass of 10~million tons.

There has long been evidence that the production of heavy elements (beyond iron)
in stars and supernovae is not sufficient to account for the observed
abundance. 
To explain this paradox, the existence of \modelNameKN\ events has been 
predicted for decades \citep{LS1974}  
to be the primary origin of heavy elements (e.g., gold, platinum),
which are formed from rapid neutron capture (r-process) nucleosynthesis.
As the neutron star material is expelled from the merger, the material undergoes 
the r-process to produce heavy neutron-rich elements. The radioactive decay of
these elements heats the material, causing it to shine a thousand times 
brighter than a nova (hence the term `kilo-novae'),
yet a KN event is still much fainter than \modelNameSNIa\ events.
\modelNameKN\ events are rare, fade rapidly, and are optically faint, 
making them difficult to find.

After decades of searching for these elusive KNe,
the LIGO-Virgo Collaboration (LVC) discovered a BNS signal from a 
gravitational wave on 2017 August 17 \citep{GW170817:PRL,GW170817:Capstone};
this landmark event is known as GW170817.
Two seconds after the LVC detection, a 
short gamma-ray burst (GRB)
signal from the same sky area was detected in space by the 
Fermi Gamma-ray Burst Monitor \citep{GW170817:GRB}.
Later that night (${\sim}11$~hr later), several teams independently
discovered the optical counterpart using ground-based telescopes;
see Fig.~2 of \citet{GW170817:Capstone}, and
\citet{GW170817:Swope,GW170817:DLT,GW170817:VISTA,GW170817:MASTER,GW170817:DECAM,GW170817:LCO}.
Over the next few months, dozens of instruments were used to
observe this event over a wide range of wavelengths, 
from radio to gamma rays.

Since the host galaxy for GW170817 was identified and has a well-measured redshift,
the combination of GW distance from LVC and \spec\ redshift was used to 
measure the Hubble constant ($H_0$) with a precision of 
${\sim}$15\% \citep{GW170817:H0}. The future prospects are excellent 
for discovering many more KN events, and using them to precisely measure $H_0$
\citep{Chen2018}.
This is of particular interest in the cosmology community because
current precise measurements of $H_0$ using a local ladder \citep{Riess2016}
and cosmic microwave background \citep{Planck2018} differ by ${\sim}8$\%,
or more than 3 standard deviations. This discrepancy has led to 
a large amount of speculation about the presence of unknown physics
in the early universe, and unknown systematic errors in these experiments
\citep{Freedman2017}.

Other science interests related to KNe include 
element abundances, the neutron star equation of state, and
formation mechanisms for compact binaries.
For GW170817, the 2~second time difference between the GW and GRB
detection shows that the graviton and photon speed are the same
to within 1 part in $10^{-15}$; this constraint results in stringent limits
on modified theories of gravity \citep{Baker2017}.

\subsubsection{ {\headerDetails}  {\modelNameKN}  }

Using a single SED time-series model to describe GW170817, 
\citet{Scolnic2018:KN} simulated KN rates in past, 
present, and future surveys, including LSST. 
We expect more diversity than this single event,
so for \acro\ we included the set of SED time-series models of BNS mergers
from \citet{Kasen2017}. These models depend on three parameters: 
ejecta mass, ejecta velocity, and lanthanide fraction.
Increasing ejecta mass results in brighter events,
increasing ejecta velocity results in shorter-lived light curves, and
increasing the lanthanide fraction results in redder events.
We do not have parameterized distributions for these parameters,
and therefore each SED was selected with uniform probability.
The rest-frame peak magnitude range is $-17$ to $-9$ ($i$ band),
compared with $-15.5$~mag for GW170817.

\medskip {\bf Rate Model:} 
A volumetric \modelNameKN\ rate of $1{\times}10^{-6}\RateUnit$ is 
estimated in \citet{Scolnic2018:KN} 
based on a compilation of rates in \citet{LIGO2016:BNSrate}.
For \acro, we increased this rate by a factor of 6 for two reasons:
to provide a sufficient training set (${\sim}100$), and
to reduce the Kaggle score change from correctly identifying 
each \modelNameKN.\footnote{Each model class has similar weight 
   in the scoring metric, and thus a \modelNameKN\ class with very few events 
   can result in a measurable score change for each new \modelNameKN\ event
   that is correctly identified. Increasing the rate was intended to limit
   the use of this scoring artifact.}
Near the end of the Kaggle competition, LVC provided rate estimates in
\citet{GW170817:RATE}, where the 90\% confidence upper limit for BNS mergers is
$3.8{\times}10^{-6}~\RateUnit$, or roughly 60\% of the rate used to simulate \acro.
\textEventGenSED{\NtemplateKN}



\subsection{  Active Galactic Nuclei ({\modelNameAGN})     }
\label{ssec:model_AGN}

\subsubsection{ {\headerOverview} {\modelNameAGN} }

An Active Galactic Nucleus (AGN) refers to the central region of a galaxy that is 
much brighter than average, and AGN are among the brightest \exgal\ sources.
It is hypothesized that AGN activity is a phase in the evolution of most galaxies, 
and is caused by a large influx of gas onto a SMBH in the center of the galaxy.
The gas influx could be from galaxy mergers 
\citep{Sanders1988,Barnes1991,Hopkins2006}, or recycled stellar material.
The associated accretion disk results in the emission of electromagnetic radiation
from radio to X-ray wavelengths.

AGN exhibit stochastic, aperiodic variability with ${\sim}10$\%
variations on timescales of weeks to years. 
This characteristic variability has been used, along with other features, 
to identify AGN in previous time-domain surveys.

Here we give a few examples of how AGN are used to study astrophysics.
The energy outflows from AGN can heat gas in the interstellar medium, 
which can reduce or stop star formation; thus AGN feedback is an important 
component in understanding galaxy evolution \citep{Silk1998}.
Next, a technique called {\it reverberation mapping} 
\citep{Blanford1982_RM,SDSS_RM_Project}
has been developed to measure the mass of the central SMBH.
The ultimate goal is to measure these masses as a function of redshift and
AGN environments, and to learn about black hole formation over cosmic time.
Finally, there have been attempts to standardize the AGN brightness
\citep{Watson2011,LaFranca2014,RL2017}  
to measure the cosmic expansion history at very high redshifts.

\subsubsection{ {\headerDetails} {\modelNameAGN} }

The LSST Project CatSim framework \citep{Connolly2010,Connolly2014}
provides a simulated volume of galaxies by applying a semi-analytic model of galaxy
formation \citep{DeLucia2006} to the Millennium $N$-body simulation
\citep{Springel2005}.  This provides us with a population of galaxies on a
$4.5 \times 4.5$~deg$^2$ patch of sky.  
The entire sky is simulated by tiling this patch over the entire celestial sphere.  
The semi-analytic model determines which galaxies contain AGN.  In its quiescent
phase, each AGN is represented by the composite AGN SED derived from SDSS
observations in \citet{Vandenberk2001}.  
As described in \citet{MacLeod2010}, SED variability is added 
in the form of a damped random walk in $\Dmb$, where $m_b$ is the
magnitude of the AGN in the requested band $b$.

Each AGN is assigned:
(i) a characteristic timescale corresponding to $\tau$ in Eq.~1 of \citet{MacLeod2010}, 
(ii)  a unique integer to seed a random number generator, and 
(iii) six structure function values (one for each LSST band)
corresponding to the $\text{SF}_\infty$ parameter in Eq.~3 of \citet{MacLeod2010}.  
For each simulated AGN \obs, a damped random walk with $\text{SF}_\infty=1$ is started
well before the start time of the survey, 
and is propagated forward to the requested \obs\ time.
The result of this random walk is  multiplied by the structure function of the requested 
LSST band to  determine $\Dmb$.
Note that only a single damped random walk is simulated for each AGN.  Any
variation in color of the AGN is solely due to the different structure function
values assigned to each LSST band, corresponding to different amplitudes in the
random walk through $\Dmb$.

\newcommand{\URLAGNrepo}{\url{http://github.com/lsst/sims\_catUtils}}
\newcommand{\AGNfile}{\tt python/../mixins/VariabilityMixin.py}
\newcommand{\AGNmethod}{\tt applyAgn}

The Python code implementing this model 
is publicly available.\footnote{See file \AGNfile\ in 
    GitHib repository \URLAGNrepo\ ({\AGNmethod} method).}

\medskip {\bf Rate Model:} 
\modelNameAGN\ were generated with an isotropic distribution on the sky.
A arbitrary total of \NgenAGN\ events were generated.
\textEventGenLC{\NtemplateAGN}


\subsection{  RR Lyrae ({\modelNameRRL})      }
\label{ssec:model_RRL}

\subsubsection{ {\headerOverview} {\modelNameRRL} }

\modelNameRRL\ are periodic variable stars from the horizontal branch 
that formed more than 10 billion years ago.
Their pulsations result in brightness variations on ${\sim}1$ day time scales, 
and their well known period-luminosity-metallicity (P-L-Z) relation makes them 
excellent distance indicators \citep{RRL_book2015}.
RRL are also used to probe star clusters, streams, 
and satellite galaxies within the Milky Way.
While RRL are useful probes within the Milky Way, their low luminosity limits
their use as \exgal\ distance indicators.

\subsubsection{ {\headerDetails} {\modelNameRRL} }

\newcommand{\RRLurl}{\texttt{https://lsst-web.ncsa.illinois.edu/sim-data/\\sed\_library/seds\_170124.tar.gz}}

The LSST Project CatSim framework \citep{Connolly2010,Connolly2014}
provides a simulated distribution of Milky Way stars based on color-space
distributions drawn from SDSS using the GalFast model of \citet{Juric2008}.
\modelNameRRL\ variability is added to each star by using color-space matching 
to assign a template light curve from \citet{Sesar2010}.
Light curves for \acro\ were selected with quiescent $r$-band magnitudes between $16.0<r<26.0$.  
The model light curves are publicly available.\footnote{\RRLurl}

\medskip {\bf Rate Model:} 
\modelNameRRL\ were generated with the \Gal\ latitude distribution in Fig.~\ref{fig:dndb}a.
An arbitrary total of \NgenRRL\ events were generated.
\textEventGenLC{\NtemplateRRL}


\subsection{  M-dwarf stellar flare  ({\modelNameMdwarf})       }
\label{ssec:model_Mdwarf}

\subsubsection{ {\headerOverview} {\modelNameMdwarf} }

Stellar flares on cool dwarf stars are anticipated to be a major source of
transients in the LSST data stream. Because flaring activity is
stochastic, potentially very energetic \citep{Kowalski2009}, and most common
on low temperature stars that may not be detected in the quiescent phase
\citep{West2011,Walkowicz2011}, 
stellar flares are expected to be discovered as transients
rather than as extensions of known variable light curves. 

Based on detailed observations of well-known flare stars \citep{Hawley2014}
and the analysis of light curves from survey data \citep{Kowalski2009,Walkowicz2011},
typical flares can range in duration from a few minutes to several tens of minutes,
and the amplitude can vary from  ${\sim}0.01$-0.1~mag,
with some extreme flares producing up to 5~mag in brightness variability.

\subsubsection{ {\headerDetails} {\modelNameMdwarf} }

We begin with a realistic distribution of cool dwarf stars on the sky, 
each with a unique light curve representing a stochastic population of stellar flares.  
This distribution is from the SDSS-based GalFast model \citep{Juric2008}, 
as served through the LSST Project's CatSim framework \citep{Connolly2010,Connolly2014}.
We include all simulated stars redder than $r-i = 0.62$ as candidate flaring dwarfs.

We simulate individual stellar flares using the empirical model of \citet{Davenport2014},
which parameterizes flares in terms of their amplitude and duration.
Light curves for individual stars are generated by assigning a realistic
random sample of flares along the duration of the simulated light curve.
This sample of flares is taken from 
\citet{Hilton:phd}  and \citet{Hilton2011}, 
who provide distributions of flare energies for  
five different classes: 
(1) early type active, 
(2) early type inactive, 
(3) mid type active, 
(4) mid type inactive, and 
(5) late type
(see  Eq.~4.2 and Table 4.3 of \citealt{Hilton:phd}).
Here ``early'' corresponds to spectral types M0-M2, 
``mid'' corresponds to spectral types M3-M5, and
``late'' corresponds to a star cooler than M5.

For each light curve, we randomly select flare times from a uniform
distribution so that the number of flares over the duration of the light
curve matches the cumulative rate of flares per hour at the minimum energy
reported in Table 4.3 of \citet{Hilton:phd}.  
For each flare time, we randomly assign a flare energy according to
\begin{equation}
   E = E_{\rm min} \times(1.0-X)^{\left(1.0/\beta\right)}~,
   \label{eq:E_Mdwarf}
\end{equation}
where $X$ is a random number  between 0 and 1,
and $E_{\rm min}$ and $\beta$ are set to values in Table~4.3 of \citet{Hilton:phd}.  
This prescription assures that the energy distribution of
flares matches that given by Table~4.3 and Eq.~4.2 of \citet{Hilton:phd}.
To avoid modeling the poorly sampled energy tail,
a flare drawn with an energy exceeding
$10^{34}$ erg is clipped to exactly $10^{34}$ erg.

Next, we determine the flare's amplitude and duration.
By studying the distributions of flares on the known flare star GJ 1243,
\citet{Hawley2014} provide a relationship between flare energy, duration, and
amplitude (see their Figure 10).  Assuming these relationships hold for all
stellar flares, we take the energy distributions from \citet{Hilton:phd} 
and convert them into flare durations by randomly drawing from
Gaussians whose mean and variance as a function of flare energy is heuristically
fit to the distribution in the middle panel of Figure 10 from \citet{Hawley2014}.  
We motivate this assumption using Fig.~16 of \citet{Chang2015}, 
which shows no significant evolution in the relationship between flare duration and energy 
as a function of flare magnitude in the population of flares observed in M37.  
Once the energy and the duration have been specified, 
the amplitude is numerically solved by assuming that the flare profile 
has the shape specified by \citet{Davenport2014}. 
To determine a flare's colors, we model each flare as a 9000~K blackbody according to \citet{Hawley2003}.

To assign spectral types to our simulated stars, we convert Table~2 of \citet{West2011}
into a probability density, $P({\rm type},r-i,i-z)$, which depends on spectral type and
stellar colors $r-i$ and  $i-z$.
Each star is assigned to the spectral type such that $P({\rm type},r-i,i-z)$ is maximized.
Finally, we assign an  ``active''  or ``inactive'' status by comparing
the star's position above the simulated galactic plane with Fig.~5 of
\citet{West2008}, which presents the fraction of stars that are magnetically
active as a function of distance above the galactic plane and drawing from the
appropriate distribution.  
Magnetic activity is not necessarily the same as flaring activity 
(the nomenclature of \citealt{Hilton2011,Hilton:phd}). 
We therefore use the bottom panel of Figure 12 of \citet{Hilton2010}, which shows both the total
distribution of flare active and magnetically active stars as a function of
distances from the galactic plane, to derive a ratio between the scale height of
flare active and magnetically active stars in the galaxy.  We use this ratio to
correct the distribution of active stars from \citet{West2008}.

\newcommand{\LSSTFlareURL}{\texttt{http://github.com/lsst-sims/MW-Flare}}
\newcommand{\DavFlareURL}{\texttt{http://github.com/jradavenport/MW-flare}}

The Python code used to generate this model is publicly 
available.\footnote{See \texttt{lsst\_sims} directory of GitHub
  repository {\LSSTFlareURL}, which is forked from {\DavFlareURL}, 
    an open-source implementation of the flare model in \citet{Davenport2014}.
} 

\newcommand{\NLIBMdwarf}{1846}
\medskip {\bf Rate Model:} 
\modelNameMdwarf\ events were generated with the \Gal\ latitude 
distribution in Fig.~\ref{fig:dndb}b.
An arbitrary total of \NgenMdwarf\ events were generated.
\textEventGenLC{\NtemplateMdwarf}
While each template light curve was generated more than 400 times,
the efficiency is only ${\sim}10$\% because of the short light curve duration,
and thus the the re-use factor in the data set is ${\sim}50$.



\subsection{ Eclipsing Binary Stars ({\modelNamePHOEBE})  }
\label{subsec:model_EB}

\subsubsection{ {\headerOverview}  {\modelNamePHOEBE} }

Eclipsing binary stars (EBs) are systems where the orbital plane is aligned with our line of sight, resulting in eclipses as the stars orbit their common center of mass. 
These systems are relatively ubiquitous: the census of \textsl{Kepler} targets revealed 
a $\sim$1--2\% occurrence rate across the sky \citep{keplerEBs2011, kirk2016}, 
with the rates increasing towards the galactic plane. 

Eclipsing binary light curves are generally easy to recognize. Provided a sufficiently high signal-to-noise ratio, eclipses provide readily distinguishable signatures in light curves: V-shaped or U-shaped flux dips during eclipses, along with the out-of-eclipse variability owing to tidal and rotational distortion of the stars known as ellipsoidal variations. The real power of EBs becomes evident when both components contribute a comparable amount of light; 
we see both components in the spectra of EBs and we call such systems double-lined 
spectroscopic binaries or SB2. 
Coupled with photometric data, SB2 systems provide us with masses and radii of individual components from first principles: Newtonian dynamics and geometry. 
SB2 systems comprise ${\sim}25$\% of all EBs, and the state-of-the-art precision of masses and radii is  ${\sim}1$\%. 
%
%
EBs are therefore indispensable astrophysical laboratories for measuring stars, 
and for providing calibration opportunities across stellar and galactic astrophysics \citep{torres2010}. 
They also serve as reliable distance gauges within our Galaxy and beyond \citep{guinan1998}.

\subsubsection{ {\headerDetails}  {\modelNamePHOEBE} }

We used Galaxia \citep{sharma2011}, a stellar population model based on the 
Besan\c con model of the Galaxy \citep{Robin2003}, 
to generate a synthetic model of single stars in our Galaxy to the depth of $r=24.5$. 

We paired coeval stars into binary systems according to the observed distributions 
in multiplicity rates, orbital period, mass ratio, and eccentricity 
\citep{raghavan2010, duchene2013, kirk2016, moe2017}. 
Other orbital properties, namely inclination, argument of periastron and semi-major axis, 
were either computed or drawn from expected theoretical distributions. 
All other physical properties (temperatures, individual masses and radii, distance, etc.) 
were inherited from the stellar components drawn from the Galaxia sample. 
The generated systems were tested for stability and unphysical or unstable systems 
were removed from the sample. 
The process is described in more detail in \citet{wells2017} 
and M.Wells \& A.Pr\v sa (2019, in preparation).
The light curves were calculated using {\tt PHOEBE} \citep{prsa2016}, 
an eclipsing binary modeling suite that supports LSST passbands.

\medskip
{\bf Rate Model:}
The Galactic latitude dependence is from Fig.~\ref{fig:dndb}a.
The overall number of generated events was arbitrarily 
chosen to be 220,000.
\textEventGenLC{\NtemplatePHOEBE}


\subsection{ Pulsating Variables Stars ({\modelNameMIRA})   }
\label{subsec:model_Mira}

\newcommand{\URLMIRACODE}{\url{http://www.mso.anu.edu.au/~mireland/codex}}
\newcommand{\URLOGLE}{\url{http://vizier.u-strasbg.fr/viz-bin/VizieR?-source=I\%2F244A}}
\newcommand{\URLRCA}{\url http://simbad.u-strasbg.fr/simbad/sim-id?Ident=R+Cas+}

\subsubsection{ {\headerOverview}  {\modelNameMIRA} }
Mira-type variables are ${\sim}1\Msun$ stars in the late stages of evolution, 
which undergo stellar pulsation. These cool red giants with radius typically 
200 times that of the sun are also very bright, often with luminosities 
that are 2000 times brighter than the sun.
\modelNameMIRA\ variables are difficult to model given the complex balance of 
pulsation, shocks, and radiation pressure in the star.

Named after the most famous example of such a star, $\omicron$Ceti, 
\modelNameMIRA\ variables are observed to be either oxygen-rich or carbon-rich. 
The chemical composition of the star affects its luminosity changes due to material 
being dredged up from the stellar interior; however the exact fundamental properties 
of \modelNameMIRA\ variables, like their mass-loss rate or metallicity, 
are hard to measure from their spectra. They vary on periods of $P\sim 330$ days, 
however their maximum brightness varies each cycle and therefore without a clear period-luminosity relationship these stars are not 
good distance indicators.

\subsubsection{ {\headerDetails}  {\modelNameMIRA} }

We model \modelNameMIRA\ variable SEDs
through the Cool Opacity-sampling Dynamic EXtended (CODEX) atmospheric model series 
for M-type (oxygen-rich) \modelNameMIRA\ variables \citep{Mira2008, Mira2011}. 
The models include self-excited pulsation with specific approximations for 
convective energy transport \cite[see ][for details]{KW2006} and employ an opacity 
sampling method for radiative transfer in local thermodynamic equilibrium.
Although these models were originally developed to explain interferometric observations of \modelNameMIRA\ variables at mid-infrared and radio wavelengths, they are still useful to produce SEDs across the optical wavelengths
covered by the LSST passbands.

A large number of reference light curves were constructed from five SED template realizations of the underlying \modelNameMIRA\ CODEX models for $\omicron$Ceti (`compact', 'extended') 
and from RCas.\footnote{\URLRCA}
These model outputs are available online.\footnote{\URLMIRACODE}
These SED fluxes were interpolated between the modeled time intervals. 
The model time ranges were clipped to ensure that only integer periods of the oscillations were included. 
For each realization, the pulsation period of the variable was randomly selected from a Gaussian distribution with a mean of $\langle P\rangle=330$ days and $\sigma=0.1\langle P\rangle $.

The light curves were generated by producing synthetic photometry from the model SED 
using the LSST passbands and the AB system. 
The distribution of $i$ band magnitudes was chosen to reflect the 
distribution from the Optical Gravitational Lensing Experiment (OGLE, described below)
and the magnitudes in the other bands were determined from relationships in the CODEX-generated SED fluxes.

{\bf Rate Model:}
The Galactic latitude dependence is from Fig.~\ref{fig:dndb}a.
The overall number of generated events is \NgenMIRA, 
and was computed from OGLE \citep[OGLE,][]{OGLE2009} 
General Catalog of Variable Stars.\footnote{\URLOGLE}
The full OGLE sample of long-period variables includes 1667 \modelNameMIRA\ 
stars along with the  photometric and astrometric properties of these stars.
We restrict the sample to have declination $\delta < 10$~deg, 
$i$ band magnitude $i> 15$, 
and Galactic extinction $E(B-V) < 3$. 
\textEventGenLC{\NtemplateMIRA}

\subsection{ Microlensing from a Single Lens ({\modelNameuLensSingle}) }
\label{subsec:model_uLensSingle}

\newcommand{\murel}{\mu_{\rm rel}}
\newcommand{\uLensMethodOne}{PyLIMA}
\newcommand{\uLensMethodTwo}{GenLens}
\newcommand{\URLPYLIMA}{\url{https://github.com/ebachelet/pyLIMA}}
\newcommand{\URLSYNPHOT}{\url{https://pysynphot.readthedocs.io/en/latest}}
\newcommand{\URLSPECLITE}{\url{https://speclite.readthedocs.io/en/latest/filters.html}}
\newcommand{\URLLSSTSTARS}{\url{https://zenodo.org/record/1136115\#.WlAF\_ktG3LE}}

\subsubsection{ {\headerOverview}  {\modelNameuLensSingle} }
\label{sss:overview_uLens1}

As a special case of gravitational lensing, microlensing occurs when a 
foreground star (the lens) 
crosses the line of sight of a more distant star (the source).
General relativity predicts that several images of the source are created. 
These images are separated by a few angular Einstein ring radii $\theta_E$:
\begin{equation}
   \theta_E = \sqrt{ {{4GM_l}\over{c^2}}(D_l^{-1}-D_s^{-1})}
\end{equation}
where $G$ is the gravitational constant, $c$ is the speed of light in vacuum, 
$M_l$ is the mass of the lens, and $D_l$ and $D_s$ are distances
to the lens and source, respectively \citep{1986ApJ...304....1P}. 
In the case of microlensing, the mass of the lens is small 
($\sim$few solar masses)
and $\theta_E$ is order of milli-arcseconds, leading to indistinguishable images, 
even with the highest resolution instrument to date.
The images are also magnified, creating a brightening of the source. 
The total magnification factor versus time, defined as $A(t)$, 
is the fundamental observable predicted by \cite{1964MNRAS.128..295R}.
In the simplest case of a single source and a single lens (both point sources),
one can derive from general relativity 
(see for example \citealt{1986ApJ...304....1P}):
\begin{equation}
  A(t) = {{u(t)^2+2}\over{u(t)\sqrt{u(t)^2+4}}}
\end{equation}
where the impact parameter $u(t)$ is the angular distance of the source from the lens,
divided by $\theta_E$.
The dependence on time ($t$) is due to the relative angular motion 
($\murel$) between the source and the lens.
Often, the impact parameter is described with three fundamental parameters:
\begin{equation}
  u(t)^2 = u_o^2+{ {(t-t_o)^2} \over{t_E^2} }
   \label{eq:mu_lens}
\end{equation}
where $u_o$ is the minimum impact parameter at the 
time of maximum magnification, $t_o$,
and $t_E = \theta_E/\murel$ is the Einstein ring crossing time.

The real observable from image analysis is the variation of the total flux 
on the line of sight:
\begin{equation}
  f_\lambda(t) = f_{s,\lambda}A(t)+f_{b,\lambda}
\end{equation}
where $f_{s,\lambda}$ is the source flux at wavelength $\lambda$, and
$f_{b,\lambda}$ is the blend flux along the line of sight not related to the lensing events. 
The blend flux is often from other stars along the line of sight,
particularly for dense fields near the Galactic center,
but  can also come from the lens itself. 
If the flux from the lens is measured, the properties of the lens (i.e. the distance and the total mass) 
are much better constrained from observations (e.g., \citealt{2018Univ....4...61B}).
A more complete review on microlensing is given in 
\citet{2012RAA....12..947M} and \citet{2018Geosc...8..365T}.

\subsubsection{ {\headerDetails}  {\modelNameuLensSingle} }
\label{sss:details_uLens1}

Two independent methods were used to generate {\modelNameuLensSingle} events:
\uLensMethodOne\ and \uLensMethodTwo.
\uLensMethodOne\ used the Gaia catalog to select source stars,
and did not include blending.
\uLensMethodTwo\ used a simulated LSST star catalog to generate a source star,
and also selected a second unlensed star. Light from the second star
altered the lensing light curve through blending.
This \uLensMethodTwo\ model was also used to model binary lenses as
described in \S\ref{subsec:model_uLensBinary}.

\medskip{\bf {\uLensMethodOne}:}
This method is based on the first open-source microlensing 
software tool\footnote{\URLPYLIMA} \citep{2017AJ....154..203B}. 
We compute $u(t)$ (Eq.~\ref{eq:mu_lens}) by selecting
$t_o$ from a uniform distribution spanning 2850 days,
$u_o$ from a uniform distribution in [0,1], and
$t_E$ from a log-normal distribution (mean$=3.1$, $\sigma=1.0$) 
that mimics the observed distribution toward the Galactic Bulge \citep{2017Natur.548..183M}. 
We neglect second order effects, such as distortion induced by the rotation of the 
Earth around the Sun, known as the microlensing parallax 
\citep{2004ApJ...606..319G}.

After computing the magnification $A(t)$ from $u(t)$,
the source and blend fluxes are needed.
As a simplification, we ignore blending from other stars.
To obtain a realistic source star magnitude distribution, 
we first select a random position in the sky from a uniform distribution in 
right ascension and declination. 
Next, we query the Gaia DR2 catalog at this position \citep{Gaia2018:DR2}
and choose a random star (with $T_{\rm eff}>3500$~K). 
From the luminosity, we derive the mass of the star using $L\sim M^{3.5}$ 
and its surface gravity using the radius measurement from Gaia. 
Using the surface gravity and effective temperature,
an artificial spectrum of this star is estimated using the models from
\cite{1993KurCD..13.....K}, and implemented with 
pysynphot.\footnote{\URLSYNPHOT}
The spectrum is transformed to AB magnitudes in the six LSST passbands using 
the speclite module.\footnote{\URLSPECLITE}
To avoid saturation in the LSST footprint, the star brightness is reduced by 4~mag.

\newcommand{\uomax}{1.67}
\medskip{\bf {\uLensMethodTwo}:}
This method uses information from known microlensing events,
and selects the source and lens from an LSST catalog\footnote{\URLLSSTSTARS}
with \bands\ magnitudes for almost 17 million simulated stars.
The most important characteristic of a microlensing event is the 
Einstein-radius crossing time, $t_E$.
For point-lens events, $t_E$ is the only quantity that can be derived from model 
fits to the light curve, which contains information about the mass of the lens.
We created a $t_E$ distribution from 24,000 microlensing event candidates that
had been discovered through the combined efforts of several survey teams
\citep{Udalski1992,Alcock1993,Bond2001}.
These observed events are close to the \Gal\ 
bulge,\footnote{Most microlens events were observed with 
    $260{<}{\rm R.A.}{<}275$~deg and  $-37{<}{\rm decl.}{<}-20$~deg}
and we make an approximation using these events to populate the entire LSST-WFD area.
The estimated $t_E$ values range from less than a day to more than 500 days.

After choosing a random $t_E$ value from the measured distribution,
we select the distance of closest approach, 
$u_o = U_{[0,1]}{\times}R_E$,  where $U_{[0,1]}$ is a random number drawn from 
a uniform distribution over $[0,\uomax]$, and $R_E$ is the Einstein radius.
In the absence of blending, $u_o$ determines the value of the peak magnification. 
The maximum value  $u_o=\uomax\, R_E$ corresponds to 
the minimum peak magnification, $A_{\rm peak}=1.1$.
Blending is included by adding flux from a second (unmagnified) star
randomly chosen from the LSST catalog.
Because we start with the value of $t_E$, we have a relationship between
the duration of each time interval in our simulation and the value of the
Einstein-radius crossing time. We therefore do not need to
separately generate values of the lens mass or of the velocities of
source star and lens.
We compute the value of the magnification every 15 minutes,  and to limit the 
output library size, we store magnitudes with changes $>0.001$~mag.
The light curve duration for each event was $14\, t_E.$

\medskip {\bf Rate Model:} 
\modelNameuLensSingle\ events were generated with the 
\Gal\ latitude distribution in Fig.~\ref{fig:dndb}a.
A total of \NgenuLensSingle\ events were generated (half for each method).
The \NtemplateuLensSingle\ light-curve models were selected with a
probability proportional to the light-curve duration.


\subsection{  Microlensing from Binary Lens ({\modelNameuLensBinary})    }
\label{subsec:model_uLensBinary}


\subsubsection{ {\headerOverview}  {\modelNameuLensBinary} }
\label{sss:overview_uLens2}


For {\modelNameuLensSingle} events (\S\ref{subsec:model_uLensSingle}), 
light curves rise gradually from baseline, are symmetric, and are
described by a simple  mathematical function.  
The majority of observed microlensing light curves have the general appearance 
expected for the \modelNameuLensSingle\ model.
Simulated \modelNameuLensBinary\ light curves, however,  exhibit great variety, 
including  asymmetries, multiple peaks, plateaus, and quasiperiodic behavior 
\citep{Mao1991,Mao1995,Guo2015}.
Caustic crossing light curves exhibit sharp variations in magnification,  
and this \modelNameuLensBinary\ feature has been commonly observed 
\citep{Udalski1994}.
Other \modelNameuLensBinary\ light curves, however, have been found
in much smaller numbers than expected \citep{Stefano2000}.
For example, only a few light curves show evidence of binary rotation, 
and the rotation in these cases is modest 
\citep{Dominik1998,Alfonso2000}.

\subsubsection{ {\headerDetails}  {\modelNameuLensBinary} }
\label{sss:details_uLens2}

Using the same \uLensMethodTwo\ method as in \S\ref{sss:details_uLens1},
the \modelNameuLensBinary\ model accounts for blending effects,
and also the orbital motion of the binary system.
We start by generating $t_E$ using the same distribution as for 
\modelNameuLensSingle\ events. 
We choose the value of $u_o$ from a uniform distribution
with a maximum value of $2\, R_E$. 
The peak magnification of binaries can be high, 
even for values of $u_o$ as large as $2\, R_E.$

Because binary lenses are more complex than point lenses,
it is necessary to select additional parameters to describe
\modelNameuLensBinary.
First is the relative transverse speed of source and lens in the 
observer's frame ($v$), selected from a uniform distribution extending from 
$15$~km~s$^{-1}$ to $105$~km~s$^{-1}$.
Second is the distance to the source, $D_S$, which is fixed at 8~kpc.
Third is the distance to the lens, $D_L$,  which was selected from a 
uniform distribution between $10$~pc and $D_S$. 
With $t_E$, $v$, $D_S$, and $D_L$,
the total mass of the binary lens system can be determined. 
To eliminate lensing from objects greatly exceeding the
largest known black hole masses,
we excluded events with $t_E > 300$~days if $D_L < 1$~kpc.

Two properties of the binary system determine the characteristics of the 
lensing light curve. One is the mass ratio, $q=M_2/M_1$, where $M_2 < M_1$. 
To model stellar binaries we select $q$ from a uniform distribution in $[0.2,1.0]$.
The second property is $\alpha=a/R_E$,  where $a$ is the separation between 
the two binary objects, and $R_E$ is the Einstein radius. 
We select $\alpha$ from a uniform distribution in $[0.2,2.0]$.
We also select a random orientation of the orbital plane
and random direction of orbital motion 
(prograde or retrograde with equal probability).

\begin{figure}
\begin{center}
 \includegraphics[scale=0.4]{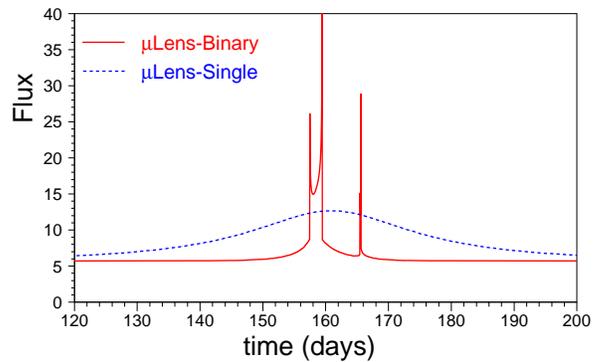}
\end{center}
  \vspace{-0.2in}
\caption{
  For a fixed source star, example microlensing light curve ($r$-band) for 
  binary lens (red curve) and single lens (blue dashed curve).
  The flux-to-mag conversion is given in Fig.~\ref{fig:LCmodel_exgal} caption.
    }
\label{fig:uLens} \end{figure}

Note that binary systems with small values of $\alpha$, large orbital periods, 
and nearly edge-on configurations can produce light curves very similar to
\modelNameuLensSingle\ light curves.
The most dramatic differences between 
\modelNameuLensSingle\ and \modelNameuLensBinary\ light curves 
occur when $a$ is comparable to $R_E$,
and Fig.~\ref{fig:uLens} illustrates this difference for the same source star.

\medskip {\bf Rate Model:} 
\modelNameuLensBinary\ events were generated with the 
\Gal\ latitude distribution in Fig.~\ref{fig:dndb}a.
An arbitrary total of \NgenuLensBinary\ events were generated.
The \NtemplateuLensBinary\ light-curve models were selected with a
probability proportional to the light-curve duration.

\subsection{Intermediate Luminosity Transients (\modelNameILOT) }

\subsubsection{ {\headerOverview} {\modelNameILOT} }
Intermediate Luminosity Transients ({\modelNameILOT}s) have peak optical luminosities 
between those of supernovae and novae, 
and display clear signs of interaction between their ejecta 
and a dense surrounding circumstellar material (CSM). 
These transients are sometimes associated with the progenitors of 
Type IIn supernovae. 

\subsubsection{ {\headerDetails} {\modelNameILOT} }

We use the \mosfit\ software package to generate the {\tt csm} model, using the {\modelNameILOT} parameter range from Table~1 in \cite{Villar2017}.
We model SEDs following \cite{chatzopoulos2012generalized}, 
and assume that the forward and reverse shocks from the ejecta-CSM interaction 
convert their kinetic energy into radiation.
This model is identical to that used for Type~IIn SNe in \S\ref{sss:SNII_details},
however, the explored parameter space (i.e., the ejecta masses and energies) 
is significantly different, leading to a distinct class of objects.
The rapid drop after ${\sim}100$~days (Fig.~\ref{fig:LCmodel_class99})
is an artifact of how \mosfit\ models the nebular phase.

\medskip {\bf Rate Model:} 
For the volumetric rate, we assume the same rate vs. redshift 
as for Type IIn: 6\% of the CC rate from \citet{CC_S15}.
\textEventGenSED{\NtemplateILOT}


\subsection{Calcium-rich Transients (\modelNameCART) }

\subsubsection{ {\headerOverview} {\modelNameCART} }

\modelNameCART\ events, as their name suggests, 
appear to be rich in Calcium based on their spectra. 
These events are a mysterious class with few members. 
They are somewhat dim compared with more traditional supernovae 
and occur far from their host galaxies \citep{Lunnan2017}.
The rapid light-curve evolution indicates a small ejecta mass 
(${\lesssim}0.5$ M$_\odot$).

\subsubsection{ {\headerDetails} {\modelNameCART} }

We model \modelNameCART\ with the \mosfit\ {\tt default} model, 
powered by the radioactive decay of Nickel.
We generate light curves empirically by 
matching \obss\ from \citet{Lunnan2017} and \citet{Milisav2017}, 
and setting uniform priors in 
ejecta mass, ejecta velocity, and nickel fraction.
Models were kept for $-18 < M_r < -14$,
the approximate luminosity range of observed CaRTs.
Since \acro\ does not include host galaxies, we do not model the
large \modelNameCART\ distances from their hosts.

\medskip {\bf Rate Model:} 
\citet{Perets2010} report a relative \modelNameCART\ rate of $(7\pm 5)$\%
of the \modelNameSNIa\ rate. We simulated a 
volumetric rate following the star formation rate from \citet{MD14},
with $R(0) = 2.3{\times}10^{-6}~\RateUnit$, or about 9\% of the
\modelNameSNIa\ rate. \textEventGenSED{\NtemplateCART}
While preparing this manuscript, we learned that
a few months before \acro\ started,
PTF had used a sample of 3 events to report a \modelNameCART\ rate that 
is $\times 5$ higher ($2.5\sigma$) than what we used in \acro\ 
\citep{CART_RATE2018}.


\subsection{Pair-instability Supernova (\modelNamePISN) }

\subsubsection{ {\headerOverview} {\modelNamePISN} }

Pair-instability Supernovae (\modelNamePISN) are thought to arise when 
low-metallicity Population III stars, with $M_*{\sim}140$-$260 M_\odot$,
reach sufficiently high core temperatures that $\gamma$-rays produce 
electron-positron pairs.
This leads to a drop in the internal pressure, resulting in a gravitational
collapse which initiates a thermonuclear explosion that obliterates the 
entire star \citep{Barkat1967_PISN,Kasen2011_PISN}.
Like some core-collapse supernovae, the optical light curves of PISNe are 
powered mainly by the radioactivity decay of $^{56}$Ni;
however, {\modelNamePISN}e typically have larger ejecta masses, 
higher kinetic energies, and much brighter luminosities 
(similar to those of {\modelNameSLSN}, \S\ref{ssec:model_SLSN}).

Observing and identifying \modelNamePISN\ light curves is challenging because
they are expected to be found at high redshift where the 
observed light curve can last several years.
Evidence for a few \modelNamePISN\ events has been reported in
\citet{GalYam2009}, \citet{Cooke2012}, and \citet{Kozyreva2018}.

\subsubsection{ {\headerDetails} {\modelNamePISN} }

We use the \mosfit\ {\tt default} model, with the \modelNamePISN\ parameter ranges 
described in \cite{Villar2017}. 

\medskip {\bf Rate Model:} 
We use a theoretically motivated function of redshift
in \citet{Pan2012_PISN} to describe the volumetric rate:
\begin{eqnarray}
    R(z) & = & [ 1.98 + 6.38z + 6.55z^2 - 4.42z^3 \nonumber \\
         &  & + 0.8312z^4 - 0.0508z^5 ]~ \RateUnitGpc\ \nonumber
\end{eqnarray}
At redshift $z=0$, the \modelNamePISN\ rate is ${\sim}10^4$ lower than the \modelNameSNIa\ rate, and thus \modelNamePISN\ detections are expected at higher redshifts.
\textEventGenSED{\NtemplatePISN}



\subsection{  Microlensing from Cosmic Strings ({\modelNameuLensSTRING})  }
\label{subsec:model_string}

\subsubsection{ {\headerOverview}  {\modelNameuLensSTRING} }

Cosmic superstrings are hypothesized to have been formed from the basic constituents of 
string theory that have been stretched to macroscopic size during the epoch of inflation. 
For a review of inflation in string theory see \cite{Baumann:2014nda}, 
and for a review of  superstring properties see \cite{Chernoff:2014cba}. 
Cosmological evolution of these entities yields a network of long, 
horizon-crossing segments and sub-horizon loops. 
For a review of this general scenario see \cite{2000csot.book.....V}.
For string tensions that are not ruled out by {\obss}, 
the loops are the dominant component of interest \citep{Chernoff:2017fll},
and are expected to cluster with dark matter when structure forms in the universe
\citep{Chernoff:2009tp}.  These loops 
have a negligible contribution to the galaxy's total mass, 
but are potentially detectable as stellar flux variation if
a source (star), lens (string) and observer are suitably aligned
\citep{Vilenkin:1981zs,Vilenkin:1984ea,Bloomfield:2013jka}.

A direct search for these fossil superstring remnants of the early
universe requires repeated flux measurements of stars \citep{Chernoff:2007pd}. 
String microlensing models predict that
the brightness of an unresolved, point-like source (star)
is magnified by exactly a factor of 2, 
which is quite distinct from other microlensing signatures.
Microlensing of stellar sources in the galaxy {\it repeats} $\sim 10^3$ times 
because the loop center of mass moves at the halo velocity
whereas the internal oscillations of the loop are relativistic.  
These distinctive features (factor of 2 enhancement, repetitions, achromatic) 
make a search for cosmic superstring loops possible, 
but the brief duration of the microlensing signal makes the search challenging.
A discovery of a string microlensing source could be used to 
determine the string tension, 
one of the fundamental theoretical unknowns in this scenario, 
and provide important information about the multiplicity of string types, another 
theoretical uncertainty.

\subsubsection{ {\headerDetails}  {\modelNameuLensSTRING} }

This \acro\ contribution provided sample light curves for
microlensed stars drawn from a stellar model of the galaxy for
a set of string tensions $\mu$ consistent with known upper limits
on that quantity \citep{Chernoff:2017fll}. The selected values were
$G \mu/c^2 = 10^{-13}$, $10^{-12}$, $10^{-11}$ or $10^{-10}$.  These
choices yielded a representative set of  templates in LSST passbands.
No attempt was made to compute the total rate of superstring 
microlensing from first principles; 
instead, an arbitrary choice of \NgenuLensSTRING\ events were generated.

A small number of stellar sources were selected from the Besan\c{c}on
galactic model \citep{Robin2003} in the direction $(\ell,b) = (323.2, -6.8)$
degrees with fluxes $22 < g, i < 24$; each source has catalog-derived
distance, kinematics, and colors. 
String loop lenses were randomly drawn from a model of the galactic distribution of
loops \citep{Chernoff:2017fll} restricted to the line of sight to the
source; each loop had tension, invariant length, orientation, phase of
oscillation, loop configuration (4 types of loop trajectories were
considered with cusps and/or kinks) and center of mass velocity
consistent with halo kinematics. The string and star combinations form
a fair sample of geometric alignments of source, lens, and
observer. 
Each alignment gives a deterministic sequence of microlensing events. 
The duration of each event and the timespan for the repetitions were calculated for
each pair. 
The timescale of a single microlensing event is proportional to string tension,
which has a  broad range of theoretical uncertainty; this timescale
ranges from less than 1 sec to hours. 
Templates for events with timescale $<3$ sec were omitted from
the catalog since the average flux enhancement is limited by the
minimum LSST exposure.  Likewise, the timescale for the full set of
$\sim 10^3$ repetitions ranged from months to thousands of years. 
We used a randomly selected portion of each template light curve, 
corresponding to the experiment's duration.

\medskip {\bf Rate Model:} 
There are no viable rate estimates for this hypothetical source;
{\NgenuLensSTRING} \modelNameuLensSTRING\ were generated and none
satisfied the 2-detection trigger (\S\ref{subsec:trigger_model}).

\section{Models-II: Photometric Redshifts from the Host Galaxy }
\label{sec:model_zphot}

\newcommand{\DM}{D_M}   
\newcommand{\NCM}{N_{\rm CM}} 
\newcommand{\zphoterr}{\delta z_{\rm phot}}
\newcommand{\Dz}{\Delta z}        
\newcommand{\sigIQR}{\sigma_{\rm IQR}}
\newcommand{\ztrue}{z_{\rm true}}
\newcommand{\zphot}{z_{\rm phot}}
\newcommand{\Ndof}{N_{\rm dof}}

For \exgal\ models, photometric redshifts are based on 
a library of galaxies characterized by  a true redshift ($\ztrue$), 
photometric redshift ($\zphot$), and photo-$z$ uncertainty ($\zphoterr$).
Here we describe the creation of this library, and \S\ref{subsec:sim_zphot} 
describes how this library is used in the simulation.
For \Gal\ models, $\zphot=0$ because we do not model random associations 
with a distant galaxy.

We use a method based on the color-matched nearest neighbors (CMNN)
photometric redshift estimator from \citet[][hereafter G18]{Graham2018}, 
which is comparable to the photo-$z$ estimators
presented by \cite{Ball2008} and \cite{Sheldon2012}. 
The CMNN estimator is not intended to provide the best photometric redshifts 
for LSST data, but was developed as an analysis tool to assess how the LSST 
survey parameters, and the projected LSST photometric depths, 
can affect the bulk quality of the photometric redshifts. 
Examples of other photo-$z$ methods used on existing data are
described in \citet{Hoyle2018_DES} and \citet{Tanaka2018_HSC}.
A future generation of photo-$z$ estimators intended for scientific analyses
is currently in development (e.g., \citealt{Speagle2016,Sadeh2016,Leistedt2017}).

The basic idea for CMNN is to select a training set of galaxies and define a 
distance metric in a five-dimensional color space. For galaxies that
are not in the training set (i.e., the test set), the color space distances
to nearest neighbors in the training set are used to determine the photo-$z$ and its \unc.

\begin{figure}
\begin{center}
 \includegraphics[scale=0.21]{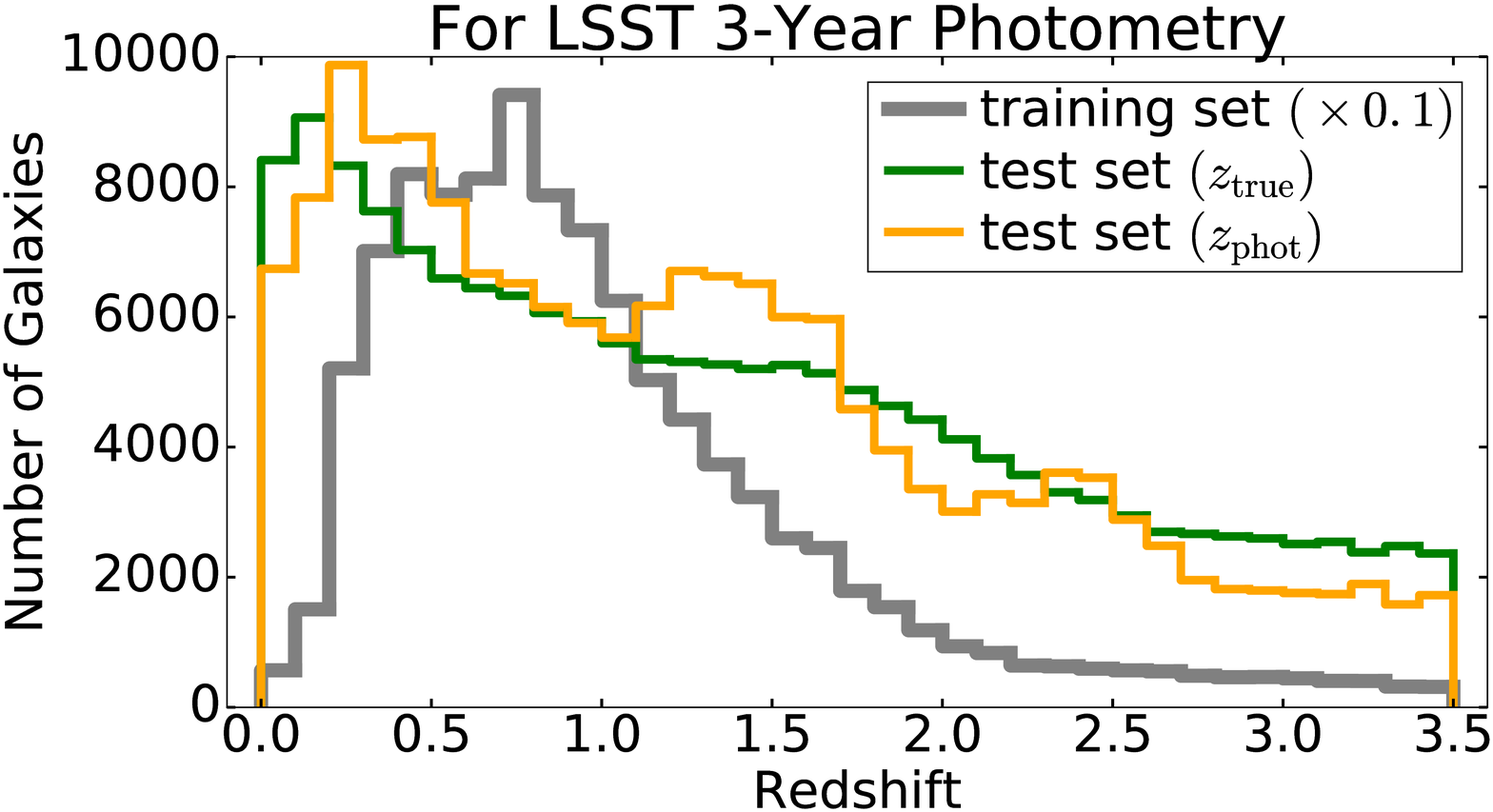}
\end{center}
  \vspace{-0.2in}
\caption{
    True (green) and photometric (orange) redshift distributions of the test set of galaxies used for \acro, 
    along with the true redshift distribution of the training set (grey; scaled by $0.1$). 
    }
\label{fig:Nz} \end{figure}

The G18 method is trained on a galaxy {\it training} set with known redshifts, 
and then we construct a {\it test} set for \acro. Note that the galaxy training and test 
sets are different than the \acro\ training and test sets in Table~\ref{tb:models}.
For both the training and test sets we use a catalog of simulated galaxies 
based on the Millennium simulation \citep{Springel2005}, constructed using the
lightcone construction techniques described by
\cite{Merson2013}\footnote{\url{http://galaxy-catalogue.dur.ac.uk}}. 
As described in G18, this catalog was designed to serve as a realistic representation 
of future LSST catalogs. The training set of galaxies is essentially the same as that
used in G18: $\sim10^6$ galaxies with a  photometric depth equivalent to a
10-year survey: $26.1, 27.4, 27.5, 26.8, 26.1, 24.9$~mag in
\bands. This represents a plausible spectroscopic sample of true redshifts  
with a realistic redshift distribution (gray line in Fig.~\ref{fig:Nz}).

The test set of galaxies that we use
as the foundation library for assigning photo-$z$ to the \acro\ samples
is significantly different from G18: 
${\sim}1.7{\times}10^5$ galaxies limited to the photometric depth after
3-years of the LSST survey: $25.4, 26.7, 26.9, 26.2, 25.4, 24.2$~mag in \bands. 
Furthermore, although the training set is drawn randomly from the catalog, 
we have randomly selected a larger
number of low-redshift galaxies in the test set (green line in Fig.~\ref{fig:Nz}).
This enhancement serves as a more appropriate library for the \acro\ sample,
and it avoids artifacts from having to re-use the same low-$z$ library galaxy
multiple times.

CMNN uses a training set of galaxies with known redshifts to estimate a
photo-$z$ for each galaxy in a test set 
(in this case, the library used for \acro). 
The simulated galaxy catalog used to generate the
library is discussed below. The CMNN estimator first identifies 
a color-matched subset of training galaxies by calculating the 
Mahalanobis distance  ($\DM$) in color space
between the test galaxy and all training-set galaxies, 
and then applies a $\DM$ threshold value that defines a good color match. 
For a given number of colors ($\Ndof)$,
the value of this threshold is set by a percent point function (PPF),
the percentage of training galaxies that have a good color match.
As an example, with ${\rm PPF}=0.68$ and $\Ndof=5$, applying a threshold
of $\DM<5.86$ will identify $68\%$ of the training galaxies that have a
good color match. Following G18, we use PPF$=0.68$ for {\acro}
and define $\NCM$ to be the number of color-matched galaxies.

The CMNN estimator randomly chooses one training-set galaxy from the
subset of those with a good color match, weighted by $\DM^{-1}$, 
and uses that training galaxy's known redshift as the test-set galaxy's photo-$z$. 
The photo-$z$ uncertainty ($\zphoterr$) for the test galaxy is the standard deviation 
in the true redshifts of the color-matched subset of training galaxies. 
In the rare cases when there are $\NCM<10$ color-matched training galaxies
in the subset, the $10$ nearest neighbors are used by default. 
The PPF value corresponding to the $10^{\rm th}$ nearest neighbor's $\DM$ is 
calculated (${\rm PPF}_{10}$), and $\zphoterr$
is multiplied by a factor of ${\rm PPF}_{10}/0.68$ to account for the degraded
quality  of the color-matched subset. 
Compared with G18, this implementation of the CMNN estimator is slightly different 
to ensure that all galaxies are assigned a photo-$z$ 
(i.e., no test galaxies fail to obtain a photo-$z$ estimate).

To improve processing speed, we have applied both the color and magnitude
preselection criteria to the full training set, as shown in \S3.3 of G18. 
The color cut has little effect, but the magnitude cut effectively works as a 
``pseudo-prior" by limiting the training set to the 20\% of training galaxies 
with an $i$-band magnitude nearest to the test galaxy's $i$-band magnitude. 
This means that all test-set galaxies with $i>25$ use the same $20\%$ of the 
faintest galaxies. 
The pseudo-prior may improve accuracy for some photo-$z$ estimates, 
but it also introduces a small redshift bias. 

\begin{figure}
\begin{center}
  \includegraphics[scale=0.4]{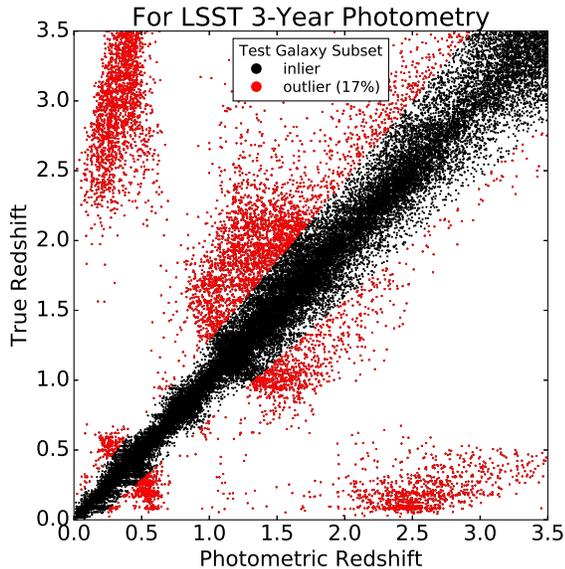}
\end{center}
  \vspace{-0.2in}
\caption{
    True vs. photometric redshift for $50,000$ randomly chosen galaxies 
    in the test set.  Outliers (defined in the text) are colored red. 
      }
\label{fig:tzpz} \end{figure}
\begin{figure}
\begin{center}
 \includegraphics[scale=0.4]{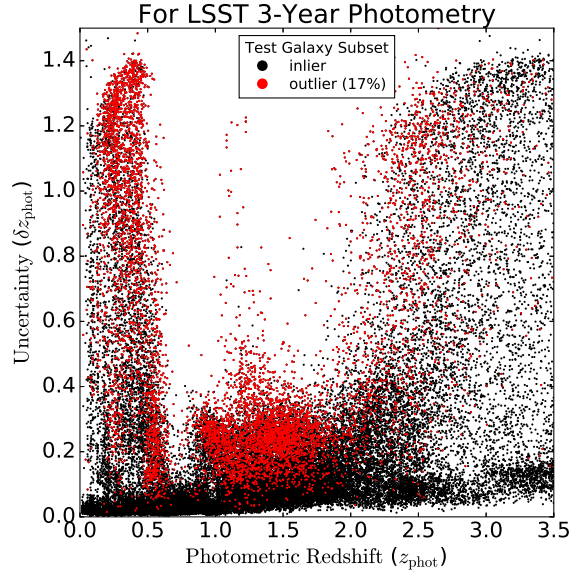}
\end{center}
  \vspace{-0.2in}
\caption{Estimated photo-$z$ uncertainty ($\zphoterr$)  vs. $z_{\rm phot}$ for a subset 
    of test set galaxies.
    As in Fig.~\ref{fig:tzpz}, outliers are colored red.    
    For $z_{\rm phot} < 0.5$, galaxies with a large uncertainty are mostly catastrophic outliers. 
     }
\label{fig:pzpze} \end{figure}

To illustrate the results of our implementation of the CMNN estimator for
generating the \acro\ photo-$z$ library, we plot $\ztrue$ vs. $\zphot$
for the test set of galaxies in Fig.~\ref{fig:tzpz}
(for visual clarity we show a random subset of $50,000$ test galaxies). 
We define the photo-$z$ residual (or error) of a test galaxy to be 
$\Dz = (\ztrue - \zphot)/(1+\zphot)$, and identify outliers 
(red points) as test-set galaxies with 
$\vert\Dz\vert > 3\sigIQR$ or $>0.06$, 
whichever is {\it larger}, where $\sigIQR$ is the robust
standard deviation in $\Dz$ over the full redshift range 
(i.e., converted from the width of the interquartile range, IQR). 
In Fig.~\ref{fig:pzpze} we show the estimated
photo-$z$ uncertainty ($\zphoterr$) 
as a function of $\zphot$, again with outliers as red points.

\begin{figure}
\begin{center}
  \includegraphics[scale=0.3]{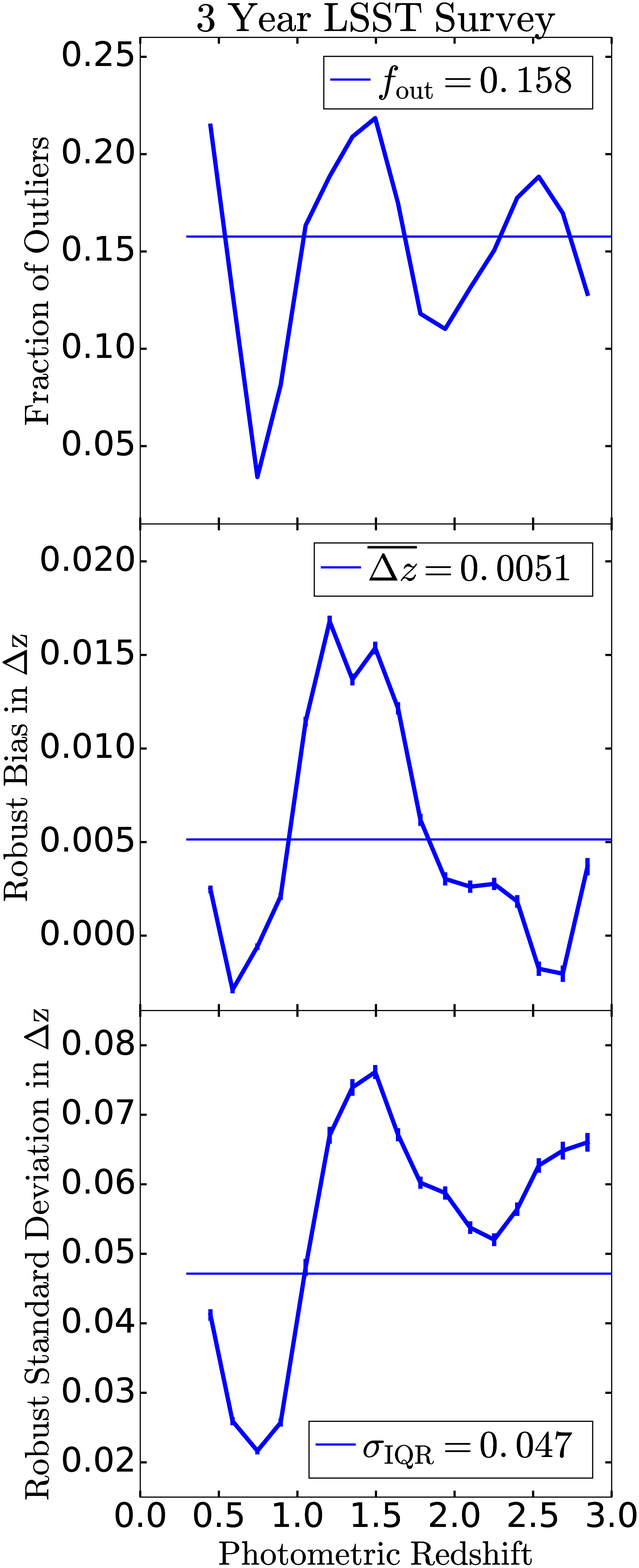}
\end{center}
  \vspace{-0.2in}
\caption{ Photo-$z$ performance measures: 
    outlier fraction (top), bias (middle), and standard deviation (bottom).
    The average value across all bins is shown as a horizontal line.  
    }
\label{fig:photoz_stats} \end{figure}

Fig.~\ref{fig:photoz_stats} shows performance summaries in bins of $\zphot$.
The top panel shows the fraction of photo-$z$ outliers; 
the fraction varies from 0.05 to 0.2 as a function of $\zphot$,
with an average of 0.158.
The middle panel shows the {\it robust} $\Dz$ bias
for test galaxies within the interquartile range (IQR; the middle 50\%);
the bias varies from $-0.005$ to $+0.015$ as a function of $\zphot$,
with an average $\Dz$ bias is $0.005$.
The bottom panel shows the robust standard deviation of $\Dz$;
it varies from 0.02 to 0.08 as a function of $\zphot$,
and the average is 0.047.

We note that the clouds of catastrophic outliers ($\vert\ztrue - \zphot\vert>2$) 
in Fig.~\ref{fig:tzpz} are quite large, which might cause trouble for 
classifiers using the \acro\ photo-$z$. 
However, Fig.~\ref{fig:pzpze} shows that the CMNN estimator produces a photo-$z$ 
uncertainty ($\zphoterr$) that is large for catastrophic outliers (as it should be).

\section{Simulation}
\label{sec:sim}

\begin{figure*}  
\begin{center}
    \includegraphics[scale=0.6]{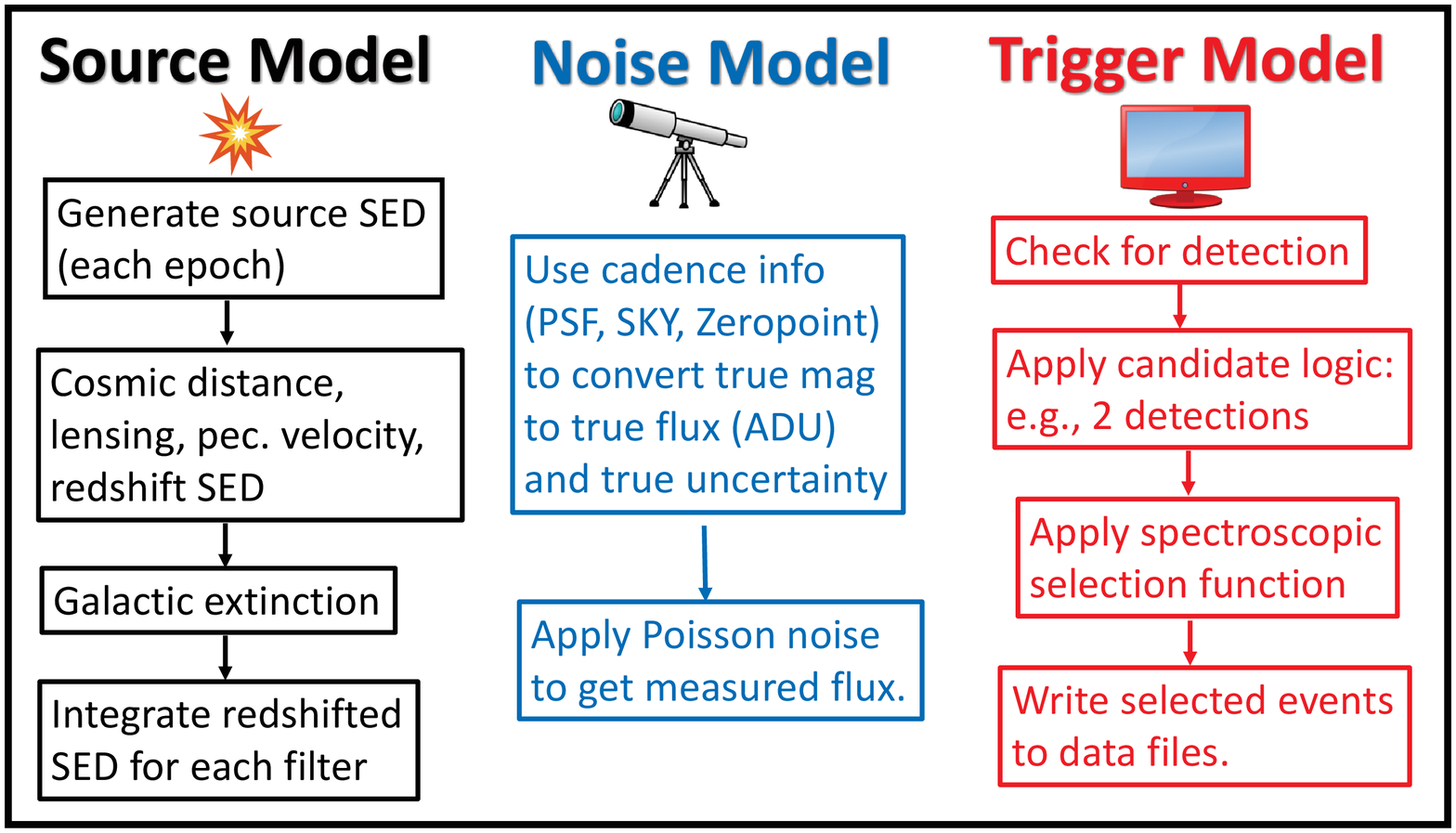}
 \end{center}
     \vspace{-0.9in}
 \caption{ Flow chart of stages in the \SNANA\ simulation of \exgal\ events. \\  \\ }
\label{fig:flowChart}  \end{figure*}

We use the simulation code from \SNANA\ \citep{SNANA};
an updated and detailed description of the simulation code is given in 
\citet[hereafter K18]{K18_SIM}. 
Here we give a brief and less technical description based on the overview
shown in Fig.~\ref{fig:flowChart}.

\begin{figure}  
  \includegraphics[scale=0.33]{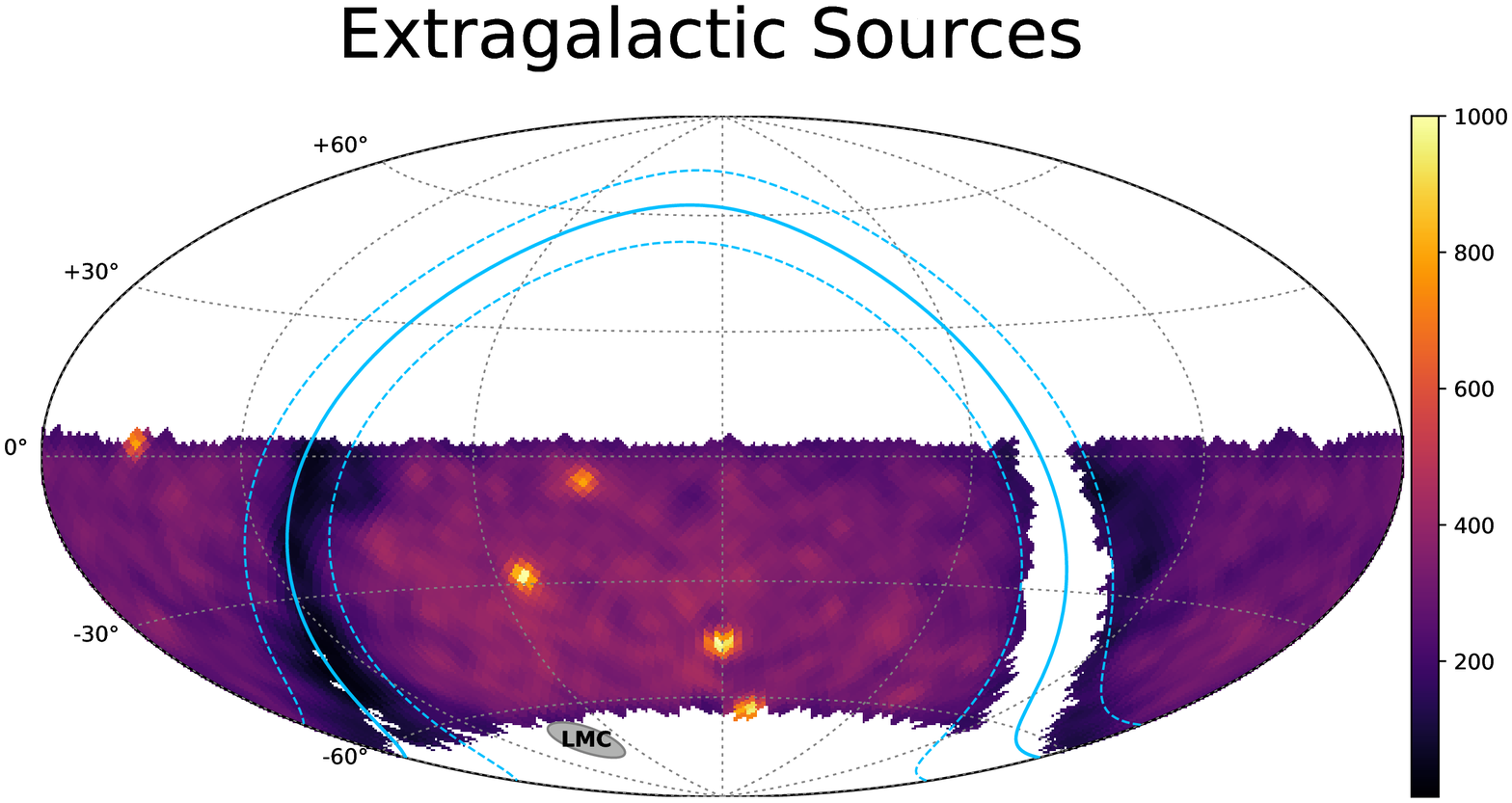}
  \includegraphics[scale=0.33]{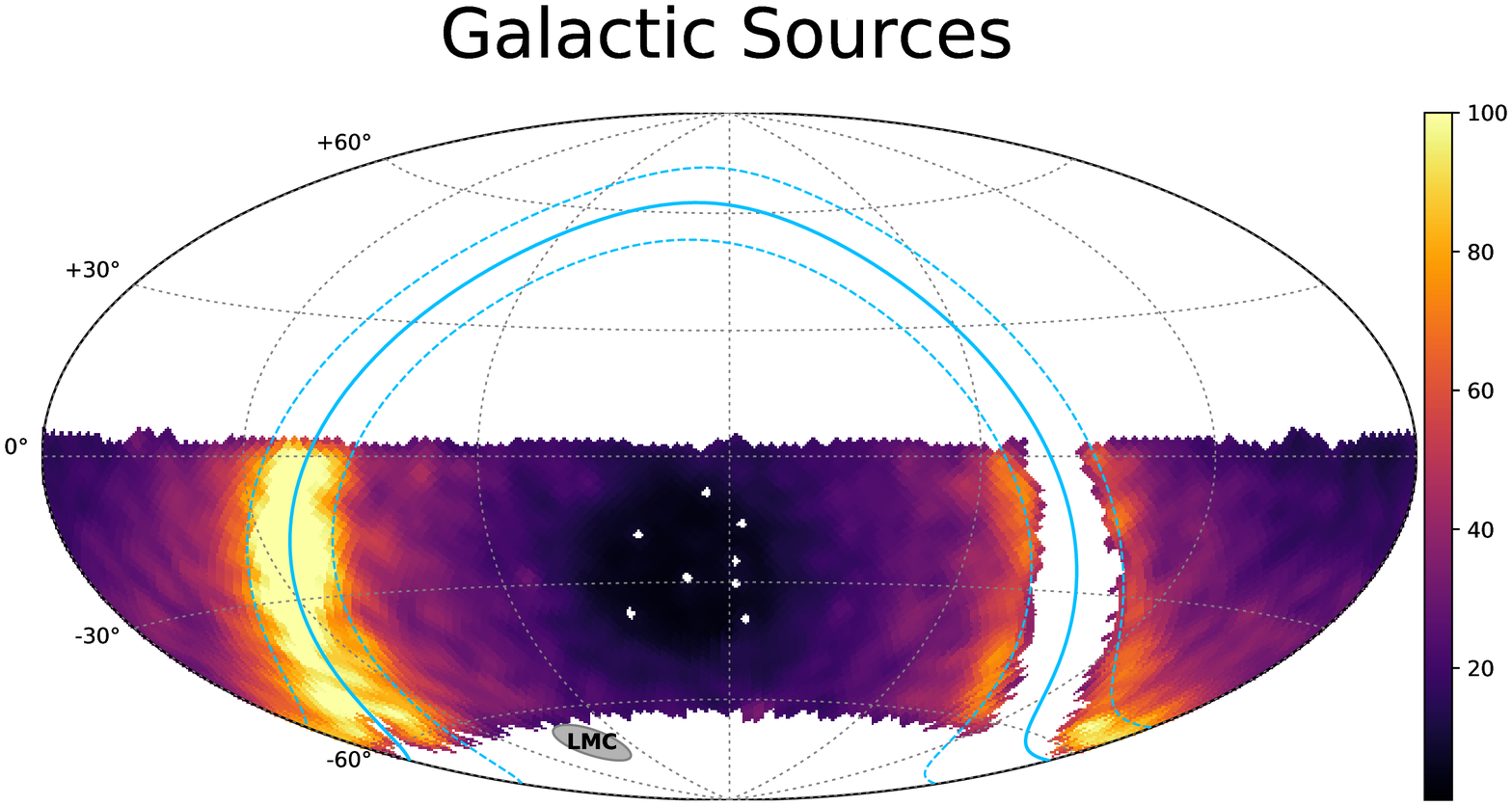}
  \caption{
     Sky maps of \acro\ events shown in Hammer-Aitoff projection. 
     The maps are generated with HEALPix NSIDE=32 
     and corresponds to a pixel size of 3.35~deg$^2$, 
     or about 1/3 of the the LSST field of view (9.6~deg$^2$). 
     Top panel is for \exgal\ sources; the 5 bright spots correspond to the DDFs,
     and the low-density bands overlap the Galactic plane (indicated with blue lines).
     Bottom panel is for \Gal\ sources. \\
  }
\label{fig:sky}  \end{figure}

\subsection{Source Model}
\label{subsec:source_model}

Here we describe the simulation stages under ``Source Model" 
in Fig.~\ref{fig:flowChart}. These stages correspond to \exgal\ models
described by rest-frame SEDs (\S\ref{sec:models_source}).
For \Gal\ models, these stages are replaced by precomputed 
magnitudes.

\subsubsection{Model Enhancements}
\label{sss:enhance}

While the models in \S\ref{sec:models_source} are packaged as libraries 
the first step of the simulation is to provide a few model enhancements to avoid
unphysical light-curve artifacts, and to include host-galaxy extinction.
The first enhancement for SED-based models is related to the finite time range, 
typically a few hundred days.  To avoid unphysical light-curve truncation, the
magnitudes are linearly extrapolated. To reduce pathologies from noisy models
at late times, the extrapolation is based on a least-squares fit to the last five days.

The next enhancement is to extrapolate fluxes into the far ultraviolet (UV) region so that
$u$ band is defined at all redshifts. The blue edge of the $u$ band is ${\sim}3000$~\AA,
and thus at a maximum \acro\ redshift of $z=3.5$, 
this band probes the SED down to a wavelength of ${\sim}670$~\AA.
The SED models typically extend down to 1000 or 2000~\AA, and therefore
the $u$ band (and sometimes $g$ band) flux is not defined at high redshifts.
Using the default SED models, these undefined passband fluxes would have been
excluded from the output data files,
and these drop-out artifacts could have been used as a feature in 
classification codes.
To avoid UV drop-out artifacts in \acro, 
and since real data will not have such artifacts,
the SED flux at the bluest defined wavelength was linearly
extrapolated down to zero flux at $500$~\AA. This extrapolation was performed
in each time bin for the SED grid. The resulting $u$-band model fluxes are 
negligible, and thus the reported $u$-band fluxes are dominated by
sky noise fluctuations.

Data-driven models are assumed to include the effects of
host-galaxy dust, which preferentially absorbs blue light and re-emits 
in the red, making the source appear redder from outside its host galaxy.
The theoretically based models ({\mosfit}-generated and {\modelNameKN})
do not include dust, and thus for these models, we include
extinction from dust described by $A_V$, the magnitude dimming at 5500~\AA.
The dimming at other wavelengths is determined by the same
color law used to describe Milky Way reddening \citep{Fitz1999}.
For all \mosfit\ models except \modelNameTDE, 
$A_V$ is selected from a ``Galactic Line of Sight" distribution (Eq.~2 of \citealt{WV07})
consisting of a Gaussian core ($\sigma=0.1$~mag) and exponential tail ($\tau=0.4$~mag).
\modelNameTDE\  are expected to be near galactic centers, 
and thus only the exponential component is used.
\modelNameKN\ are expected to have large kicks, and thus the weight of the exponential component reduced by a factor of two.

\newcommand{\TLSST}{T_{\rm LSST}}
\newcommand{\TuLens}{T_{\mu\rm Lens}}

The final enhancement is related to the probability of selecting nonrecurring 
\Gal\ events from the library of model light curves;
this includes \modelNameuLensSingle\ and \modelNameuLensBinary\ models.
Events with longer duration are more likely to overlap with the LSST survey time,
and we therefore select microlensing light curves with probability proportional
to $\TLSST + \TuLens$, where $\TLSST$ is the survey duration (3 years)
and $\TuLens$ is the duration of the microlensing event. 
Events with $\TuLens \ll \TLSST$ have very nearly equal probabilities,
while events with $\TuLens \gg \TLSST$ have selection probability 
roughly proportional to $\TuLens$.
For recurring \Gal\ events, each model light curve is selected with uniform
probability.

\subsubsection{\Exgal\ Source Model and Photon Voyage to Earth}
\label{sss:sim_photons}

The  \exgal\ models in \S\ref{sec:models_source} describe the SED at a 
distance of 10~pc from the source. Here we describe the 
the lower 3 panels under ``Source Model' in Fig.~\ref{fig:flowChart}:
simulation of the photons' journey through an expanding universe, 
through the Milky Way to the top of Earth's atmosphere,
and through the LSST passbands.

For distance $D$ within our Galaxy, the apparent brightness
of a source follows the well known inverse-square law, $1/D^2$.
For a source outside our Galaxy, the effect of cosmic distance in an expanding universe 
is characterized by replacing $D$ with a luminosity distance ($D_L$),
which depends on cosmological model parameters as defined in Eq.~2 of K18. 
For \acro, we used the following parameters to compute $D_L$:
matter density $\OM=0.30$, dark energy density $\OL=0.70$, and
dark energy equation of state parameter $w=-1$.
The apparent magnitude of each \exgal\ source includes a distance modulus ($\mu$)
defined as $\mu = -2.5\log(10{\rm pc}/D_L^2)$, and therefore the intrinsic
brightness for each model is defined at a distance of 10~pc.

As the light travels to earth, there are weak lensing effects in which
the gravitational potential from galaxies near the light path trajectory 
can magnify or demagnify the source. 
To model this effect, a convergence distribution is determined from a 
900~deg$^2$ patch of the MICECAT $N$-body simulation \citep{MICE2015}.
The magnification probability distribution is asymmetric and increases with redshift 
(\S5.4 of K18);
the root mean square (rms) of the magnification is roughly $0.05$ times the redshift.

Extragalactic sources reside in galaxies, and these galaxies have random
``peculiar velocities'' with respect to the cosmological redshift. The simulation
selects a random velocity from a Gaussian distribution with 
$\sigma_v=300$~km/s (\S5.5 of K18),
which is equivalent to a redshift error of $\sigma_z = 0.001$.

As the light enters the Milky Way, it travels through dust similar
to the dust from the host galaxy. Instead of using the $A_V$ distribution
from the host galaxy, we use a map of the color excess, $E(B-V)$,
that has been precisely measured at every sky location \citep{SFD98}.
The extinction is given by $A_V = R_V \times E(B-V)$ where $R_V=3.1$.
The color law from \citet{Fitz1999} is used to determine the extinction 
as a function of wavelength. 
The \unc\ on $E(B-V)$ is $0.16{\times}E(B-V)$.

As the light reaches the top of our atmosphere, the  redshifted SED is integrated over
the wavelength range for each \bands\ filter, resulting in six {\it true} fluxes.
The conversion of true flux into measured flux is described below in 
\S\ref{subsec:noise_model}.

\subsection{Noise Model}
\label{subsec:noise_model}

Here we describe the simulation stages under ``Noise Model" 
in Fig.~\ref{fig:flowChart}. These stages are applied identically
to both the \exgal\ and \Gal\ models.

\subsubsection{Model of Observing Conditions and Cadence}
\label{sss:OpSim}

As described in \S\ref{sec:LSST}, we simulate two components of the survey. 
The first component is the Wide-Fast-Deep (WFD), covering 17,950~deg$^2$ 
(44\% of the entire sky).
The second component is the set of Deep-Drilling Fields (DDF), 
which includes 5 telescope pointings covering 47.6~deg$^2$.
For events passing the trigger (\S\ref{subsec:trigger_model}),
the \acro\ sky densities for both the WFD and DDF are shown 
as HEALPix maps\footnote{\URLHEALPIX} \citep{HEALPIX2005}
in Fig.~\ref{fig:sky}.
Note that the \exgal\ sources have nearly uniform coverage over the
WFD area, while the Galactic events cluster much closer to the Galactic plane.

The sky coverage and sequence of observations are adapted from a baseline cadence 
published by LSST using the Operations Simulator, hereafter referred to as ``\OPSIM''
\citep{LSST_OPSIM,LSST_OPSIM2,LSST_OPSIM3}.
\OPSIM\ includes a prototype scheduler that queues LSST \obss\ to optimize 
science programs (\S\ref{sec:LSST}) while accounting for a 
high-fidelity model of the telescope, and also accounting for real-time environmental 
factors such as weather, seeing, clouds, sky brightness, and maintenance downtime.
We use \OPSIM\ version 3 that was available when the \acro\ data set was generated.
The following discussion is for the 10-year survey, which was truncated
to the first 3 years for the \acro\ simulation.

\OPSIM\ incorporates observing conditions with a time-dependent model of
seeing, cloudiness, a dark sky spectrum, 
a model to compute the contribution of the moon to the sky brightness~\citep{Krisciunas1991},
and a model of twilight behavior.
The seeing model is based on two years of data at Cerro Pachon 
recorded every 5 minutes. The cloudiness was measured 
at Cerro Tololo during the same time period,
with 0 being completely clear and 1 being completely cloudy.
These data are repeated to cover the 10-year duration of LSST, 
and nearest-neighbor interpolation is used to determine values between 
measurements.
A global open-shutter time constraint is applied to match the 10-year 
duration of LSST.
Exposures of 30~seconds are scheduled when the sun altitude is below 
$-12$~deg (horizon is at 0~deg), the airmass is below $1.5$, 
and the distance to the moon (regardless of the phase) is greater than $30$~deg. 

At any given time, the choice of science program and sky location is based on a 
greedy optimization algorithm  to maximize the number of visits during a 
fixed block of time, and to also minimize slew time.
The optimization algorithm tracks the fraction of completed visits for each science
program in each spatial region, and computes the ideal fraction based on a 
uniform temporal distribution of visits. Priorities are adjusted to make the
observed fractions match more closely to the ideal fractions.
In addition, redder bands are given higher priority during twilight,
bluer bands are preferred during dark time (no moon), and
long temporal gaps are avoided for the transient science program.
Finally, dithers up to 1.75~deg \citep{Krughoff2016}
are added to each WFD visit to  cover chip gaps, spatially smooth the coadded depth, 
and randomize location and orientation dependence 
to reduce systematic biases in measurements of large scale structure and weak lensing
(eg. see \citealt{Carroll2014, Awan2016, Marshal2017}). 
The final output is a publicly available LSST \obs\ database\footnote{\URLOPSIM}
of telescope pointings that includes  
sky coordinates, time of observation, bandpass, and
quantities characterizing the observational conditions including  
airmass, point spread function (PSF), sky brightness, and $5\sigma$ depth.

For the public LSST \obs\ database used for \acro, the sky-brightness model 
was in development and tends to be overly conservative. 
We therefore applied an improved sky-brightness model from \citet{Yoachim2016}, 
where the largest changes are in the $z$ and $y$ bands 
during twilight.\footnote{In the public \OPSIM\ output, $z$ and $y$ 
   band sky brightness during twilight is constant with no variations.}
On average, the resulting \obs\ depths for \acro\ are a few tenths of a mag deeper
compared to using the public database.

The final step is to translate the \OPSIM\ output \citep{Biswas2019}
into an \obs\ library
for the \SNANA\ simulation (\S\ref{subsec:noise_model}), and to truncate
the 10-year survey to the first 3 years for \acro. 
While an average observing season is ${\sim}6$ months,
the sharp 3-year cutoff results in some short seasonal fragments
in the third year. To reduce classification difficulties from
this season-truncation artifact, seasonal fragments less than
30 days were not used.

The \OPSIM\ translation results in a 3-year \obs\ history for \NSKYLOCWFD\ 
random sky locations within the WFD footprint;
for DDF, \NSKYLOCDDF\ sky locations are used.
The number of sky locations is a compromise between dense
sky sampling and library size. 
The library sky density is ${\sim}2.8$ locations per square-degree,
and thus for the 117 million generated events (Table~\ref{tb:models}),
each sky location is re-used an average of ${\sim}2,300$ times.
For the 3.5~million events passing the trigger, each sky location
is re-used an average of ${\sim}70$ times.
After co-adding \obss\ within each night and passband,
the average number of nightly 
passband visits\footnote{Observing each of the \bands\ passbands in one 
    night counts as 6 passband visits.}
over 3 years is \NOBSAVGWFD\ for WFD and \NOBSAVGDDF\ for DDF.

\subsubsection{Instrumental Flux and Noise}
\label{sss:sim_noise}

Starting from the top of the atmosphere, the simulation would ideally trace the 
light through the atmosphere and the LSST instrument. While such simulations
have been used for the LSST design (e.g., \citealt{Tyson2014,Peterson2015,Rowe2015}), 
they are very CPU intensive, especially for the 100 million sources that 
were generated for \acro.
Here we compute the observed flux and uncertainty from an ``\obs\ library''
(\S6.1 of K18), which  consists of random sky locations and a list of \obss\
at each location. For each observation the following information is included:
modified Julian date (MJD), passband, sky noise, 
size of PSF, and zero point. 
The observation library for the \SNANA\ simulation was created by translating the 
\OPSIM\ output described in \S\ref{sss:OpSim}.
For a given true flux at the top of the atmosphere, the \obs\ properties are used
to compute the measured flux and uncertainty as described in 
\S6.3 of K18.

In addition to modeling the flux from the source, we also model flux 
in the reference images, also commonly called template images.
The fluxes in the data files are flux differences, $\DF$, 
corresponding to ``search$-$reference" fluxes that are expected to be produced 
from the LSST image-subtraction pipeline
\citep{LSST2018_IMPROC}.\footnote{While we make assumptions about images, 
    no images were used in these simulations.}
For \exgal\ transients, we assume that the source brightens after the 
reference images have been acquired, and therefore the reference flux is 
exactly equal to the search-image flux without the source.
$\DF$ is therefore positive when the source is bright,
and includes Gaussian sky noise. When the true source flux is zero, 
$\DF$ reduces to a Gaussian distribution centered at zero flux.
For the \Gal\ and \modelNameAGN\ models, the reference flux is modeled 
as an average of 4 snapshots taken over 4 consecutive days prior to the 
start of LSST operations. The resulting $\DF$ is therefore negative or positive.

The observed fluxes are determined in units of photoelectrons ({\tt GAIN=1}).
We simulate the common practice of ``forced photometry,'' 
where for each detected object the fluxes are measured 
at all previous and future \obss\ at the same location.
Simulated forced photometry means that for each object 
satisfying the \acro\ trigger (\S\ref{subsec:trigger_model}), 
all \obss\ are recorded for the
entire 3-year duration, regardless of the signal-to-noise ratio (S/N). 
The calibrated fluxes and \uncs\ in the data files are on a common \SNANA\ 
zero-point of 27.5. The calibrated flux does not correspond to a physical unit,
but was arbitrarily chosen during the SDSS-II Supernova analysis \citep{K09}, 
before \SNANA\ existed,  so that the calibrated SDSS flux has approximate
units of photoelectrons. 

Previous analyses have reported anomalous flux scatter from bright host galaxies; 
for DES, see Figs~9-10 of \citet{Kessler2015} and Fig.~5 of \citet{SMP};
for Pan-STARRS-1 (PS1), see Fig.~3 of \citet{Jones2017}.
For \acro, we ignored these effects.

\subsection{Trigger Model}
\label{subsec:trigger_model}

Here we describe the simulation stages under ``Trigger Model" 
in Fig.~\ref{fig:flowChart}. These stages are applied identically
to both the \exgal\ and \Gal\ models.

Monitoring transient activity for every CCD pixel is not practical,
and therefore we follow the common practice of using a
{\it software trigger} to reduce the pixel data into catalog of objects 
with time-varying brightness. 
The trigger initiates photometric measurements on all previous and 
future \obss\ for each object in the catalog.
For the \acro\ trigger, we use assumptions based on the 
DES supernova program (DES-SN) described in \citet[hereafter K15]{Kessler2015}.

In our \acro\ simulation, a transient source must satisfy a trigger 
to be written out to the data files.
While LSST plans to identify single-epoch detections for asteroid searches,
we impose a transient trigger intended to reject moving objects by
requiring 2 detections separated by at least 30 minutes. 
While asteroids with slow proper motions can still satisfy this trigger, 
our simulation does not include such objects.
A detection is a group of CCD pixels with excess flux compared
with a reference image, and a flux profile consistent with the PSF. 
We do not simulate pixel data, but instead use a DES-SN detection model 
from K15, which is based on analyzing artificial point sources 
overlaid on CCD images.

This model describes the detection
\eff\ as a function of true S/N (Fig.~8 of K15), $\SNRtrue$, 
which is computed from the true source brightness, sky noise, and PSF. 
The detection efficiency is 50\% at $\SNRtrue{\sim}5$, and is nearly 100\% at $\SNRtrue{\sim}10$.  In addition to using these \eff\ curves, we also
required $\SNRtrue>3$ to avoid spurious detections on very 
low-S/N \obss.\footnote{\label{fn:SNR} While using $\SNRtrue$ for choosing detection
probability is valid, the $\SNRtrue>3$ requirement is a subtle mistake; 
this cut should have been applied to measured SNR.
This mistake does not cause leakage in \acro, but might have resulted in 
subtle data/simulation discrepancies if real data had been available.} 

DES-SN only detected positive fluxes, meaning that the flux was required to be 
larger than the presurvey template flux.
To allow for variables and longer-lived transients in LSST, a \acro\ detection
is based on the absolute value of S/N, and thus increasing and decreasing
fluxes (with respect to template flux) have the same probability of being detected.

\subsection{ Spectroscopically Confirmed Training Subset}
\label{subsec:sim_train}

\newcommand{\fourMOST}{{\sf 4MOST}}
\newcommand{\effSpec}{\epsilon_{\rm spec}}

A small subset of triggered events were flagged as \specy\ identified and used for the
training set.  We loosely model the training set based on 
\fourMOST,\footnote{{\fourMOST}: 4-m Multi-Object Spectroscopic Telescope 
({\tt https://www.4most.eu/cms})}
a \spec\ instrument with 2400 fibers that is currently under construction,
and a proposed ``Time Domain Spectroscopic Survey'' \citep{TIDES2019}.

We used the DES characterization of the \spec\ identification efficiency ($\effSpec$)
as a function of peak $i$ band magnitude, $m_i$.
Specifically, we used Fig.~4 of K18 where $\effSpec$ falls to 50\% 
at around $m_i\simeq 22$~mag, but shifted the curve by 0.2~mag to reflect
an assumption that the \fourMOST\ spectrograph will be 0.2~mag deeper than 
spectrographs used for DES.
The anticipated \fourMOST\ improvements include better seeing and fiber efficiency. 
The expected number of \spec\ classifications for \fourMOST\ is below $10^4$,
and thus in addition to the $i$-band dependent $\effSpec$,
prescale fractions in Table~\ref{tb:spec_prescale} were also applied.
A prescale of 0.008, for example, means that that the $\effSpec$ curve
(Fig.~4 of K18) is multiplied by 0.008. 
The total number of events in the training set is \NtrainTOTAL, and almost 3/4 of the
training set events are in the WFD. The largest classes in the training set are
\modelNameSNIa\ (30\%) and \modelNameSNII\ (15\%).

\begin{table} 
\caption{
  Prescale Fractions Applied to \Spec\ Identification Efficiency
  }     
\begin{center}
\begin{tabular}{ | l | l | l | }
\tableline 
   Model          &   \multicolumn{2}{c|}{Prescale for}  \\
   Num: Name  &   DDF   &  WFD  \\
    \hline
   \modelNumSNIa: \modelNameSNIa\     & 0.500 & 0.003 \\
   \modelNumSNII: \modelNameSNII\       &  1.000 & 0.008 \\
   \modelNumSNIbc: \modelNameSNIbc\  & 1.000 & 0.008 \\
   \modelNumbg: \modelNamebg\              & 1.000 & 0.008 \\
   \modelNumSNIax:  \modelNameSNIax\           & 1.000 & 0.008 \\
   \modelNumKN: \modelNameKN\           & 1.000 & 1.000 \\
   \modelNumSLSN: \modelNameSLSN\  & 1.000 & 0.008 \\
   \modelNumTDE: \modelNameTDE\      & 1.000 & 0.200 \\
   \modelNumAGN: \modelNameAGN\     & 1.000 & 0.008 \\
  \modelNumRRL: \modelNameRRL\                    & 0.300 & 0.0008 \\ 
  \modelNumMdwarf: \modelNameMdwarf\           & 1.000 & 0.008 \\
  \modelNumPHOEBE: \modelNamePHOEBE\    &  1.000 & 0.008 \\
  \modelNumMIRA: \modelNameMIRA\                   & 0.500 & 0.020 \\
  \modelNumuLensSingle: \modelNameuLensSingle\ & 0.500 & 0.100 \\
   \hline
\end{tabular}
\end{center} 
  \label{tb:spec_prescale}
\end{table}

\subsection{Photometric Redshift from Host Galaxy}
\label{subsec:sim_zphot}

\newcommand{\zsource}{z_{\rm source}}
\newcommand{\zphotran}{z_{\rm phot,ran}}

For \exgal\ transients, we use the photometric redshift library described in \S\ref{sec:model_zphot},
where each galaxy includes a true redshift ($\ztrue$), photometric redshift ($\zphot$),
and photometric redshift \unc\ ($\zphoterr$). 
For a true source redshift generated from the rate model,
$\zsource$, the simulation picks a random galaxy 
from the library satisfying
\begin{equation}
    | \ztrue - \zsource| < 0.01 + 0.05 \zsource~,
\end{equation}
where the tolerance provides a broader distribution of selected galaxies.
Defining $\zphotran$ as the photometric redshift associated with the randomly selected galaxy,
the photometric redshift for the source is
\begin{equation}
    \zphot = \zphotran + (\zsource-\ztrue)~,
\end{equation}
and $\zphoterr$ is the reported 68\% confidence \unc.
For more details, see \S6.2 of K18.

\subsection{ Spectroscopic Redshifts from Host Galaxy}
\label{subsec:sim_zspec}

\begin{figure} 
\begin{center}
   \includegraphics[scale=0.4]{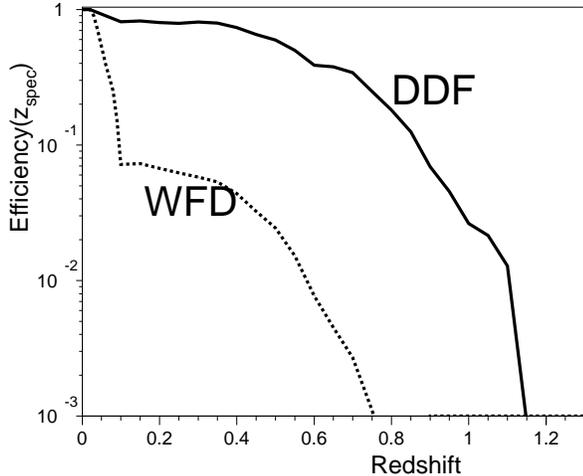}
\end{center}
\vspace{-0.2in}
  \caption{
   Efficiency vs. redshift for measuring an accurate \spec\ redshift from the host galaxy.
   The DDF spectra are stacked from multiple visits, and thus the
    DDF \eff\ is higher compared to WFD.
  }
\label{fig:eff_zspec}  \end{figure}

Using the \fourMOST\ instrument capabilities as a guide, 
we expect to obtain accurate \spec\ redshifts ($\zSpec$) for a small fraction 
of host galaxies.
Figure~\ref{fig:eff_zspec} shows the $\zSpec$ \eff\ vs. redshift for the WFD and DDF,
and these \eff\ curves were used in the simulation to select a random subset 
of 140,000 events (4\% of sample)
with a $\zSpec$ measurement. 
Both efficiencies are set to one at very low redshift ($z<0.02$)
to avoid photo-$z$ artifacts, and for $z>0.02$ the \eff\ drops to 
connect with the {\fourMOST}-based \eff.
The WFD \eff\ drops to 1\% around $z\sim 0.5$.
The DDF \eff\ drops to 1\% around $z\sim 1.1$;
many repeat DDF visits are stacked, resulting in higher \eff\ compared with WFD.

\subsection{ Special Features for {\acro} }
\label{subsec:sim_features}

Here we describe a few simulation features that were implemented specifically for \acro.
First, all fluxes are corrected for the measured \Gal\ extinction to prevent
professional astronomers from using their domain knowledge to gain an advantage 
over the Kaggle community.
The correction is based on measured $E(B-V)$,  which includes random 
Gaussian scatter about the true value with $\sigma = 0.16E(B-V)$. 
We also require $E(B-V) < 3$ to limit the extinction values.

The next issue concerns the default option of reporting fluxes only within a 
well-defined time window to limit the output data volume.  To prevent participants 
from using the duration of reported fluxes as a classification feature, 
we report a flux and \unc\ for every LSST \obs\ over the 3 year duration.
An approximate light-curve duration, however, can be obtained from
a boolean flag that was set for each detection (\S\ref{subsec:trigger_model}).

For LSST, the CCD pixels are expected to saturate for signals above
$\sim 10^5$ photoelectrons, which corresponds to sources brighter
than about $16^{\rm th}$~mag. The \acro\ team decided that accounting
for saturated \obss\ is an unnecessary distraction for the challenge,
and therefore we simulated a saturation level at $12^{\rm th}$~mag,
or 4~mag brighter than nominal.
In addition, we removed the small number of saturated \obss\ that remained.

Finally, the information released about \acro\ did not mention 
that the publicly available \SNANA\ simulation code was used to 
generate the data files. 
Nonetheless, we took the following precautions: 
removed {\acro}-related indicators from the \SNANA\ source code 
and documentation (e.g., name of cluster, names of rare models, etc...),
never publicly mentioned the name of the ``Midway'' computer 
cluster\footnote{\tt https://rcc.uchicago.edu} 
used to generate the data files,
and protected all {\acro}-related files on the Midway cluster.

\subsection{ Re-using Model SEDs and Light Curves }
\label{subsec:sim_reuse}

For most of the models, the number of generated events greatly exceeds
the number of SED time series for \exgal\ models,  
or light curves for \Gal\ models.
As an extreme case, \NgenPHOEBE\ events were generated for 
the \modelNamePHOEBE\ model,
but there are only \NtemplatePHOEBE\ model light curves to choose from.
When an SED time series or light curve is re-used, 
the simulated light curve is different for several reasons beyond
random Poisson noise.
For \exgal\ events (except AGN), each event has a different 
redshift, host-galaxy extinction, sky location, and cadence. 
For \Gal\ events (and AGN), each event has a different
sky location, cadence, initial phase, and reference-image flux.

\subsection{ Validation }
\label{subsec:sim_validate}

All of the models combined include roughly a million SEDs,
and a non-trivial task was to validate the simulated output, 
and in particular to minimize unphysical artifacts from software bugs and 
from artifacts in the newly created model libraries. 

Our primary tool was visual inspection of simulated light curves,
a task shared among a dozen astronomers from the {\acro}-validation team.
Artifacts were either fixed by the model developer, or fixed with upgrades
to the \SNANA\ simulation. The other main validation technique was
to inspect distributions of sky density, redshift, and luminosity functions.
Kaggle  performed their own internal tests, and they found an
interesting artifact that we could not explain: 
among the \NSIMOBS\ million observations included in the data set, 
they found about 100 pairs of duplicate fluxes.
The duplicates were from different classes, and only occurred when the
true flux is zero.

Our final, and perhaps most important validation, is that the \SNANA\ simulation has 
been previously used in numerous published measurements of cosmological parameters
where the simulation accurately predicts distributions observed in the data.
Example data-simulation comparisons have been shown for
DES (Fig.~7 in \citealt{Brout2018}),
PS1 (Fig.~4 in \citealt{Jones2017} and Fig.~7 in \citealt{Pantheon}), and
SDSS \& SNLS (Fig.~1-2 in \citealt{Kessler2013}).

While preparing this manuscript, we identified \NMISTAKE\ mistakes in the simulation:
(1) rate discontinuity at $z{=}1$ (\S\ref{sss:details_SNIa}), 
(2) lack of variation in \modelNamebg\ model (\S\ref{sss:91bg_details}),
(3) luminosity function too narrow for {\modelNameSNII}-NMF 
    (\S\ref{sss:SNII_details}), 
(4) included \modelNameSLSN\ models fainter than $-21$~mag (\S\ref{sss:model_SLSN}),
(5) incorrect \modelNameMdwarf\ $b$-dependence in the DDF 
  (footnote~\ref{fn:Mdwarf} in \S\ref{sec:models_source}), 
      and
(6) applied incorrect $\SNRtrue>3$ requirement for the trigger
  (footnote~\ref{fn:SNR} in \S\ref{subsec:trigger_model}).
These mistakes were identical in the training and test sets and therefore
did not cause leakage, and these mistakes did not compromise our 
goals for the Kaggle competition.
However, such mistakes could have led to subtle data/simulation discrepancies
if there had been real data to make such comparisons.

\subsection{ Light Curves from \acro\ Data Set}
\label{subsec:simLC}

\begin{figure}  
      \includegraphics[scale=0.4]{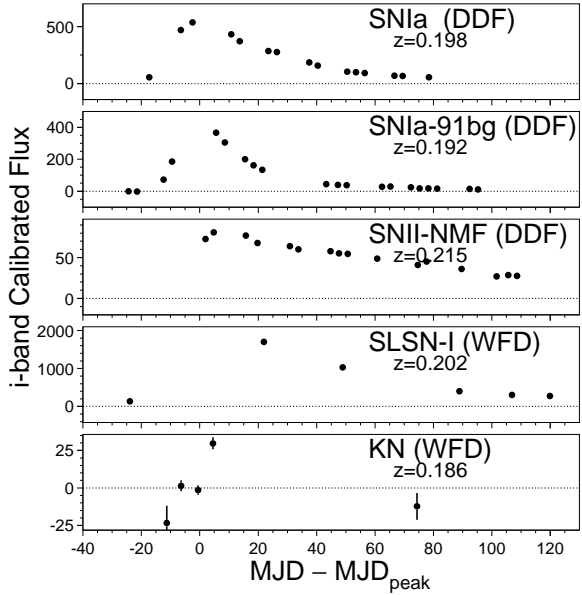}
  \caption{  
   Example \exgal\ light curves in $i$-band. 
   The model name, observing mode (DDF or WFD) and redshift 
   are shown on each panel. 
   The MJD axis is shifted so that zero corresponds to peak bolometric flux.
   The dotted horizontal line through zero is to guide the eye.
   Each light curve was selected with redshift $z{\sim}0.2$
   to visually compare flux and width.
   Note that the WFD light curves (lower two panels) have significantly fewer 
   \obss\ compared with the DDF light curves (upper three panels).
   }
\label{fig:exgal_LC}  \end{figure}

\begin{figure}  
      \includegraphics[scale=0.4]{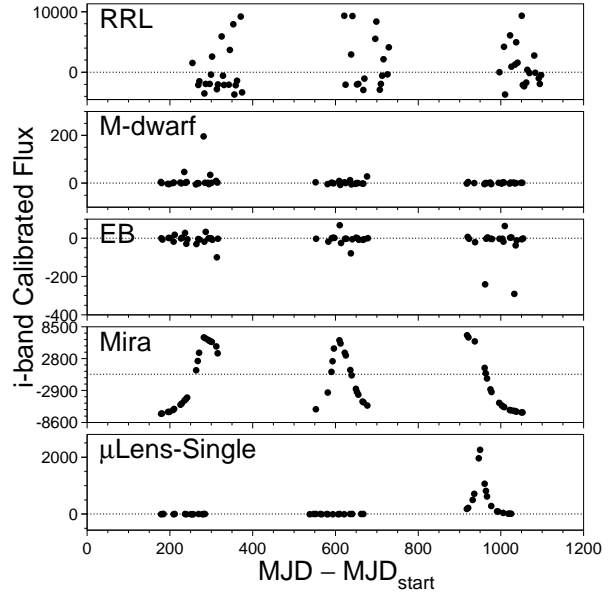}
  \caption{  
   Example \Gal\ light curves in $i$-band (DDF only). 
   The model name is shown on each panel.
   The MJD axis is shifted so that zero is the start of LSST observing.
   The dotted horizontal line through zero is to guide the eye.
   }
\label{fig:Gal_LC}  \end{figure}

A few \exgal\ $i$-band light curves from the \acro\ data set 
are shown in Fig.~\ref{fig:exgal_LC}, spanning a range of 170~days.
The first three panels show different supernova classes in the DDF,
where the different transient time-scales are visually apparent.
The lower 2 panels show \modelNameSLSN\ and \modelNameKN\ light curves 
in the WFD survey.
Although the DDF area is much smaller than WFD, 
note the vastly superior cadence in the DDF.
Negative fluxes are due to small (or zero) source flux  combined with 
Poisson sky fluctuations resulting in smaller sky level in the search image 
compared with the template used for image subtraction.
A few \Gal\ $i$-band light curves from the \acro\ data set (DDF) 
are shown in Fig.~\ref{fig:Gal_LC}, spanning the entire 3 years.
Note that the \modelNameuLensSingle\ model is a transient
with finite duration. 
Large negative fluxes are due to the source flux in the presurvey 
template image being larger than the flux measured in the search image.

\subsection{\acro\ Data Files }
\label{subsec:sim_data}

The \acro\ data files are described on the Kaggle platform
(footnote~\ref{fn:kaggle}), and here is a brief summary.
The metadata for each event include:
integer object identifier, 
sky coordinates (R.A., decl.),
host-galaxy photo-$z$ and its \unc\ (\S\ref{subsec:sim_zphot}),
host-galaxy \spec\ redshift for a small fraction of events (\S\ref{subsec:sim_zspec}),
distance modulus computed from the photo-$z$, and
Galactic extinction estimate, $E(B-V)$ (\S\ref{sss:sim_photons}).
For the training set (\S\ref{subsec:sim_train}), the model class is also provided.
The following information is provided for each {\obs}:
integer object identifier,
Modified Julian Date (MJD),
passband ({\bands}),
measured flux, flux \unc, and
boolean flag for detection (\S\ref{subsec:trigger_model}).

\section{Discussion \& Conclusion}
\label{sec:discuss}

There are two primary products resulting from the \acro\ challenge.
The first product is a set of \NCLASSTOT\ models of transients and variables
within a unified analysis framework, an enormous leap over previous simulations.
These models are publicly available in \citet{PLASTICC_MODEL_LIBS},
and each model is packaged as a separate library so that any 
simulation code can be applied.

The second product is a new set of classification techniques
(R.Hlo\v{z}ek et al. 2019, in preparation).
The winning method was based on augmenting the training set
by degrading well-measured light curves; this increased the training set 
from 8,000 to 270,000 events. Next, the light curves were smoothed
with a Gaussian process method, and 200 features were extracted.
Finally, the features were trained with a machine-learning method
called ``Light Gradient Boosting Machine'' (LightGBM).
Shortly after the end of the challenge, Kaggle participants shared their methods,
codes, and training products. Using different machine-learning methods 
on the augmented training set, participants obtained classification scores 
better than the original winning score, an encouraging sign that 
a combination of methods can significantly improve classification.

To improve classification beyond \acro, 
improvements are needed for both the models and the simulation.
In the spirit of the original community call for models, 
we invite improvements to existing models, and development of
new models that were not part of \acro. 
A critically needed simulation improvement is to associate
\exgal\ transients to host-galaxy properties.
It is also important to replace DES observing properties,
e.g., detection efficiency versus S/N and \spec\ selection vs. peak $i$-band magnitude,
with a characterization of LSST and expected \spec\ follow-up. 
Finally, LSST's difference-imaging pipeline may result in photometric light curves
with artifacts that have not been included in \acro. 
These artifacts include catastrophic outliers, excess noise on bright galaxies,
and saturated \obss.
To test the robustness of classifiers to photometric artifacts,
it is important to characterize the LSST difference-imaging pipeline and 
model this behavior in the \SNANA\ simulation.

The \acro\ challenge has provided a unique opportunity to combine efforts
from a wide range of astronomical communities, 
and also from Kaggle's data challenge community where more than 1000
people participated. The \acro\ data set was designed to reflect our best
understanding of the universe and the LSST instrumental performance.
This effort has resulted in significant improvements in simulation and analysis tools, 
which will be critical to address the scientific challenges of the LSST era.

 \section{ Acknowledgements }
 \label{sec:Ack}
 
This paper has undergone internal review in the LSST Dark Energy Science Collaboration; 
we are grateful to internal reviewers
Rachel Mandelbaum, Simon Krughoff, and Peter Nugent
for their valuable input to this manuscript.
We are also grateful to the anonymous referee for a thorough review.

This work was supported in part by the 
Kavli Institute for Cosmological Physics at the University of Chicago 
through grant NSF PHY-1125897 and an endowment from the 
Kavli Foundation and its founder Fred Kavli.
R.K. is supported by DOE grant DE-AC02-76CH03000,
and NASA grant NNG17PX03C.
This work was completed in part with resources provided by the 
University of Chicago Research Computing Center.
This research at Rutgers University (M.D., S.W.J.) is supported by 
DOE award DE-SC0011636 and NSF award AST-1615455.
D.O.J.\ is supported by a Gordon and Betty Moore Foundation
postdoctoral fellowship at the University of California, Santa Cruz.
S.G acknowledges the support from the FCT project PTDC/FIS-AST/31546/2017.
D.C. acknowledges the John Templeton Foundation New Frontiers Program grant
No. 37426 (University of Chicago) and FP050136-B (Cornell University); 
NSF grant No. 1417132.
A.J.C acknowledges support by the U.S. Department of Energy, Office of
Science, under Award Number DESC-0011635, and partial support from the
Washington Research Foundation and the DIRAC Institute.
S.A.R. and J.D.R.P. acknowledge support from the National Aeronautics and Space
Administration through a NASA EPSCoR Research Infrastructure Development award
administered by the South Carolina Space Grant Consortium.

We thank Kaggle for selecting PLAsTiCC to run under the Research Competition category,
providing expertise and guidance throughout the implementation of this project 
and for funding the main prizes. 
We also thank the entire data science community responsible for solidifying the 
tradition of data challenges and, specially, 
all the Kaggle participants who engaged in \acro. 
\acro\ was funded through LSST Corporation grant award \#2017-03 and 
administered by the University of Toronto.

The DESC acknowledges ongoing support from the Institut National de Physique Nucl\'eaire et de Physique des Particules in France; the Science \& Technology Facilities Council in the United Kingdom; and the Department of Energy, the National Science Foundation, and the LSST Corporation in the United States.  DESC uses resources of the IN2P3 Computing Center (CC-IN2P3--Lyon/Villeurbanne - France) funded by the Centre National de la Recherche Scientifique; the National Energy Research Scientific Computing Center, a DOE Office of Science User Facility supported by the Office of Science of the U.S.\ Department of Energy under Contract No.\ DE-AC02-05CH11231; STFC DiRAC HPC Facilities, funded by UK BIS National E-infrastructure capital grants; and the UK particle physics grid, supported by the GridPP Collaboration.  This work was performed in part under DOE Contract DE-AC02-76SF00515.

We thank Carlton Baugh and the The Institute for Computational Cosmology
for access to their simulated mock catalogs.
This material is based upon work supported in part by the National Science
Foundation through Cooperative Agreement 1258333 managed by the Association
of Universities for Research in Astronomy (AURA), and the Department of
Energy under Contract No. DE-AC0276SF00515 with the SLAC National
Accelerator Laboratory. Additional LSST funding comes from private
donations, grants to universities, and in-kind support from LSSTC
Institutional Members.

\bigskip
{\bf ~~~~Author Contributions are Listed Below}
\smallskip

R.~Kessler:  co-lead project, simulation software, generate \acro\ data set, 
  data curation, three models (SNIa, SNII, SNIbc), writing, 
  project administration, and supervision. 
G.~Narayan:  KN \& microlens models, data curation, validation, 
  Kaggle-participant liaison, editing, project administration, and supervision. 
A.~Avelino: two microlens models and writing. 
E.~Bachelet: single-lens microlensing model and writing.
R.~Biswas: \obs\ library, editing, project administration, and supervision.
D.~F.~Chernoff: string-microlens model and writing. 
A.~Connolly: host-galaxy photo-$z$ model.
M.~Dai: \modelNameSNIax\ model and validation. 
S.~Daniel: three models (AGN, M-dwarf, RR Lyrae) and writing. 
R.~Di~Stefano: two microlens models and writing. 
M.~R.~Drout: SNIbc model and writing. 
L.Galbany: validation, SNIa-91bg \& SNII-NMF models, writing, and editing. 
S.Gonz\'alez-Gait\'an: SNIa-91bg \& SNII-NMF models, writing, and editing. 
M.L.Graham: host-galaxy photo-$z$ model and writing.
R.~Hlo\v{z}ek: co-lead project, funding acquisition, Mira model, writing, editing,
   project administration, and supervision. 
E.~E.~O.~Ishida: Kaggle liaison, validation, editing, 
   project administration, and supervision. 
J.~Guillochon: seven models (IIn, Ibc, SLSN-I, TDE, ILOT, CaRT, PISN). 
S.~W.~Jha: SNIax model and editing. 
D.~O.~Jones: SN1a-91bg model and editing.
K.~Mandel: SNIa model.
A.~O'Grady: Mira Model.
C.~M.~Peters: validation and editing.
J.~R.~Pierel: SNII \& SNIbc models.
K.~A.~Ponder: validation and editing.
A.~Pr\v{s}a: eclipsing binary model and writing.
S.~Rodney: SNII,SNIbc models.
V.~A.~Villar: seven models (IIn, Ibc, SLSN-I, TDE, ILOT, CaRT, PISN) and writing.

\appendix

%

\section{Overview of \mosfit\ Code}
\label{app:mosfit}
To produce SEDs for transient sources we use \mosfit: 
the Modular Open-Source Fitter for Transients, 
a Python-based package which generates Monte Carlo ensembles of semi-analytical, 
one-zone SED models\footnote{\URLMOSFIT} \citep{guillochon2018mosfit}. 
Utilizing a number of potential energy sources (e.g., $^{56}$Ni decay), 
\mosfit\ can simulate SEDs for a number of astrophysical transients across a 
broad range of parameters. 
This is done by segmenting the model into a number of ``modular" components: 
e.g., the input energy source, diffusion, a photosphere, etc.
See Fig.~3 of \citealt{guillochon2018mosfit} 
for example kilonovae and superluminous supernovae models.


\bibliographystyle{apj}
\bibliography{PLASTICC_sims.bib}  


  \end{document}